\documentclass{aastex631}
\usepackage{float}
\usepackage{enumitem}
\usepackage{amsmath, amssymb}
\usepackage{soul}
\usepackage{tabularx}
\usepackage{upgreek}
\usepackage{natbib}
\usepackage{subfigure}
\usepackage{afterpage}
\usepackage{graphicx}
\usepackage{hyperref}

\begin{document}

\title{Morphologies of Bright Complex Fast Radio Bursts with CHIME/FRB Voltage Data}

\author[0000-0001-9855-5781]{Jakob T. Faber}
  \affiliation{Department of Physics, McGill University, 3600 rue University, Montr\'eal, QC H3A 2T8, Canada}
  \affiliation{Trottier Space Institute, McGill University, 3550 rue University, Montr\'eal, QC H3A 2A7, Canada}
  \affiliation{Cahill Center for Astronomy and Astrophysics, MC 249-17 California Institute of Technology, Pasadena CA 91125, USA}
\author[0000-0002-2551-7554]{Daniele ~Michilli}
  \affiliation{MIT Kavli Institute for Astrophysics and Space Research, Massachusetts Institute of Technology, 77 Massachusetts Ave, Cambridge, MA 02139, USA}
  \affiliation{Department of Physics, Massachusetts Institute of Technology, 77 Massachusetts Ave, Cambridge, MA 02139, USA}
\author[0000-0001-7348-6900]{Ryan Mckinven}
  \affiliation{Department of Physics, McGill University, 3600 rue University, Montr\'eal, QC H3A 2T8, Canada}
  \affiliation{Trottier Space Institute, McGill University, 3550 rue University, Montr\'eal, QC H3A 2A7, Canada}
\author[0000-0003-0607-8194]{Jianing Su}
  \affiliation{Department of Physics, McGill University, 3600 rue University, Montr\'eal, QC H3A 2T8, Canada}
  \affiliation{Dunlap Institute for Astronomy \& Astrophysics, University of Toronto, 50 St.~George Street, Toronto, ON M5S 3H4, Canada}
\author[0000-0002-8912-0732]{Aaron B.~Pearlman}
  \affiliation{Department of Physics, McGill University, 3600 rue University, Montr\'eal, QC H3A 2T8, Canada}
  \affiliation{Trottier Space Institute, McGill University, 3550 rue University, Montr\'eal, QC H3A 2A7, Canada}
\author[0000-0003-0510-0740]{Kenzie ~Nimmo}
  \affiliation{MIT Kavli Institute for Astrophysics and Space Research, Massachusetts Institute of Technology, 77 Massachusetts Ave, Cambridge, MA 02139, USA}
  \affiliation{Department of Physics, Massachusetts Institute of Technology, 77 Massachusetts Ave, Cambridge, MA 02139, USA}
\author[0000-0002-7164-9507]{Robert A.~Main}
  \affiliation{Department of Physics, McGill University, 3600 rue University, Montr\'eal, QC H3A 2T8, Canada}
  \affiliation{Trottier Space Institute, McGill University, 3550 rue University, Montr\'eal, QC H3A 2A7, Canada}
\author[0000-0001-9345-0307]{Victoria Kaspi}
  \affiliation{Department of Physics, McGill University, 3600 rue University, Montr\'eal, QC H3A 2T8, Canada}
  \affiliation{Trottier Space Institute, McGill University, 3550 rue University, Montr\'eal, QC H3A 2A7, Canada}
\author[0000-0002-3615-3514]{Mohit Bhardwaj}
  \affiliation{Department of Physics, McGill University, 3600 rue University, Montr\'eal, QC H3A 2T8, Canada}
  \affiliation{Trottier Space Institute, McGill University, 3550 rue University, Montr\'eal, QC H3A 2A7, Canada}
  \affiliation{Department of Physics, Carnegie Mellon University, 5000 Forbes Avenue, Pittsburgh, 15213, PA, USA}
\author[0000-0002-2878-1502]{Shami ~Chatterjee}
  \affiliation{Department of Astronomy and Cornell Center for Astrophysics and Planetary Science, Cornell University, Ithaca NY 14853, USA}
\author[0000-0002-8376-1563]{Alice P.~Curtin}
  \affiliation{Department of Physics, McGill University, 3600 rue University, Montr\'eal, QC H3A 2T8, Canada}
  \affiliation{Trottier Space Institute, McGill University, 3550 rue University, Montr\'eal, QC H3A 2A7, Canada}
\author[0000-0001-7166-6422]{Matt Dobbs}
  \affiliation{Department of Physics, McGill University, 3600 rue University, Montr\'eal, QC H3A 2T8, Canada}
  \affiliation{Trottier Space Institute, McGill University, 3550 rue University, Montr\'eal, QC H3A 2A7, Canada}
\author[0000-0003-3734-8177]{Gwendolyn ~Eadie}
  \affiliation{Dunlap Institute for Astronomy \& Astrophysics, University of Toronto, 50 St.~George Street, Toronto, ON M5S 3H4, Canada}
  \affiliation{Department of Statistical Sciences, University of Toronto, 700 University Ave., Toronto, ON M5G 1Z5, Canada}
\author[0000-0002-3382-9558]{B.~M.~Gaensler}
  \affiliation{Dunlap Institute for Astronomy \& Astrophysics, University of Toronto, 50 St.~George Street, Toronto, ON M5S 3H4, Canada}
  \affiliation{David A.~Dunlap Department of Astronomy \& Astrophysics, University of Toronto, 50 St.~George Street, Toronto, ON M5S 3H4, Canada}
  \affiliation{Present address: Division of Physical and Biological Sciences, University of California Santa Cruz, Santa Cruz, CA 95064, USA}
\author[0000-0003-2739-5869]{Zarif Kader}
  \affiliation{Department of Physics, McGill University, 3600 rue University, Montr\'eal, QC H3A 2T8, Canada}
  \affiliation{Trottier Space Institute, McGill University, 3550 rue University, Montr\'eal, QC H3A 2A7, Canada}
\author[0000-0002-4209-7408]{Calvin Leung}
  \affiliation{MIT Kavli Institute for Astrophysics and Space Research, Massachusetts Institute of Technology, 77 Massachusetts Ave, Cambridge, MA 02139, USA}
  \affiliation{Department of Physics, Massachusetts Institute of Technology, 77 Massachusetts Ave, Cambridge, MA 02139, USA}
\author[0000-0002-4279-6946]{Kiyoshi W.~Masui}
  \affiliation{MIT Kavli Institute for Astrophysics and Space Research, Massachusetts Institute of Technology, 77 Massachusetts Ave, Cambridge, MA 02139, USA}
  \affiliation{Department of Physics, Massachusetts Institute of Technology, 77 Massachusetts Ave, Cambridge, MA 02139, USA}
\author[0000-0002-8897-1973]{Ayush ~Pandhi}
  \affiliation{Dunlap Institute for Astronomy \& Astrophysics, University of Toronto, 50 St.~George Street, Toronto, ON M5S 3H4, Canada}
  \affiliation{David A.~Dunlap Department of Astronomy \& Astrophysics, University of Toronto, 50 St.~George Street, Toronto, ON M5S 3H4, Canada}
\author[0000-0002-9822-8008]{Emily Petroff}
  \affiliation{Perimeter Institute for Theoretical Physics, 31 Caroline Street N, Waterloo, ON N25 2YL, Canada}
\author[0000-0002-4795-697X]{Ziggy Pleunis}
  \affiliation{Dunlap Institute for Astronomy \& Astrophysics, University of Toronto, 50 St.~George Street, Toronto, ON M5S 3H4, Canada}
\author[0000-0001-7694-6650]{Masoud Rafiei-Ravandi}
  \affiliation{Department of Physics, McGill University, 3600 rue University, Montr\'eal, QC H3A 2T8, Canada}
  \affiliation{Trottier Space Institute, McGill University, 3550 rue University, Montr\'eal, QC H3A 2A7, Canada}
\author[0000-0003-3154-3676]{Ketan R.~Sand}
  \affiliation{Department of Physics, McGill University, 3600 rue University, Montr\'eal, QC H3A 2T8, Canada}
  \affiliation{Trottier Space Institute, McGill University, 3550 rue University, Montr\'eal, QC H3A 2A7, Canada}
\author[0000-0002-7374-7119]{Paul Scholz}
  \affiliation{Dunlap Institute for Astronomy \& Astrophysics, University of Toronto, 50 St.~George Street, Toronto, ON M5S 3H4, Canada}
  \affiliation{Department of Physics \& Astronomy, York University}
\author[0000-0002-6823-2073]{Kaitlyn Shin}
  \affiliation{MIT Kavli Institute for Astrophysics and Space Research, Massachusetts Institute of Technology, 77 Massachusetts Ave, Cambridge, MA 02139, USA}
  \affiliation{Department of Physics, Massachusetts Institute of Technology, 77 Massachusetts Ave, Cambridge, MA 02139, USA}
\author[0000-0002-2088-3125]{Kendrick Smith}
  \affiliation{Perimeter Institute for Theoretical Physics, 31 Caroline Street N, Waterloo, ON N25 2YL, Canada}
\author[0000-0001-9784-8670]{Ingrid Stairs}
  \affiliation{Department of Physics and Astronomy, University of British Columbia, 6224 Agricultural Road, Vancouver, BC V6T 1Z1 Canada}
\newcommand{\allacks}{
FRB research at WVU is supported by an NSF grant (2006548, 2018490).
%
%
%
FRB research at UBC is supported by an NSERC Discovery Grant and by the Canadian Institute for Advanced Research. The CHIME/FRB baseband system is funded in part by a Canada Foundation for Innovation John R. Evans Leaders Fund award to IHS.
The Dunlap Institute is funded through an endowment established by the David Dunlap family and the University of Toronto. B.M.G. acknowledges the support of the Natural Sciences and Engineering Research Council of Canada (NSERC) through grant RGPIN-2022-03163, and of the Canada Research Chairs program.
M.B is a McWilliams fellow and an International Astronomical Union Gruber fellow. M.B. also receives support from the McWilliams seed grant.
A.M.C. is supported by an NSERC Doctoral Postgraduate Scholarship. 
M.D. is supported by a CRC Chair, NSERC Discovery Grant, CIFAR, and by the FRQNT Centre de Recherche en Astrophysique du Qu\'ebec (CRAQ).
G.M.E. is supported by an NSERC Discovery Grant RGPIN-2020-04554, and a Canadian Statistical Sciences Institute (CANSSI) Collaborative Research Team Grant.
B.M.G. acknowledges the support of the Natural Sciences and Engineering Research Council of Canada (NSERC) through grant RGPIN-2022-03163, and of the Canada Research Chairs program.
V.M.K. holds the Lorne Trottier Chair in Astrophysics \& Cosmology, a Distinguished James McGill Professorship, and receives support from an NSERC Discovery grant (RGPIN 228738-13), from an R. Howard Webster Foundation Fellowship from CIFAR, and from the FRQNT CRAQ.
C.L. is supported by NASA through the NASA Hubble Fellowship grant HST-HF2-51536.001-A awarded by the Space Telescope Science Institute, which is operated by the Association of Universities for Research in Astronomy, Inc., under NASA contract NAS5-26555.
K.W.M. holds the Adam J. Burgasser Chair in Astrophysics and is supported by an NSF Grant (2008031,  2018490).
A.P. is funded by the NSERC Canada Graduate Scholarships -- Doctoral program.
A.B.P. is a Banting Fellow, a McGill Space Institute~(MSI) Fellow, and a Fonds de Recherche du Quebec -- Nature et Technologies~(FRQNT) postdoctoral fellow.
Z.P. is a Dunlap Fellow.
P.S. was a Dunlap Fellow.
K.S. is supported by the NSF Graduate Research Fellowship Program.
A.P.C is a Vanier Canada Graduate Scholar.
}

\correspondingauthor{Jakob T. Faber}
\email{jfaber@caltech.edu}

\singlespace

\begin{abstract}

We present the discovery of twelve thus far non-repeating fast radio burst (FRB) sources, detected by the Canadian Hydrogen Intensity Mapping Experiment (CHIME) telescope. These sources were selected from a database comprising $\mathcal{O}(10^3)$ CHIME/FRB full-array raw voltage data recordings, based on their exceptionally high brightness and complex morphology. Our study examines the time-frequency characteristics of these bursts, including drifting, microstructure, and periodicities. The events in this sample display a variety of unique drifting phenomenologies that deviate from the linear negative drifting phenomenon seen in many repeating FRBs, and motivate a possible new framework for classifying drifting archetypes. Additionally, we detect microstructure features of duration $\lesssim$ 50 $\mu s$ in seven events, with some as narrow as $\approx$ 7 $\mu s$. We find no evidence of significant periodicities. Furthermore, we report the polarization characteristics of seven events, including their polarization fractions and Faraday rotation measures (RMs). The observed $|\mathrm{RM}|$ values span a wide range of $17.24(2)$ - $328.06(2) \mathrm{~rad~m}^{-2}$, with linear polarization fractions between $0.340(1)$ - $0.946(3)$. The morphological properties of the bursts in our sample appear broadly consistent with predictions from both relativistic shock and magnetospheric models of FRB emission, as well as propagation through discrete ionized plasma structures. We address these models and discuss how they can be tested using our improved understanding of morphological archetypes.

\end{abstract}


\vspace{1cm}
\section{Introduction}\label{sec:intro}

Fast radio bursts \citep[FRBs;][]{Lorimer2007} are a class of highly luminous, predominantly extragalactic radio transients with durations of a few nanoseconds to seconds \citep{Petroff2019, Cordes2019, Majid2021, Nimmo2022, chime2022}. 
The origin of FRBs is unknown. The 
Canadian Hydrogen Intensity Mapping Experiment Fast Radio Burst (CHIME/FRB) Project \citep{chime2021} has
detected
$\mathcal{O}(10^{3})$ FRB sources of a variety of signal strengths, morphologies and durations, of which approximately 3\% show repeating behavior \citep{chime2018, chime2019, fonseca2020, chime2020b, Pleunis2021, chime2023}. 

Many models have been proposed to explain the origin of FRBs. While initially one-off FRBs appeared to suggest cataclysmic progenitor channels such as compact object mergers \citep{Kashiyama2013, Mingarelli2015, Keane2016}, or collapsing neutron stars \citep{Falcke2014}, the discovery of repeating FRBs argued for longer-lived progenitors, such as magnetars \citep[e.g.][]{lyubarsky2014,Beloborodov2017,Margalit2020} 
or giant pulses from conventional radio pulsars \citep{Lyutikov2016}. 
More exotic source models that have been proposed include cosmic strings and primordial black holes \citep[see][for a comprehensive review of FRB source models]{Platts2018}. While the aforementioned models succeed in accounting for some observed FRB properties, none explain them all, and coincident multi-wavelength \citep[e.g., gamma and X-ray bursts;][]{Scholz2017, Scholz2020, Pearlman2023} or multi-messenger emissions \citep[e.g., neutrinos;][]{Aartsen2020} that could in principle constrain such models remain elusive.


Magnetars have shown great promise as source candidates for FRBs.
Young magnetars embedded in their birth supernova remnants or wind nebulae can, for instance, possess the plasma environments for synchrotron maser emission via relativistic shocks \citep{lyubarsky2014, Beloborodov2017, Metzger2019}. These models generally account for some of the observed features in repeating FRBs, such as their luminosities, time scales and spectra \citep{lyubarsky2014, Masui2015, Margalit2018, Beloborodov2019, Wadiasingh2019, Margalit2020, Lyubarsky2020}.
Moreover, the detection of $\sim$MJy radio bursts from the Galactic magnetar SGR 1935+2154 \citep{chime2020b,Bochenek2020, Kirsten2020} lends strong supporting evidence for the magnetar model.
The detection of periodic activity windows in some repeating FRBs has been used to claim that FRBs may originate from rotating or precessing magnetars \citep[e.g.][]{chime2020,Rajwade2020,Cruces2020,Levin2020,Zanazzi2020}, sparking numerous multi-wavelength observational campaigns \citep{Scholz2016, Scholz2020, Pearlman2023}. 
CHIME/FRB detected a sub-second periodicity in a single event, which favors models that invoke magnetospheric emission from a neutron star \citep{chime2022}.

Another major advancement in the field is the publication of the first large sample of FRBs observed in a single survey with uniform selection effects by CHIME/FRB \citep{chime2021}. The catalog contains 536 FRBs detected between 2018 July 25 and 2019 July 1, including 62 bursts from 18 previously reported repeating sources, enabling detailed studies of the FRB population and its properties. 
Differences in morphology and spectra between repeating FRBs and apparent non-repeaters,
previously hinted at in more limited samples \citep{Scholz2016, chime2019, Hashimoto2020}, were made
clear by the large numbers of events in the catalog \citep{Pleunis2021}, suggesting potentially distinct emission mechanisms or circumburst environments. 

With the advent of high-time resolution surveys like CRAFT \citep{Macquart2010, Cho2020}, and the baseband raw-voltage storing capabilities of radio telescopes including CHIME/FRB, FRBs have been successfully resolved down to micro- and even nanosecond timescales \citep{Majid2021, Nimmo2021, Nimmo2022, Snelders2023}. Detections of FRB substructure at and below the microsecond scale enable powerful constraints on both emission and propagation physics by resolving subtle time-frequency variations. Time-frequency variations have been characterized in detail for a variety of repeating sources, including FRB 20121102A \citep{Hessels2019, Platts2021}, FRB 20180916B \citep{Sand2022, Sand2023}, FRB 20180301A \citep{Kumar2023}, and others. When such variations are bright, it may be possible 
to distinguish between ``propagation'' effects imposed by intervening plasmas along the line of sight (e.g., time delays due to lensing, or scattering tails) and ``intrinsic'' effects that arise from the emission mechanism itself (e.g., nanosecond-duration bursts from magnetic reconnection or beam-driven instabilities). 

In this paper, we present a high-time resolution analysis of the time-frequency properties of twelve thus far non-repeating FRBs detected by CHIME/FRB, one of which appears in the first CHIME/FRB catalog \citep[FRB 20190425A; see][]{chime2021}. We also report their polarization properties. The FRBs in our sample exhibit morphologies of compelling complexity on $\mu$s timescales with high signal-to-noise (S/N). We leverage the diversity of these morphological characteristics to evaluate a number of FRB source models, with an emphasis on magnetospheric and relativistic shock emission scenarios from magnetars. We also discuss the role that propagation effects from interstellar plasma structures may have in shaping their morphologies.

In \S\protect\ref{sec:observations} we describe the CHIME/FRB baseband analysis system and present the full burst sample.
In \S\protect\ref{sec:analysis} we describe several morphological archetypes present in the sample, particularly as they relate to time-frequency drifting. We also present a new series of fitting techniques to characterize both drifting and microstructure in complex FRBs (also described in \S\protect\ref{apA} and \S\protect\ref{apB}), and search for periodicities.
In \S\protect\ref{sec:discussion}, we explore the implications of these possible new archetypes with respect to FRB emission models and propagation effects, highlighting the relevance to relativistic shock and magnetospheric scenarios, as well as plasma lensing. 
Finally, we summarize and draw conclusions
in \S\protect\ref{sec:conclusion}, and discuss how our sample affects our understanding of non-repeating FRBs. Polarization measurements are reported in \S\protect\ref{apC:pol}.

\section{Observations \& Burst Sample}\label{sec:observations}

\subsection{The CHIME/FRB Real-Time Detection Pipeline and Baseband Analysis System}

Investigations into the underlying physics of FRB emission and propagation have historically been limited by the unavailability of baseband raw voltage data. CHIME/FRB overcomes this limitation by way of a 
real-time detection and analysis pipeline that 
records coherent electric field data measured by the full array (baseband raw voltages) upon a detection trigger \citep{Michilli2021}. Combined with the high detection rate of CHIME/FRB, this provides a wealth of high-resolution data (spectral and temporal) with full-Stokes parameters. 

Located at the Dominion Radio Astrophysical Observatory near Penticton, BC, the CHIME\footnote{See \url{www.chime-experiment.ca}} telescope consists
of four 100-m $\times$ 20-m cylindrical reflectors with North-South orientations, each bearing 256 dual-polarization feeds along the focal line that operate between 400 and 800 MHz. 

The raw voltages measured by the telescope are amplified, digitized, and channelized by an FPGA-based F-engine. Data from the F-engine are sent to the GPU-based X-engine where 1024 digital sky beams are formed \citep{ng2017}, the data from which are sampled at 0.983 ms time resolution with 16k frequency channels. See \citet{chimeoverview} for details.

The CHIME/FRB detection and analysis pipeline operates in two stages:

\begin{enumerate}[label = \textnormal{\Roman*.}]
    \item \underline{\textit{Real-time Detection}:} The first stage is the real-time detection pipeline, which runs on a dedicated cluster of 128 nodes and searches for short-duration, dispersed peaks in a total intensity data stream across 1024 independently formed beams. Upon detection of an FRB, the data are buffered and stored.
    
    The F-engine uses a 4-tap polyphase filter bank (PFB) to produce a spectrum with 1024 channels (each 390 kHz wide) at a time resolution of 2.56 $\mu$s. The baseband data are quantized with 8-bit accuracy, resulting in a data rate of 6.5 Tb/s. A memory buffer allows the storage of 35.5 seconds of baseband data at a given time. From the moment a signal arrives at the telescope, the real-time pipeline can process the event and trigger a baseband dump in about 14 seconds, leaving a usable data buffer of about 20 seconds. For a detailed overview of this first stage, refer to \citet{chime2018}. 

    \item \underline{\textit{Baseband Processing}:} The second stage is the baseband processing pipeline, which is designed to be highly scalable. The pipeline can run automatically on new events that are identified by the real-time detection pipeline, or be manually triggered to process specific events.
    
    The baseband system is configured to store approximately 100 ms of data around an FRB detection. This allows time-frequency features and propagation effects such as scattering and scintillation to be resolved. A grid of overlapping beams is formed around the initial position of the FRB candidate (i.e., beamforming), and the signal intensity is mapped and fitted against the telescope response to obtain a refined position with subarcminute precision. For a complete description of this second stage, refer to \citet{Michilli2021}.
    
\end{enumerate}

Finally, coherent dedispersion is performed to remove the dispersive effects induced by free electrons along the line of sight. This is done by applying frequency-dependent phase correction to each frequency channel, removing intrachannel smearing and preserving the temporal resolution of the signal. The phase-preserving nature of coherent dedispersion also enables the reconstruction of the polarization state of the signal. By combining the two orthogonal linear polarizations measured by each antenna, we can quantify the rotation of the polarization angle due to the interaction of the electromagnetic wave with a magnetized plasma along the line of sight, or Faraday rotation measure (RM), defined as:

\begin{equation}\label{eq:RM}
\mathrm{RM} = 8.1 \times 10^5 \int_{0}^{s}\left(\frac{n_e\left(s\right)}{\mathrm{cm}^{-3}}\right)\left(\frac{B_{\|}\left(s\right)}{\mathrm{G}}\right)\left(\frac{ds}{\mathrm{pc}}\right) = \frac{d \psi}{d \lambda^2} \quad\left[\mathrm{rad} ~\mathrm{m}^{-2}\right]
\end{equation}
where $B_{\|}$ and $n_e$ are the parallel component of the magnetic field with respect to the electromagnetic field and free electron number density along a line of sight element $ds$, $\psi$ is the polarization angle, and $\lambda$ is the observing wavelength. 
Once de-rotated, the intrinsic polarization angle (PA) of the source can be obtained. Since CHIME/FRB does not calibrate for polarization, however, we can only measure zero-mean-relative deviations in the PA. We can also measure the polarization fractions, which are the ratios of the linearly and circularly polarized intensities to the total intensity of the signal. For a more detailed description of the CHIME/FRB polarization pipeline, see \citet{Mckinven2021}.

\subsection{Burst Sample}\label{subsec:burstsamp}

\begin{figure}[h!]
  \centering
  \subfigure{\includegraphics[height=0.34\textwidth, width=0.24\textwidth]{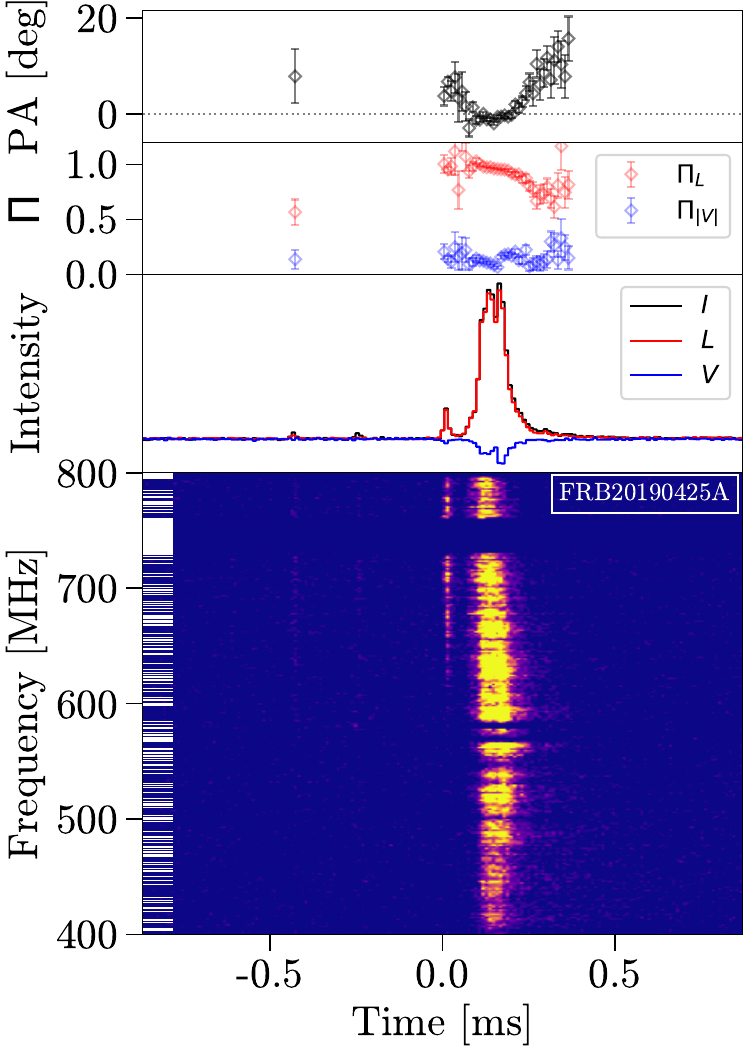}}
  \subfigure{\includegraphics[height=0.34\textwidth, width=0.24\textwidth]{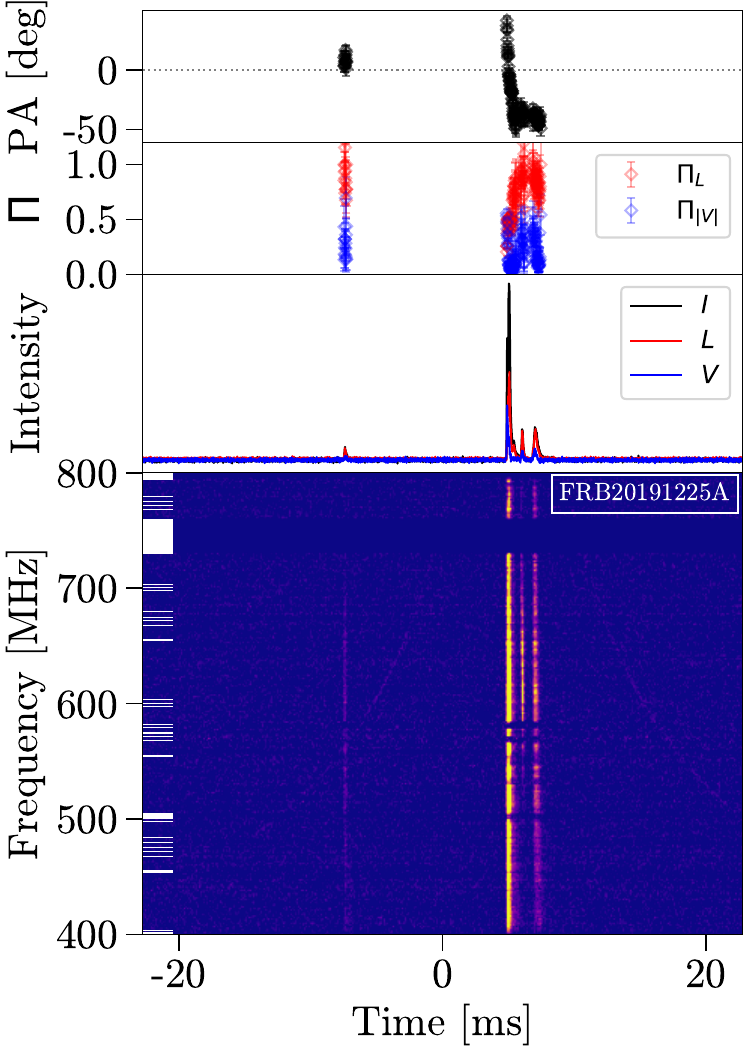}}
  \subfigure{\includegraphics[height=0.34\textwidth, width=0.24\textwidth]{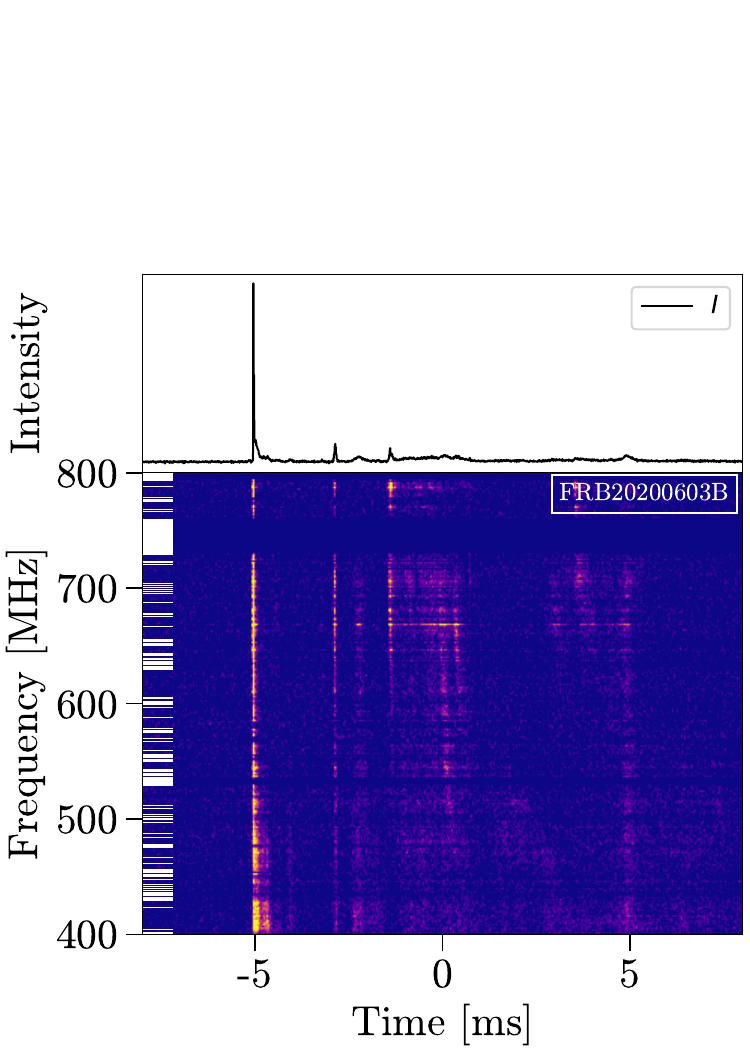}}
  \subfigure{\includegraphics[height=0.34\textwidth, width=0.24\textwidth]{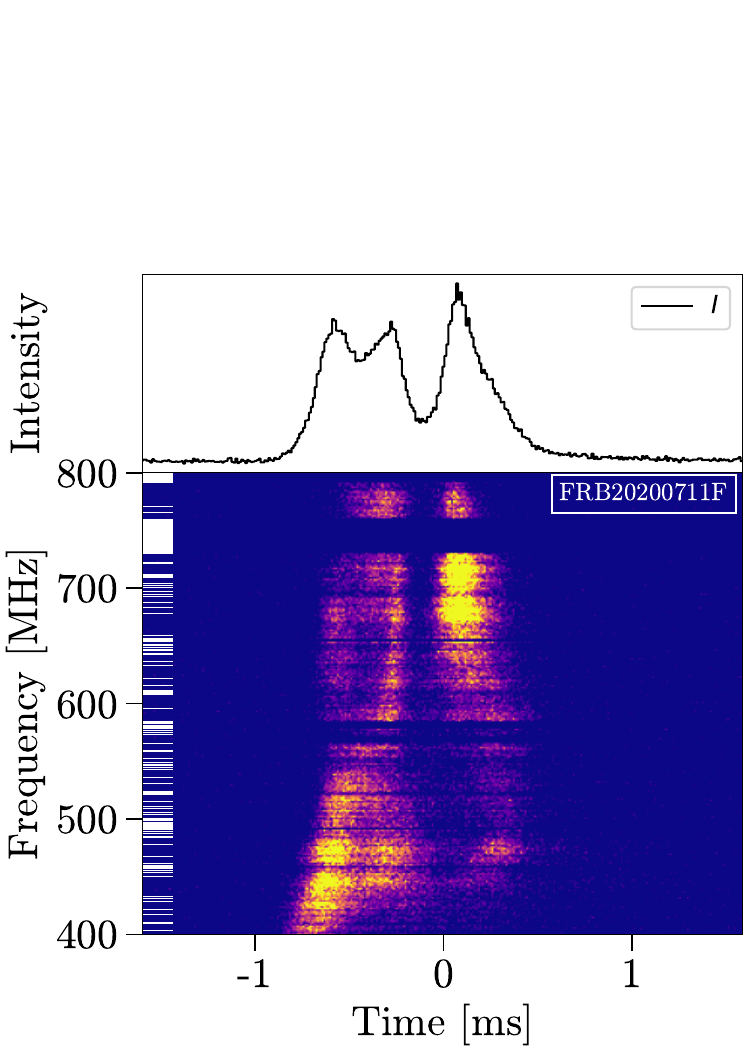}}
  \subfigure{\includegraphics[height=0.34\textwidth, width=0.24\textwidth]{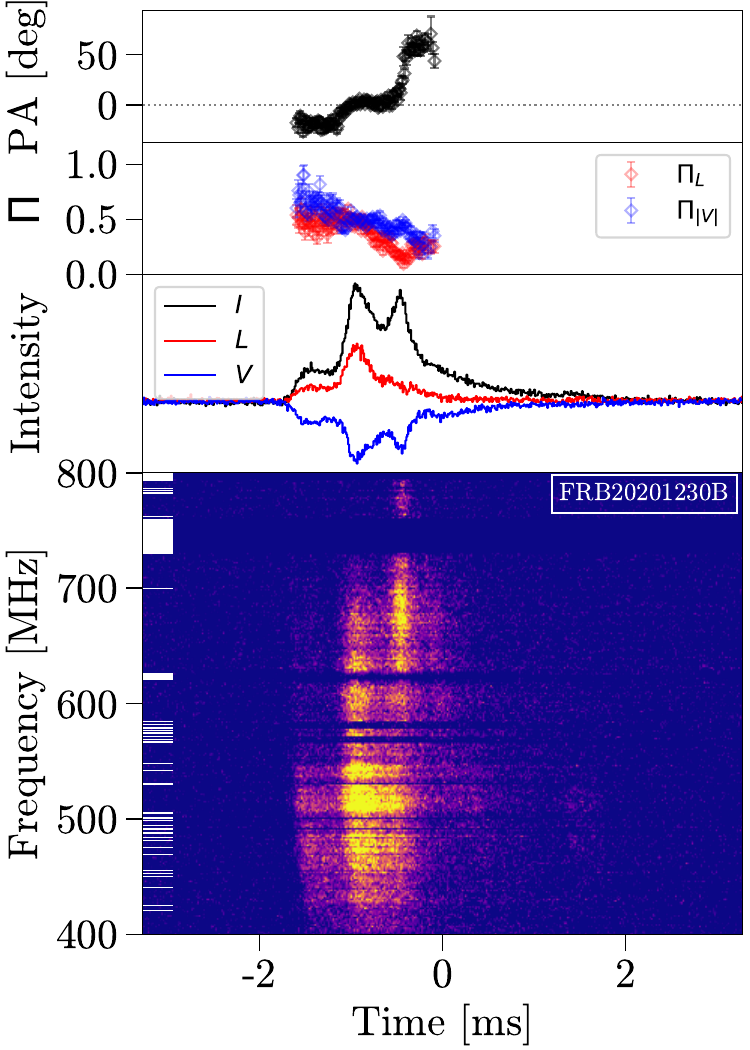}}
  \subfigure{\includegraphics[height=0.34\textwidth, width=0.24\textwidth]{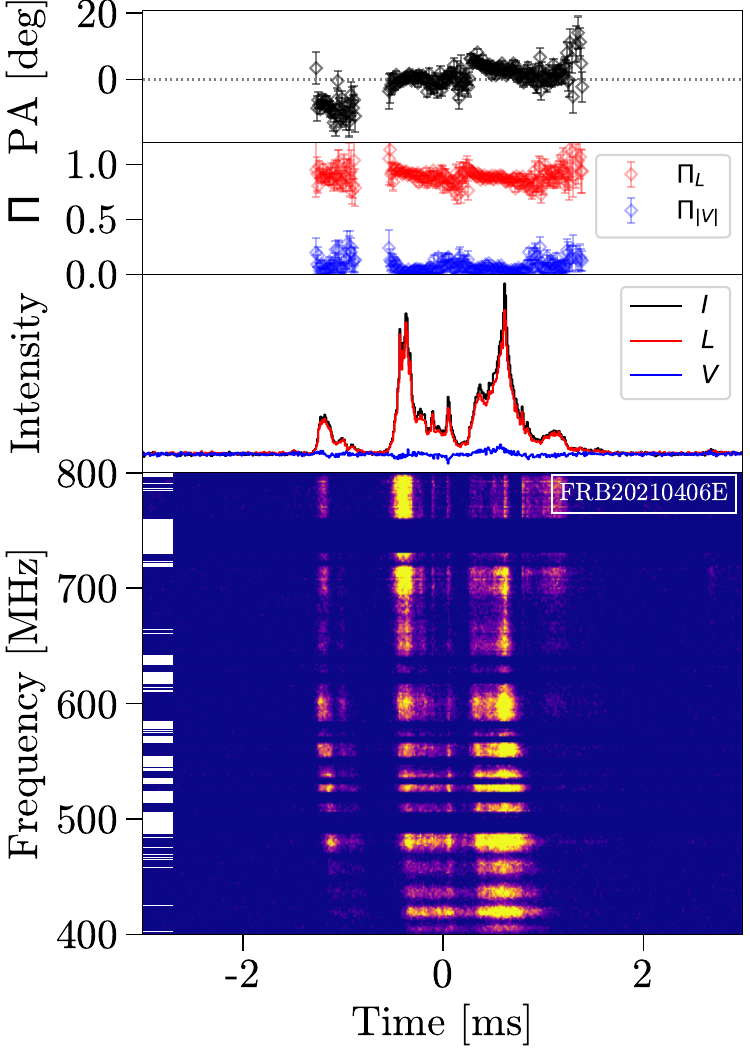}}
  \subfigure{\includegraphics[height=0.34\textwidth, width=0.24\textwidth]{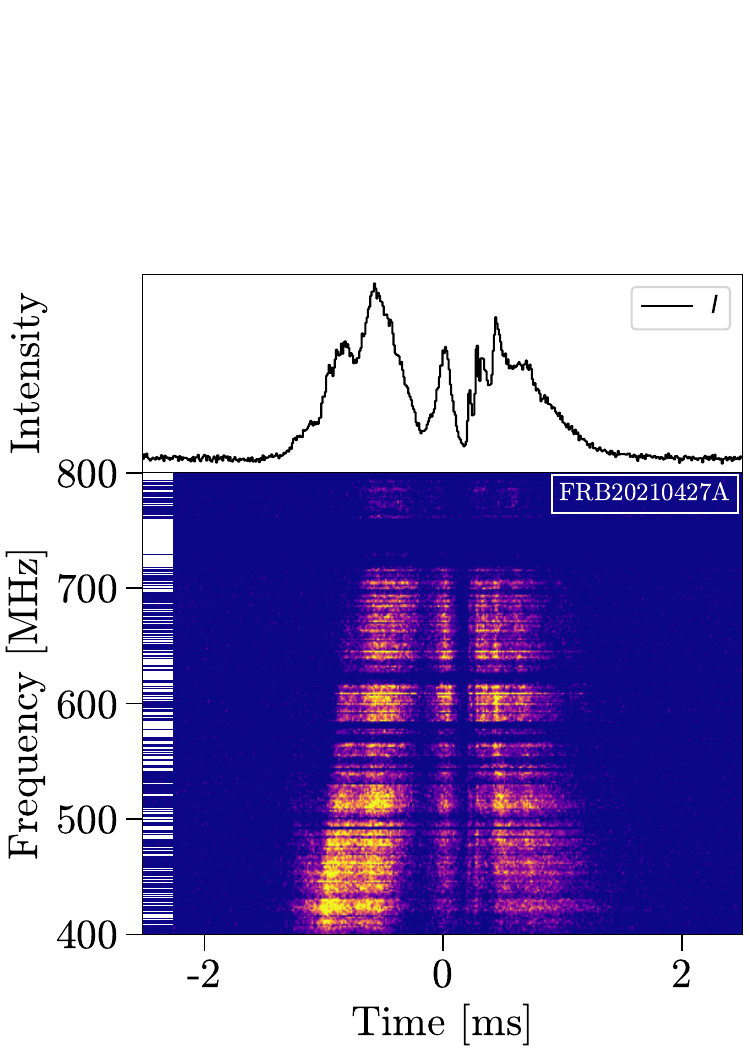}}
  \subfigure{\includegraphics[height=0.34\textwidth, width=0.24\textwidth]{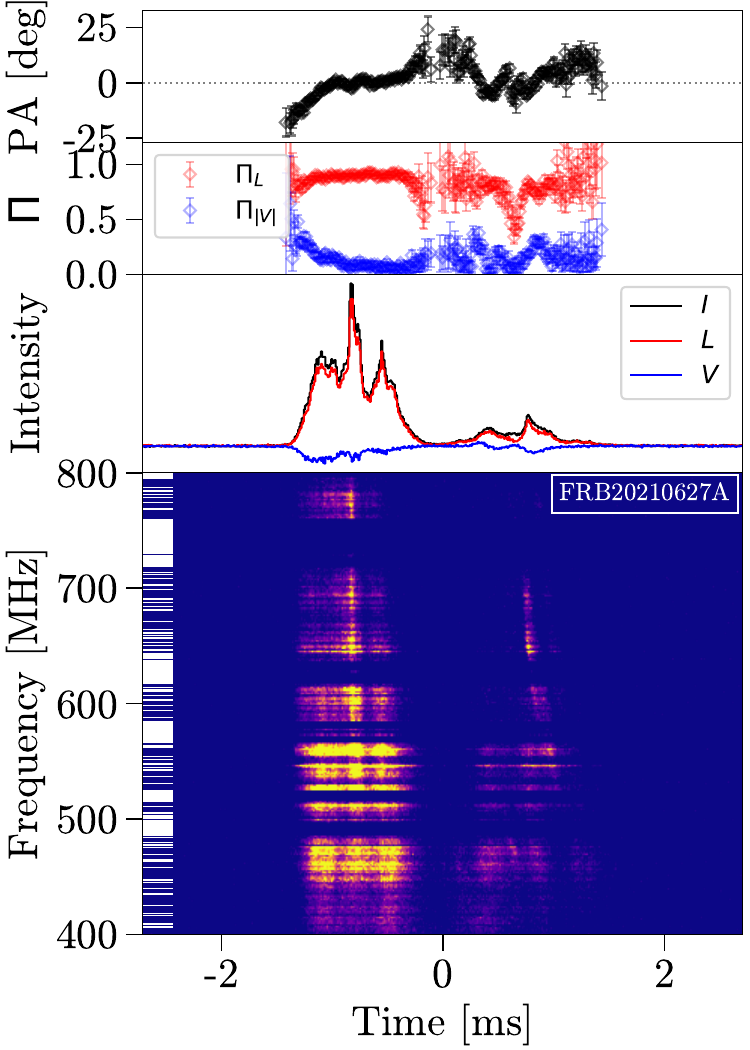}}
  \subfigure{\includegraphics[height=0.34\textwidth, width=0.24\textwidth]{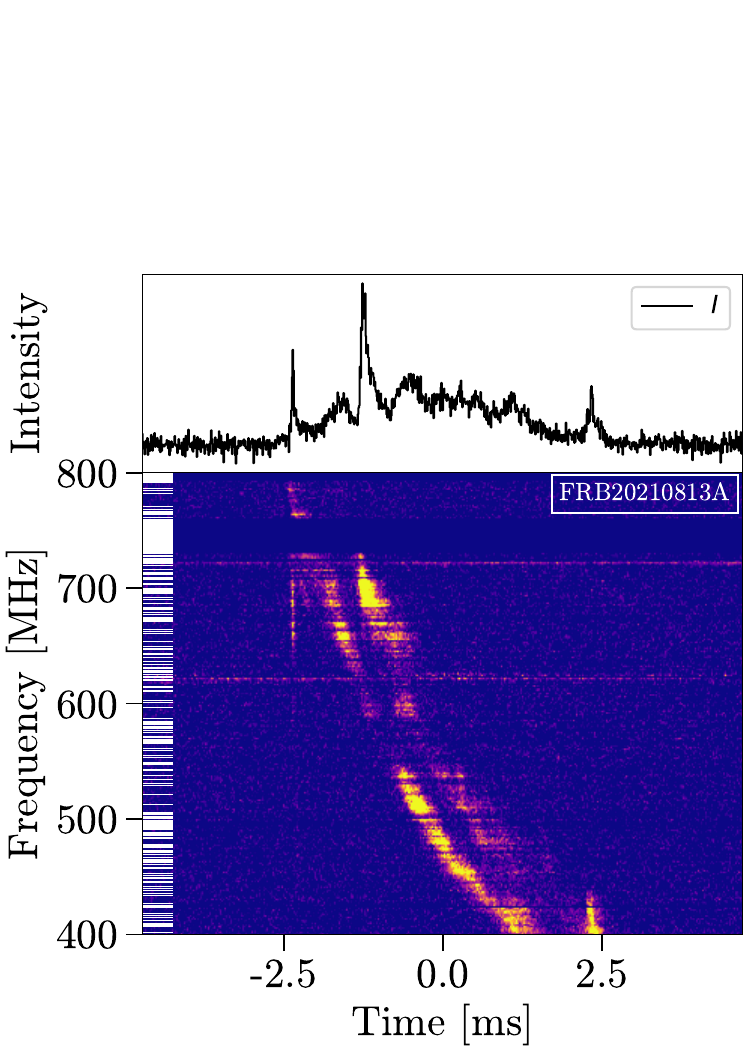}}
  \subfigure{\includegraphics[height=0.34\textwidth, width=0.24\textwidth]{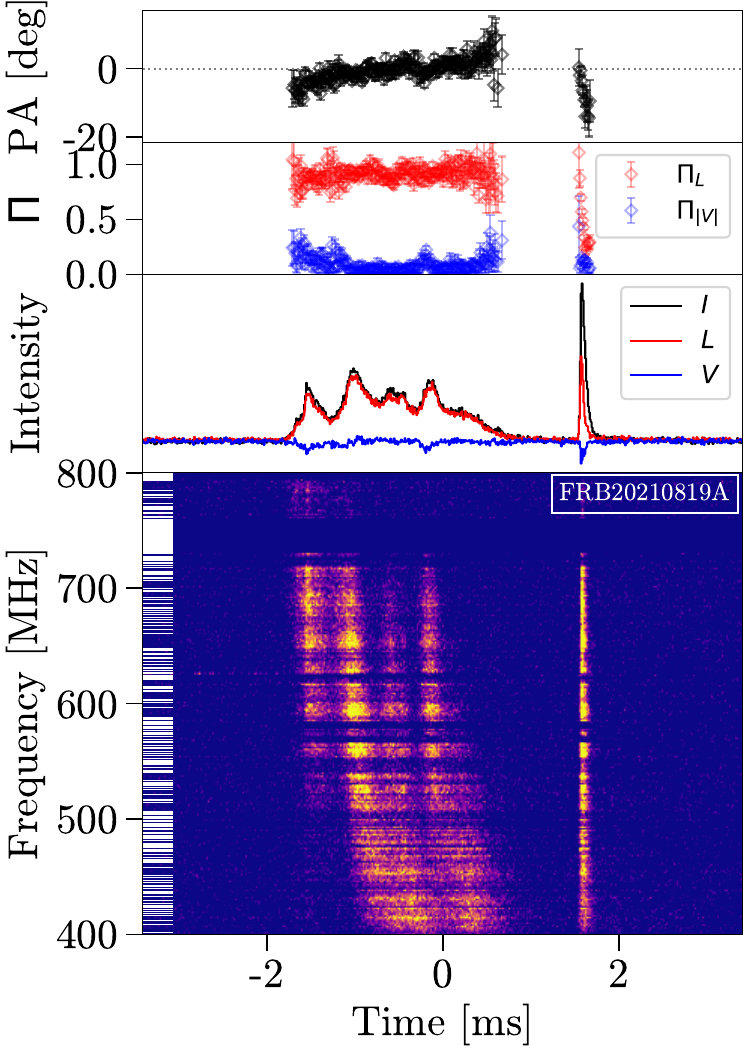}}
  \subfigure{\includegraphics[height=0.34\textwidth, width=0.24\textwidth]{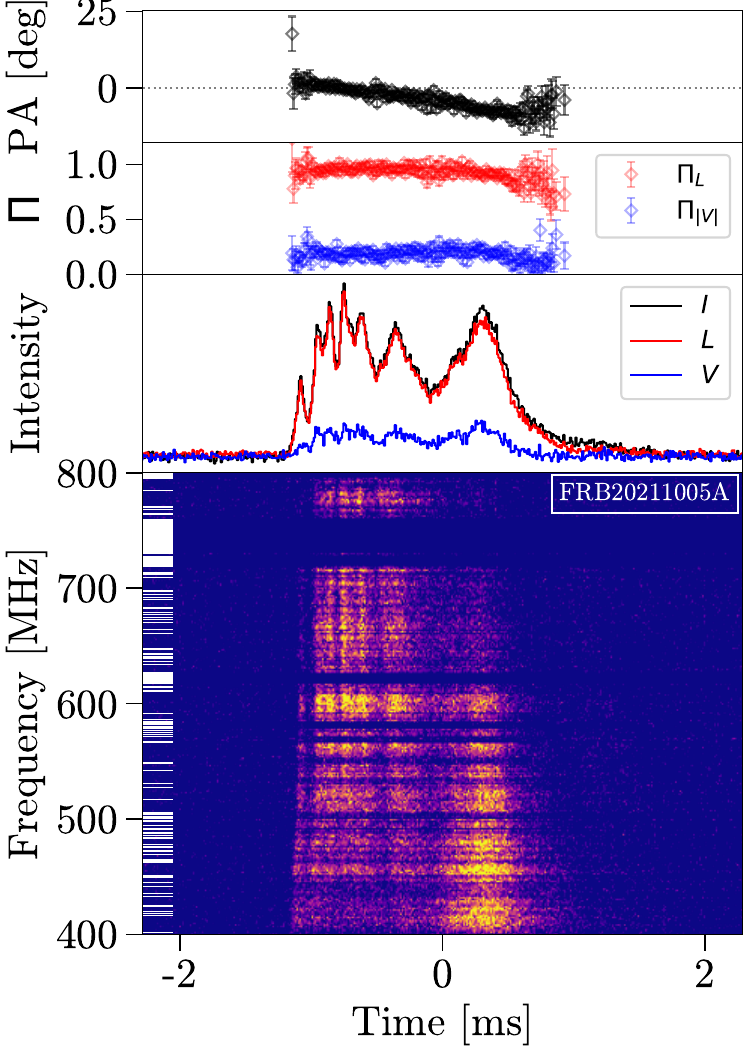}}
  \subfigure{\includegraphics[height=0.34\textwidth, width=0.24\textwidth]{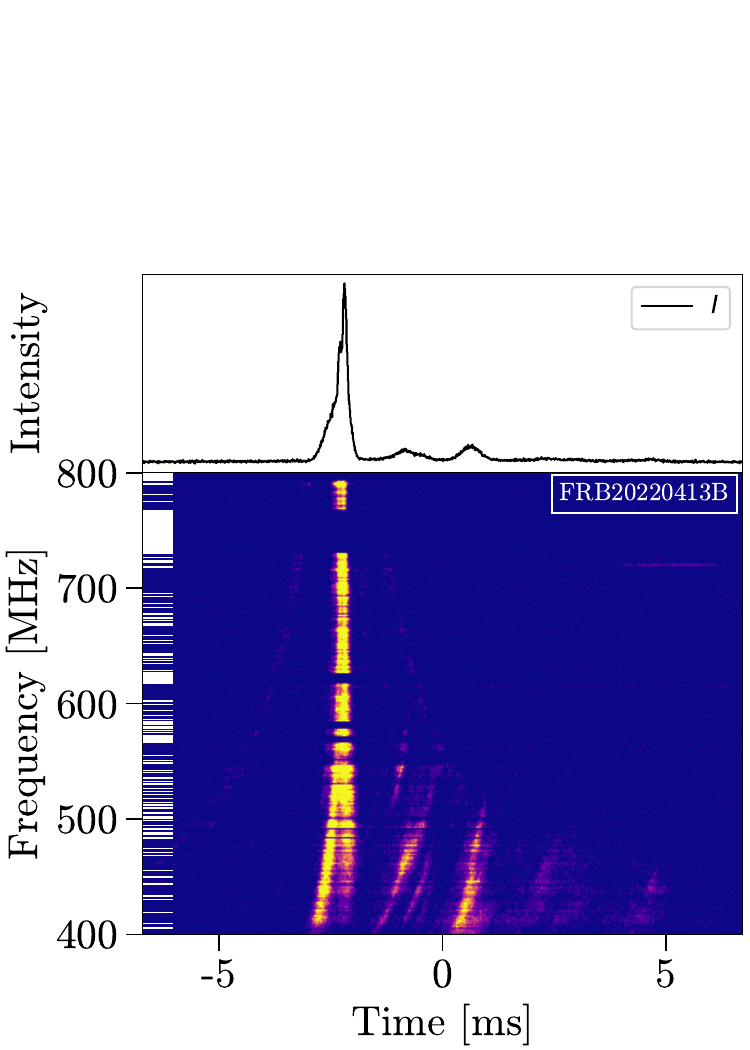}}
\caption{Dynamic spectra for the 12 bursts reported on in this work. In each panel, the bottom sub-panel shows signal intensity as a function of time and frequency. The second sub-panel from the bottom shows the frequency-integrated burst profile (timeseries). Each burst has been coherently dedispersed to a DM measured by \texttt{DM\_phase}. \textit{4-paneled figures}: the upper two sub-panels show the cumulative polarization position angle (PA) in degrees relative to a zero-mean $(\mathrm{PA} - \langle$PA$\rangle)$ and the fractional linear (\textit{L}) and circular (\textit{V}) degrees of polarization, respectively (see \S\protect\ref{apC:pol}). We omit polarization data for FRBs 20200603B, 20200711F, 20210427A, 20210813A, and 20220413B, as we were unable to confidently constrain rotation measures (RMs) due to spurious $QU$-fits. The bursts show a variety of complex morphologies, discussed in \S\protect\ref{sec:analysis}.}
  \label{fig:burstfig}
\end{figure}

The bursts in our sample are shown in Figure \ref{fig:burstfig}. These bursts were selected based on their brightness (S/N $\gtrsim$ 30 in baseband) and unusual morphological characteristics that differ noticeably from those currently observed in most FRBs --- including, but not limited to atypical time-frequency drifting, large numbers of sub-bursts, and time-variable properties. Figure \ref{fig:burstfig} shows both dynamic spectra as well as full polarimetric data for select bursts, as described in detail in \S\protect\ref{sec:analysis}.

\begin{table*}[h!]
\caption{
Burst properties, including: baseband localizations; signal-to-noise ratios (S/N), coherent power-maximizing dispersion measures (DM) obtained using \texttt{DM\_phase}; excess dispersion measures ($\langle\mathrm{DM}_{\mathrm{excess}}\rangle$) inferred from the NE2001 Galactic electron density model \protect\citep{Cordes2002}; and full burst durations ($\Delta t$).}   
\label{table:localizations}     
\footnotesize 
\begin{center}
\begin{tabular*}{\textwidth}{c c c c c c c c c c}         
\hline\hline 
{TNS Name} & {MJD} & RA (J2000) & $\sigma_\text{RA} (\arcsec)$ & Dec (J2000) & $\sigma_\text{Dec} (\arcsec)$ & S/N & DM (pc cm$^{-3}$) & $\mathrm{DM}_{\mathrm{ex}}$ (pc cm$^{-3}$) & {$\Delta t$ (ms)} \\ 
\hline
FRB 20190425A & 58598.44993 & 17$^h$02$^m$41$^s$ & 11 & +21\degr34\arcmin34\arcsec & 11 & 57.6 & 128.1279(3) & 79.4  & 0.65(1)                 \\
FRB 20191225A       & 58842.69064 & 15$^h$28$^m$37$^s$ & 11 & +85\degr29\arcmin32\arcsec & 11 & 65.1 & 683.9113(1) & 634.5 & 14.24(5)                       \\
FRB 20200603B & 59003.05128 & 10$^h$09$^m$37$^s$ & 24 & +71\degr34\arcmin48\arcsec & 13 & 31.8 & 295.0828(4) & 253.9 & 10.23(4)                                     \\
FRB 20200711F & 59041.03444 & 12$^h$10$^m$41$^s$ & 11 & +48\degr23\arcmin48\arcsec & 11
& 175.1 & 527.6773(3) & 500.7 & 1.25(2)  \\
FRB 20201230B & 59213.84582 & 19$^h$00$^m$44$^s$ & 23 & +26\degr01\arcmin15\arcsec & 12
& 101.1 & 256.1293(4) & 95.0 & 2.93(3)                           \\
FRB 20210406E & 59310.09213 & 19$^h$08$^m$26$^s$ & 23 & +71\degr20\arcmin16\arcsec & 12
& 100.7 & 355.2626(3) & 297.1 & 3.64(2)                                    \\
FRB 20210427A & 59331.58050 & 20$^h$30$^m$33$^s$ & 23 & +79\degr15\arcmin54\arcsec & 11
& 88.8 & 268.4785(4) & 204.0 & 2.13(1)                                    \\
FRB 20210627A & 59392.57575 & 00$^h$13$^m$28$^s$ & 23 & +00\degr24\arcmin43\arcsec & 11
& 85.7 & 299.158(4) & 267.5 & 2.32(2)                \\
FRB 20210813A & 59439.63073 & 04$^h$31$^m$15$^s$ & 24 & +39\degr55\arcmin57\arcsec & 12
& 61.9 & 399.264(7) & 237.0 & 2.95(9)                                     \\
FRB 20210819A & 59445.91005 & 11$^h$44$^m$03$^s$ & 23 & +30\degr00\arcmin40\arcsec & 12
& 124.9 & 362.142(2) & 342.0 & 3.46(3)                                            \\
FRB 20211005A & 59492.94871 & 15$^h$44$^m$01$^s$ & 23 & +20\degr43\arcmin26\arcsec & 12
& 73.7 & 226.1078(3) & 198.4 & 1.85(7)                                     \\
FRB 20220413B & 59682.48227 & 16$^h$49$^m$03$^s$ & 23 & +66\degr58\arcmin56\arcsec & 11
& 29.0 & 115.723(2) & 74.4 & 7.54(8)                                         \\
\hline\hline \\
\end{tabular*}
\end{center}
\label{table:burstprop}
\end{table*}

\section{Analysis \& Results}\label{sec:analysis}

\begin{figure}[ht!]
    \centering
    \includegraphics[width = 0.45\textwidth]{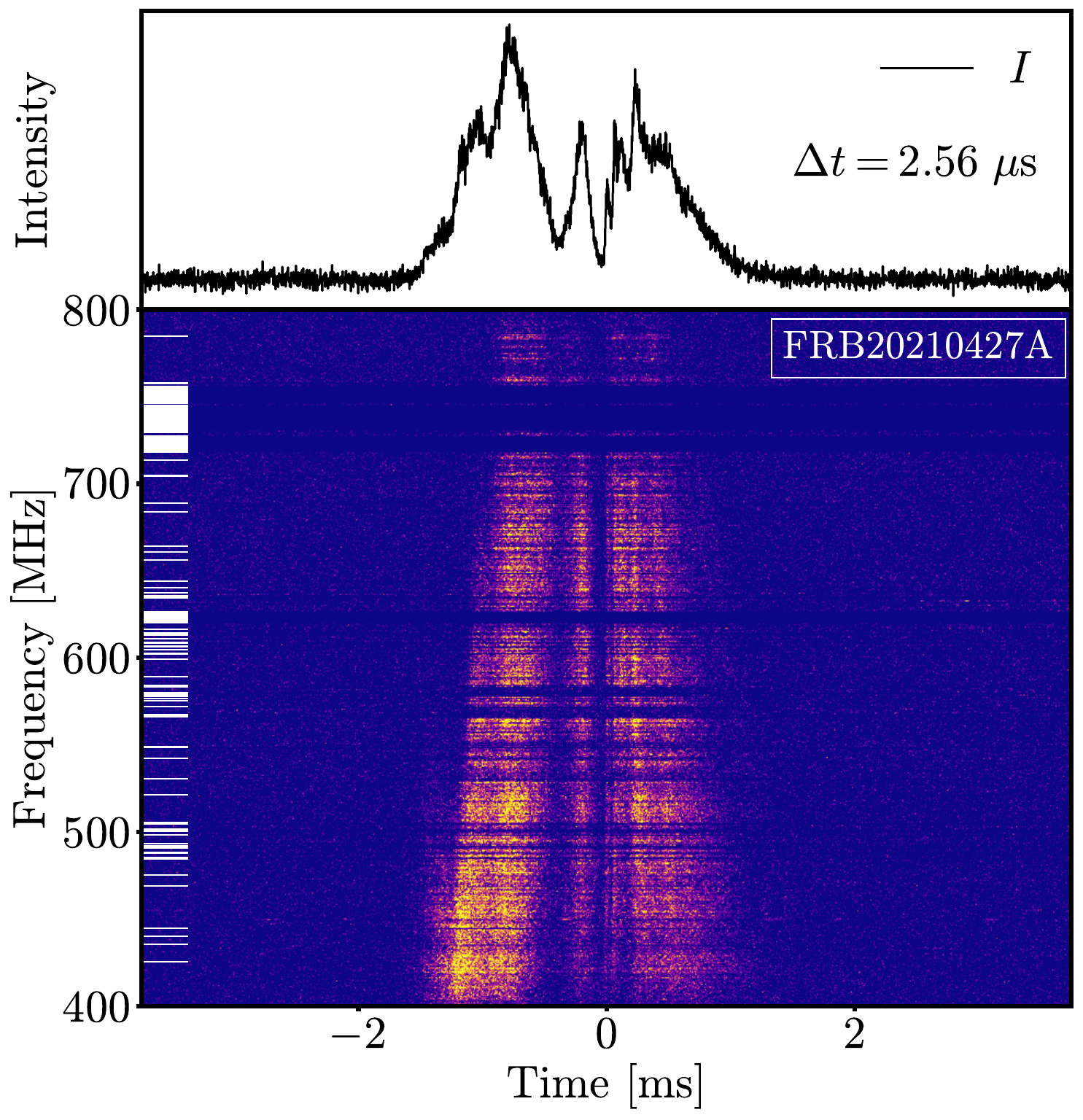}
    \includegraphics[width = 0.45\textwidth]{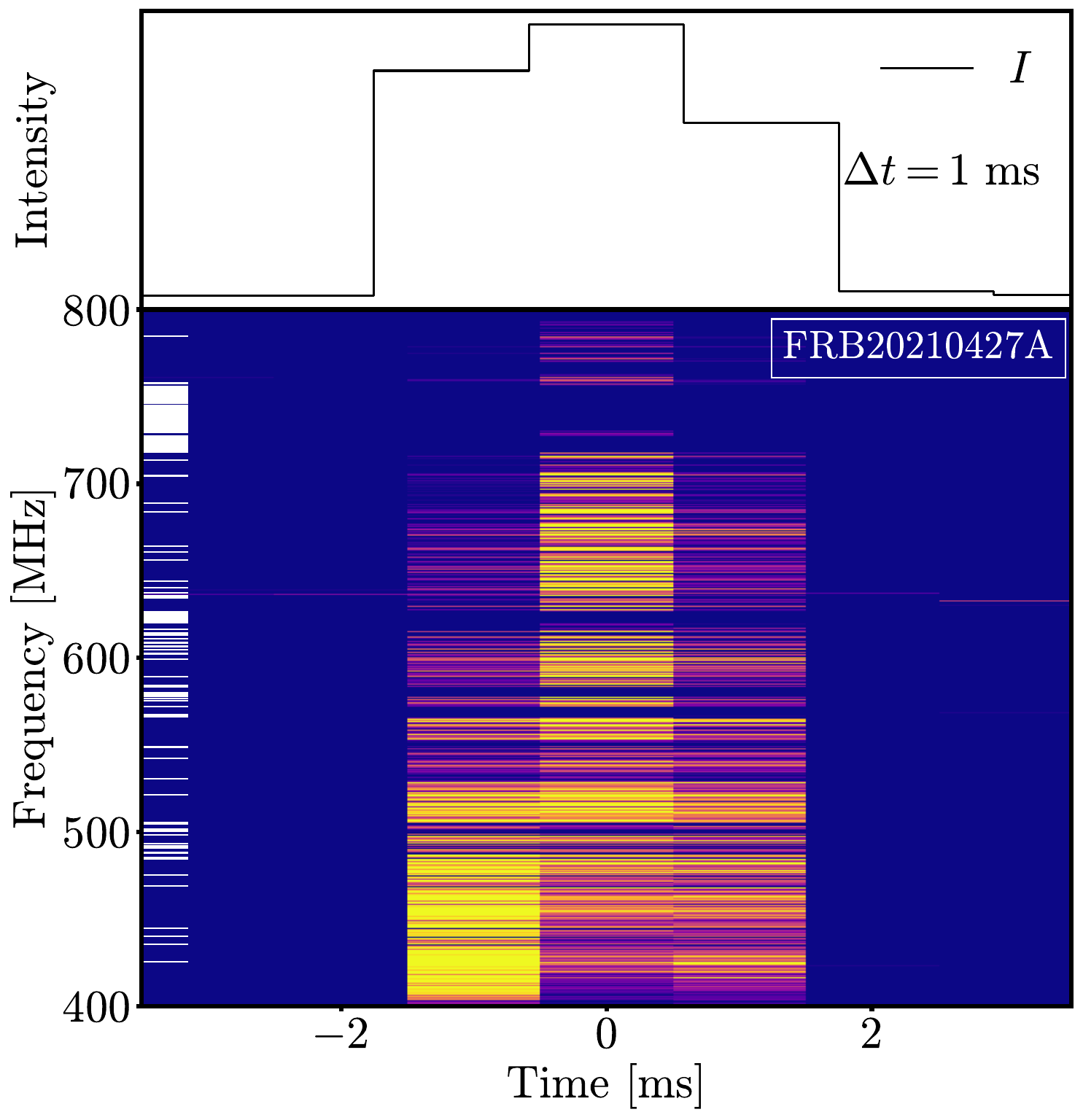}
    \caption{A comparison of the full $2.56~ \mu s$-resolution dynamic spectrum of FRB 20210427A with the same dynamic spectrum downsampled to 1 ms \protect\citep[the resolution of intensity data stored by the CHIME/FRB backend, see][]{chime2021}, reducing the resolvable sub-bursts to a simple component.}
    \label{fig:highreslores}
\end{figure}

Here we present various properties of the twelve bursts shown in Figure \ref{fig:burstfig}. 
Figure \ref{fig:highreslores} shows the dramatic difference in the appearance of a burst at the nominal time- and frequency-resolution of the CHIME/FRB search pipeline versus that available via baseband raw voltage data.

All DM measurements quoted in this work were obtained using \texttt{DM\_phase}\footnote{\url{https://github.com/danielemichilli/DM\_phase}}, which searches for the DM that maximizes the coherent power of the signal across the observing band in time. 
In addition to being dispersed by cold plasma along the line of sight, many of the bursts in this sample exhibit additional frequency ``drifting'' behavior, where the flux densities of either individual sub-bursts or burst envelopes show unique variations in arrival time across the band, after the full burst has been coherently dedispersed to a nominal DM. We attempt to characterize and measure the various forms of drifting present in the sample in \S\protect\ref{subsec:measurearchetypes}. Additionally, we search for and measure microstructure in select events with distinctly narrow features in \S\protect\ref{subsec:micro} and \S\protect\ref{apB}. We also search for periodicities in \S\protect\ref{subsec:quasiperiod}. Polarimetry performed on a sub-set of the bursts for which reasonable RM fits could be achieved is outlined in \S\protect\ref{apC:pol}.

\subsection{Time-Frequency Drifting Archetypes}\label{subsec:measurearchetypes}

\begin{figure}[ht!]
    \subfigure{\includegraphics[width = 0.24\textwidth]{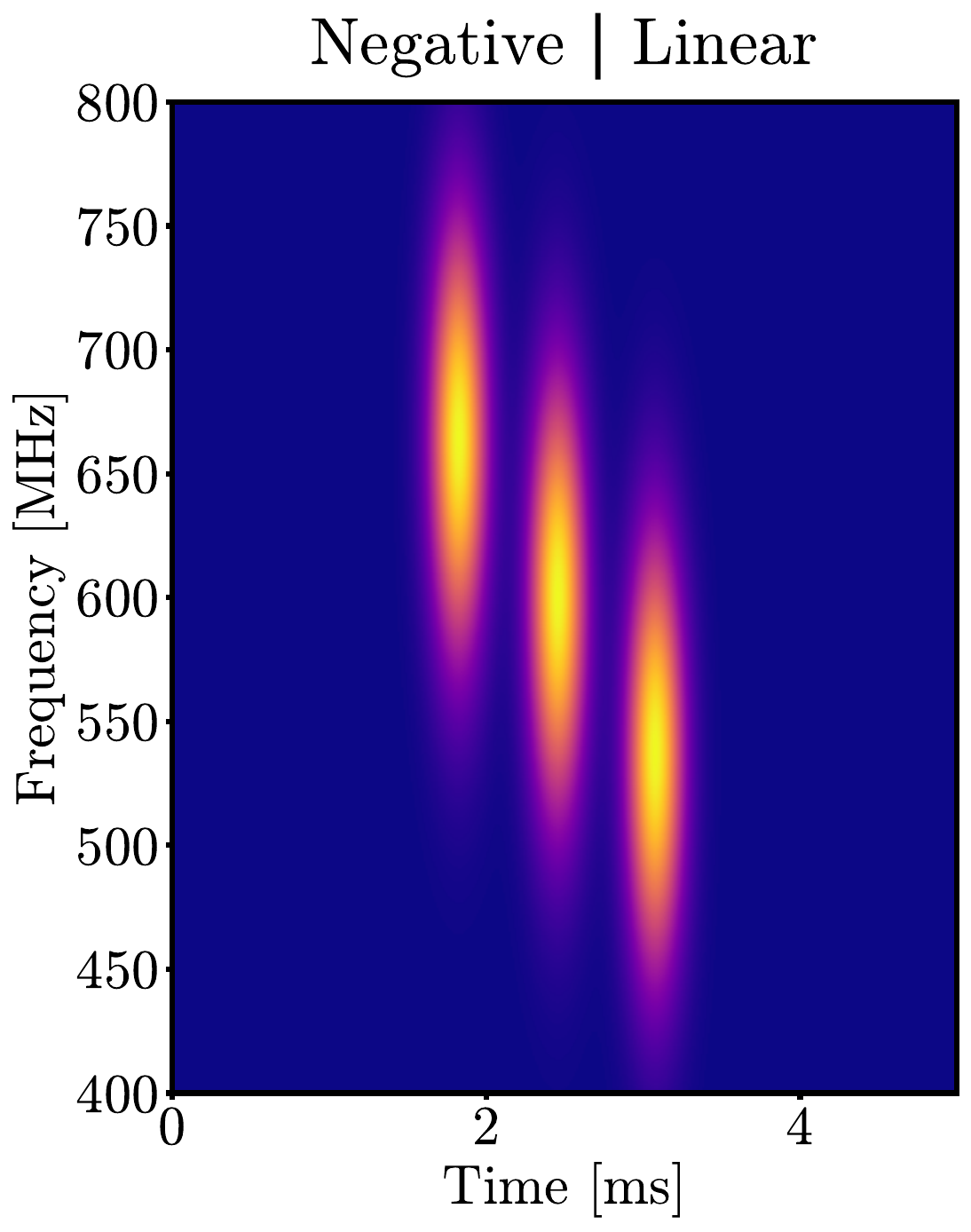}}
    \subfigure{\includegraphics[width = 0.24\textwidth]{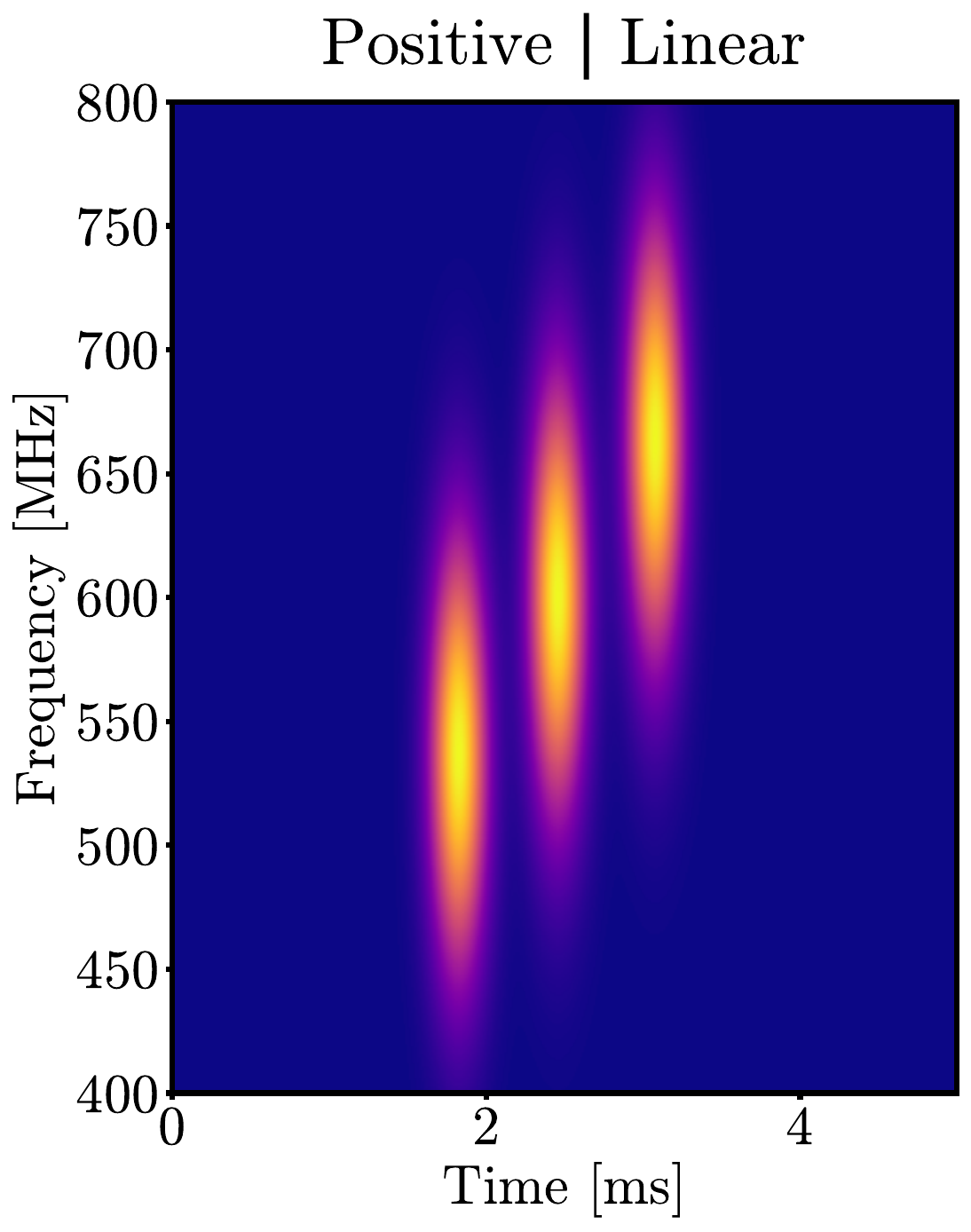}}
    \subfigure{\includegraphics[width = 0.24\textwidth]{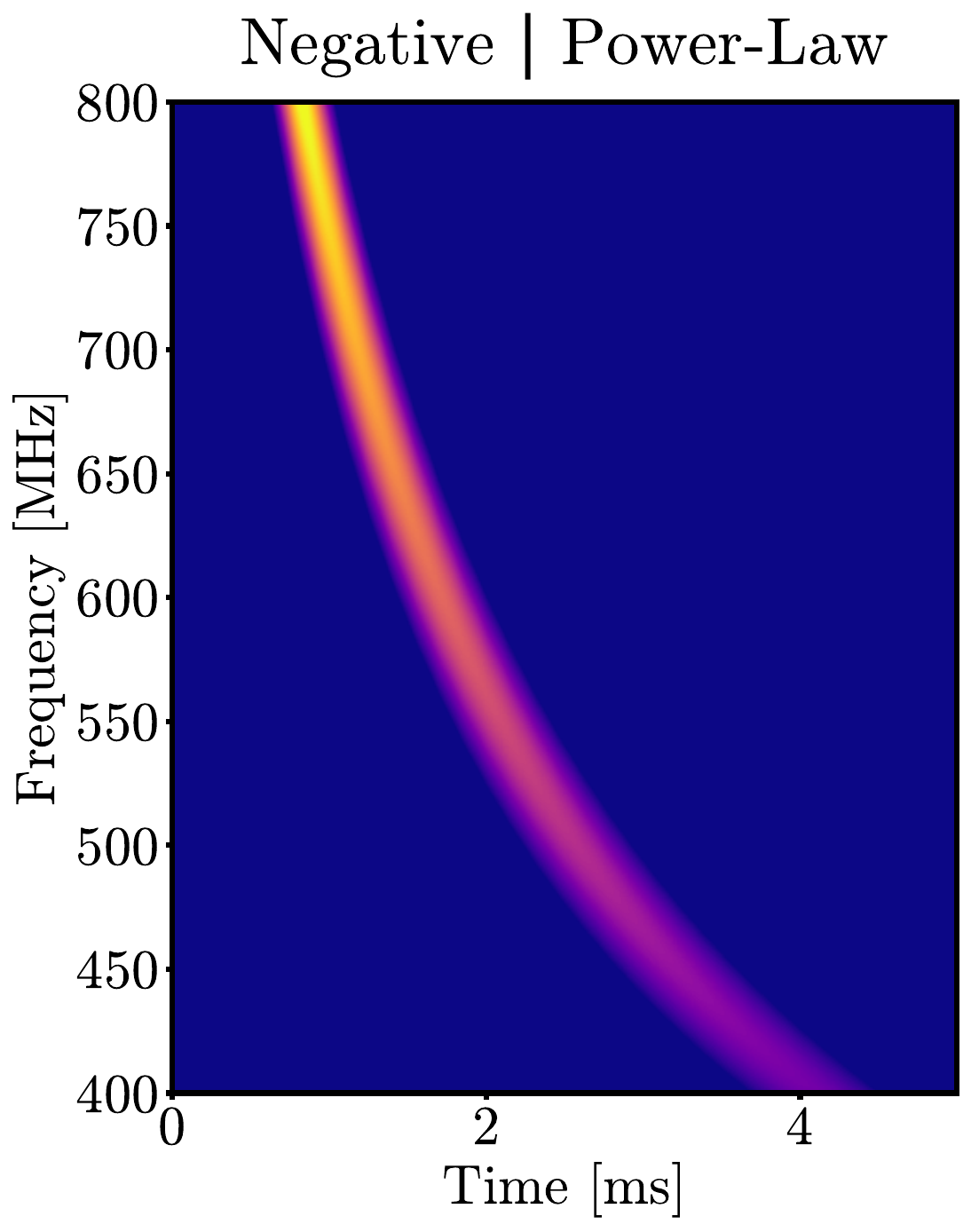}}
    \subfigure{\includegraphics[width = 0.24\textwidth]{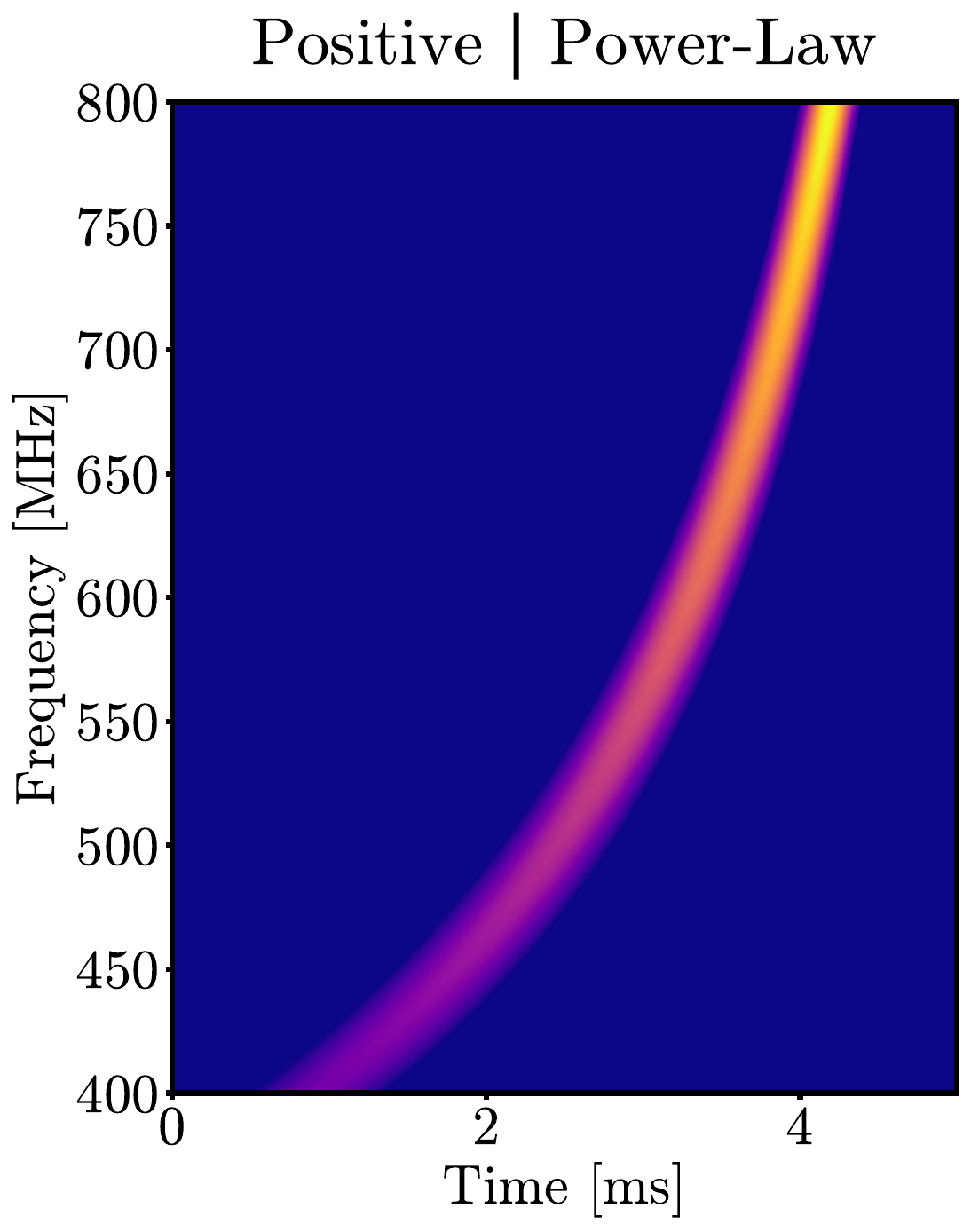}}
    \caption{Simulated (idealized) representations of dynamic spectra for the four morphological archetypes outlined in \S\protect\ref{subsec:measurearchetypes} and \S\protect\ref{subsec:archetypes}. From left to right we show examples of linear negative, linear positive, power-law negative, and power-law positive drifting. Each of the sub-bursts in these representations are modeled as 2D Gaussians for simplicity.}
    \label{fig:archetype_examples}
\end{figure}

The events shown in Figure \ref{fig:burstfig} suggest several possible categories or ``archetypes" of drift phenomenology:
linear negative drifting, linear positive drifting, power-law negative drifting, and power-law positive drifting. We illustrate these archetypes by simulating idealized burst spectra for each drifting scenario in Figure~\ref{fig:archetype_examples}. We report our measurements of both linear drift rates and power-law drift indices, as outlined below, in Table \ref{table:spectempprop}. Next, we discuss our bursts with these archetypes in mind.

\begin{table*}[ht!]
\caption{Drifting properties for three events that offer the most salient examples of both linear and power-law drifting. The power-law drift indices are measured in two ways: $(1)$ assuming a relativistic shock emission mechanism \protect\citep[fitting for $\beta$;][]{Metzger2022} as described in \S\protect\ref{subsec:nonlinnegdrift}, and $(2)$ via a heuristic, incoherent alignment algorithm (fitting for $\gamma$) described in \S\protect\ref{subsec:nonlinposdrift}.}
\centering  
\begin{tabular*}{1.\textwidth}{@{\extracolsep{\fill}}c c c c c c c c c} 
\hline\hline                       
{TNS Name} & Drift Rate (MHz ms$^{-1}$) & Drift Index ($\beta$) & Drift Index ($\gamma$) \\   
\hline   
FRB 20201230B & +285(50) & ... & ...  \\
FRB 20210813A & ... & $+0.76^{+0.07}_{-0.06}$, $+0.77^{+0.05}_{-0.07}$ & $+$1.2(1)  \\
FRB 20211005A & ... & ... & $+$3.9(5) \\
\hline\hline  \\                              
\end{tabular*}
\label{table:spectempprop}
\end{table*}

\subsubsection{Linear Negative Drifting}\label{subsec:linnegdrift}

The ``linear negative drifting'' archetype describes the case where a burst envelope or series of sub-bursts drift downwards in frequency along a single negative slope in time (also referred to as the ``sad trombone'' effect). First observed by \citet{Gajjar2018} in FRB 20121102A, it has become the most common drifting archetype in FRBs, having since been observed in events from a variety of other sources as well \citep{Hessels2019, Rajwade2020, Platts2021, Sand2022, Hewitt2023}. Despite being the most common, however, it is not obviously present in our sample. This is a selection effect, in part, as we intended to select for events with uncommon morphologies. It is also the case that the most salient instances of linear negative drifting typically occur in repeating sources, whereas the sources present here are thus far non-repeating. The only event that appears to be drifting linearly with a negative slope, is FRB 20210819A. By eye, this appears to be only subtly present in the lower half of the band, though it does not clearly match the archetype as seen in other FRB sources. For this reason, we do not focus on this particular archetype in detail here.

\subsubsection{Power-Law Negative Drifting}\label{subsec:nonlinnegdrift}

The ``power-law negative drifting'' archetype describes the case where an individual sub-burst, or entire burst envelope, drifts downwards in frequency as a power-law in time after it has been dedispersed to a nominal DM. The events in our sample that exhibit this behavior are FRB 20210813A, FRB 20200603B, FRB 20210427A, FRB 20210627A, and FRB 20211005A.

The most common example of power-law negative drifting is cold-plasma dispersion, which follows the well-understood frequency-dependent time-delay $t_{d} \propto \nu^{-2}$. The only cases where this would appear noticeable after coherent dedispersion has been performed are when events contain sub-bursts with unique DM values. The only event that clearly exhibits non-dispersive power-law negative drifting, however, is FRB 20210813A. As such, we will focus on this event here and reserve a study of DM variability for \S\protect\ref{subsec:nonlinposdrift}.

We assume a nominal DM of 399.264(7) pc cm$^{-3}$ for FRB 20210813A, based on the coherent power-maximizing DM measured for the narrowest leading feature in the spectrum, again using \texttt{DM\_phase}. In \S\ref{subsec:micro}, we measure the width of this feature to be only $\Delta t = 22.3(5) ~\mu s$ (see Table \ref{table:microgauss}), hence it provides 
good
constraining power for the DM. We find that the drifting behavior observed in this burst after coherent dedispersion is not a consequence of cold plasma dispersion, which would show a frequency-dependent time-delay in agreement with $t_{d} \propto \nu^{-2}$. To confirm this, we perform a modified form of incoherent dedispersion on the burst spectrum for a range of power-law indices, and search for the index that best aligns the intensity across the full band in time, effectively maximizing S/N in the timeseries. With this method, we find a frequency-dependent time-delay relation closer to $t_{d} \propto \nu^{-1.2(1)}$. This procedure is described in more detail in \S\protect\ref{subsec:nonlinposdrift}.

\citet{Metzger2022} have proposed a toy model for FRB time-frequency structure to explain precisely this phenomenon, with the primary intention of resolving the morphological dichotomy between one-off and repeating FRBs observed by CHIME/FRB \citep{Pleunis2021, chime2021}. According to the model,
an FRB is generated by a release of energy from a stellar-mass compact object, triggering a relativistic shock that, upon expansion into a magnetized upstream medium described by a power-law density profile, produces synchrotron maser emission. 
However, the model can
also accommodate 
coherent magnetospheric emission mechanisms that lead to a similar power-law evolution of the SED over time, including magnetic reconnection \citep{Philippov2019, Lyubarsky2020}, and curvature radiation \citep{Beloborodov2017, Ghisellini2018}. 

The model defines the flux density of a burst with multiple components as a function of frequency and time as:

\begin{equation}\label{eq:Ftfull}
    F_\nu(t)=\sum_i F_i(t) \exp \left[-\frac{\left(\nu-\nu_{\mathrm{c}, i}(t)\right)^2}{\Delta \nu_{\mathrm{c}, i}(t)^2}\right]
\end{equation}
where $F_i(t)$ is the frequency-integrated flux density of the $i$-th spectral component (or sub-burst), $\nu_{\mathrm{c}, i}(t)$ is the central frequency of the $i$-th component, and $\Delta \nu_{\mathrm{c}, i}(t)$ is the spectral width of the $i$-th component. The evolution of the central frequency in time $\nu_{\mathrm{c}} (t)$ is defined:

\begin{equation}\label{eq:nuc}
    \nu_{\mathrm{c}}(t)=\nu_0\left(1+\frac{t}{t_0}\right)^{-\beta}
\end{equation}
where $\nu_0$ is the central frequency of the SED at a characteristic timescale $t_{0}$, and $\beta$ is the power-law index that determines the rate at which the SED drifts across the band. The frequency-integrated flux \citep[see Eq. 6 in][]{Metzger2022} is defined:

\begin{equation}\label{eq:Ft}
    \begin{aligned} F_{i}(t) & \approx \frac{2}{\sqrt{\pi}} \frac{F_0}{\chi \nu_0}\left(1+\frac{t}{t_0}\right)^{(\beta+\mu-\alpha)} = \frac{2}{\sqrt{\pi}} \frac{F_0}{\chi \nu_0}\left(\frac{\nu_{\mathrm{c}, i}(t)}{\nu_0}\right)^{(\alpha-\mu-\beta) / \beta}
    \end{aligned}
\end{equation}
where $F_0$ is the initial peak flux density, $\alpha$ is the decay in time of the frequency-integrated flux, $\mu$ describes how $\Delta \nu_{\mathrm{c}}$ evolves with $\nu_{\mathrm{c}}$ within the emission envelope, and $\chi$ is a dimensionless parameter that controls the intrinsic bandwidth of the burst. In total, this model is described by seven free parameters per component ($i$): $\nu_0$, $F_0$, $t_0$, $\alpha$, $\mu$, $\beta$, and $\chi$. 

\begin{figure}[ht!]
    \subfigure{\includegraphics[width = 0.33\textwidth]{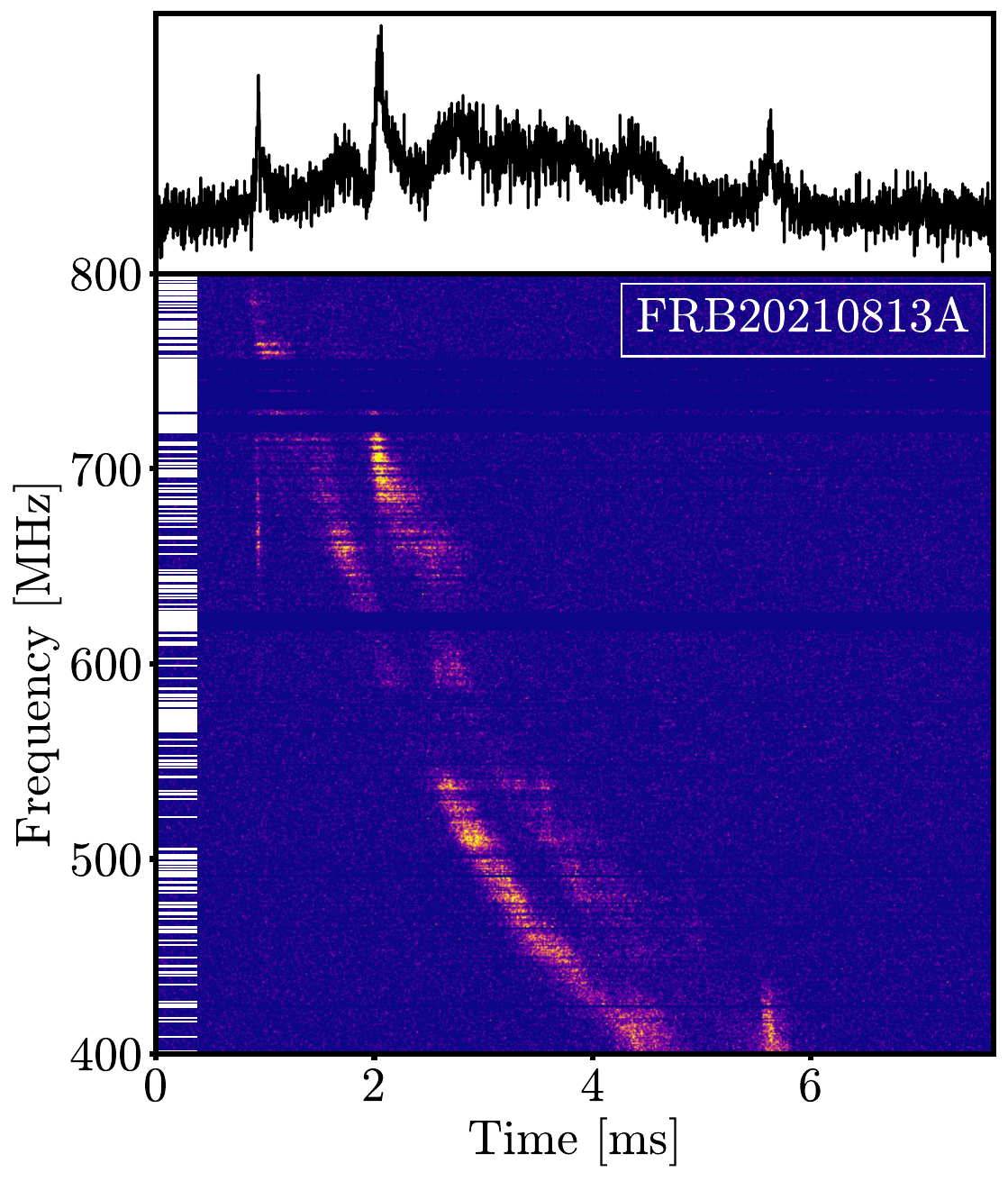}}
    \subfigure{\includegraphics[width = 0.33\textwidth]{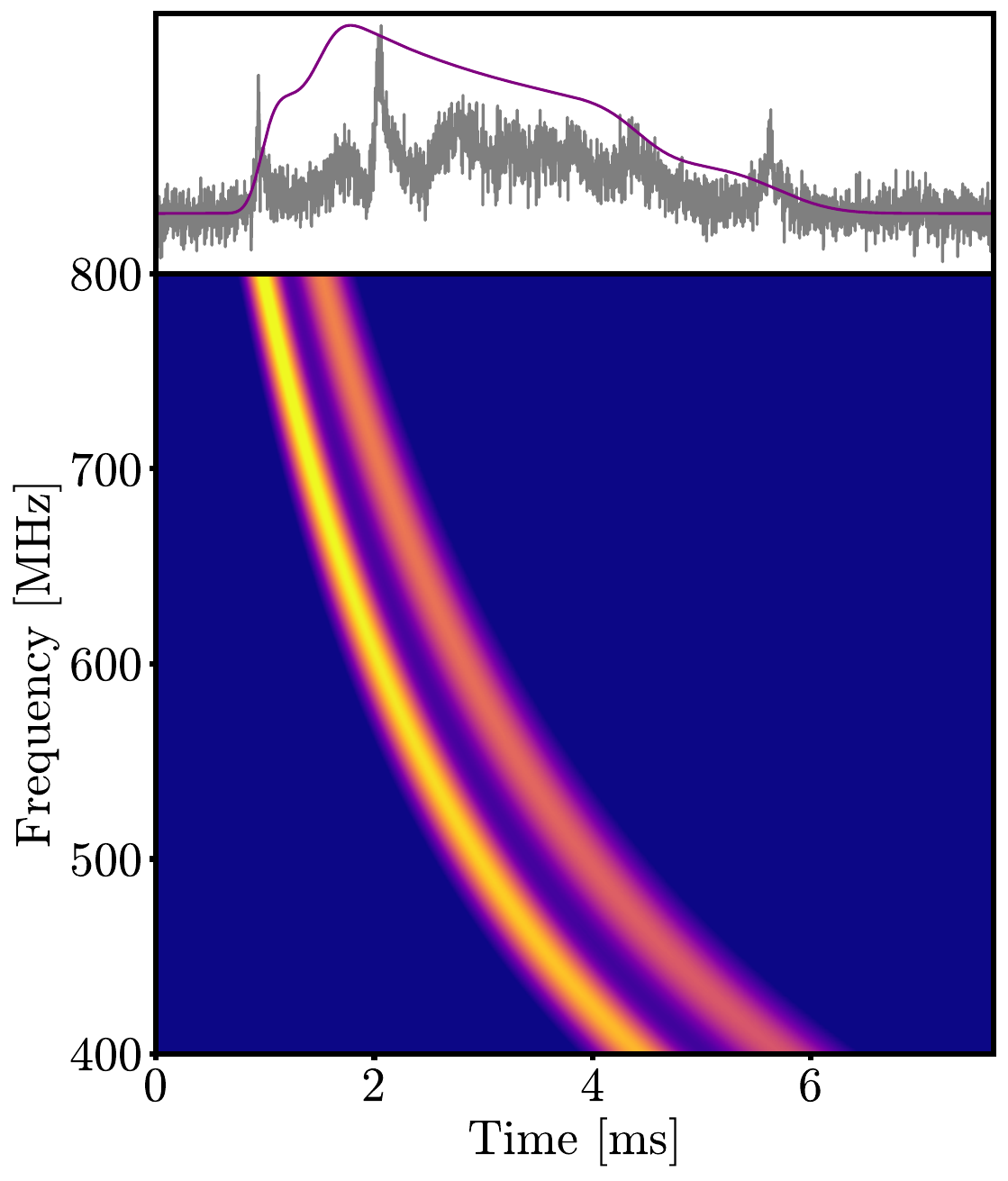}}
    \subfigure{\includegraphics[width = 0.33\textwidth]{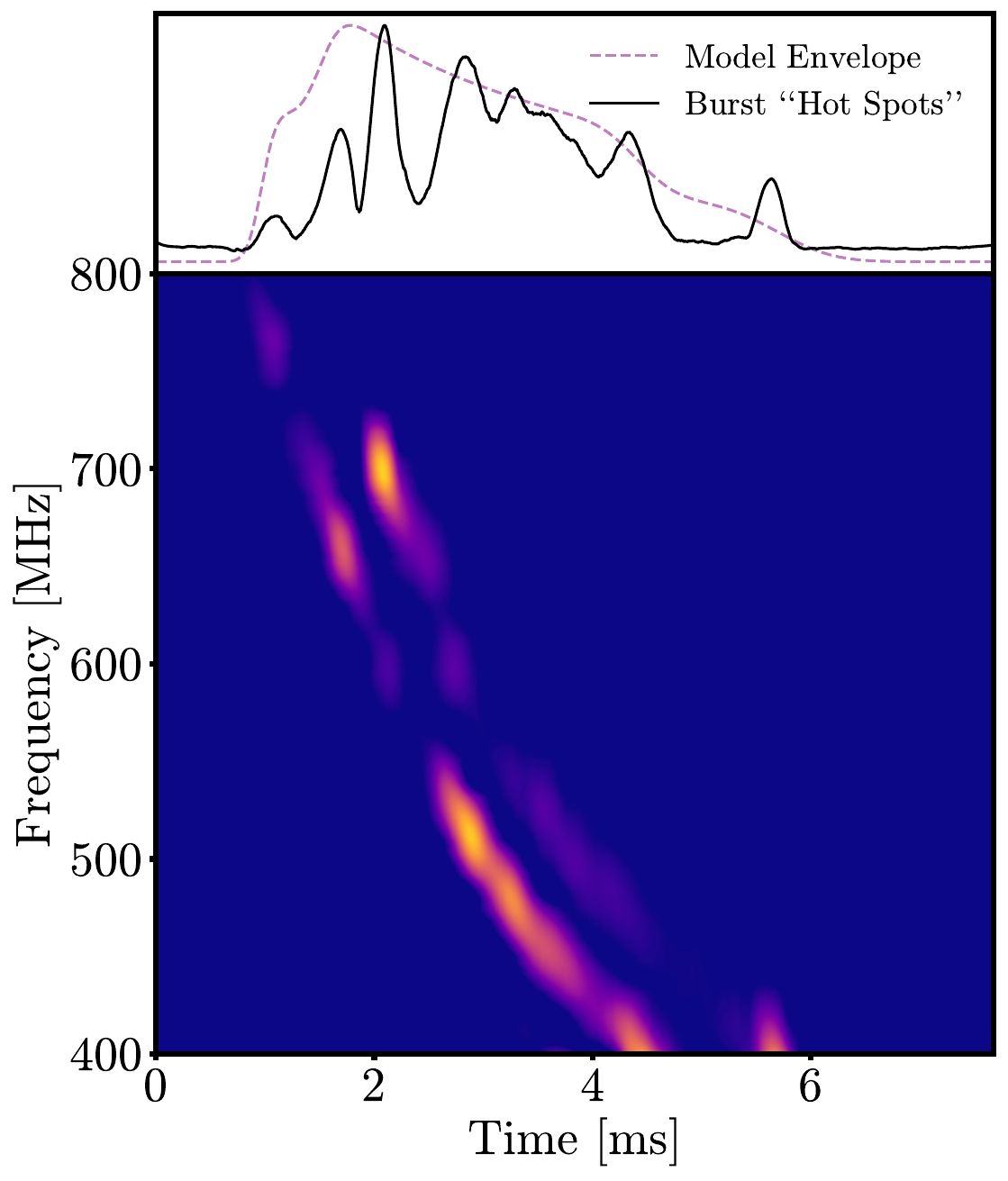}}
    \caption{\textit{Left}: The dynamic spectrum of FRB 20210813A. \textit{Middle}: The fitted two-component flux density model, defined in Eq. \ref{eq:Ftfull}, in accordance with the toy model proposed by \protect\citet{Metzger2022}. \textit{Right}: The real data are multiplied by the model and smoothed to reveal ``hot spots,'' where, if the structure is indeed produced by propagation effects, we would expect the greatest phase gradients in the lensing structure to appear, producing caustic patterns within the emission envelope.}
    \label{fig:metzgermodel}
\end{figure}

We fit the 2D flux density model defined in Eq. \ref{eq:Ftfull} by \citet{Metzger2022} to event FRB 20210813A, as it exhibits the most salient instance of power-law negative drifting in our sample, as well as complex temporal substructure. We use the \texttt{dynesty} \citep{Speagle2020} nested sampler as implemented in the \texttt{bilby}\footnote{https://pypi.org/project/bilby/} Python package \citep{Ashton2019} to estimate the best-fit parameters of Eq. \ref{eq:Ftfull} for FRB 20210813A, assuming uniform priors. Since we are mainly concerned with the drifting index $\beta$ and not the full SED, however, we can reduce the degrees of freedom by fixing values for $F_0=[1, 1]$, $\alpha=[1, 1]$, $\mu=[0, 0]$, \citep[consistent with prototypical values used by][]{Metzger2022}. We find that the best-fit values for the remaining burst parameters are $\nu_0 = [1.21^{+0.09}_{-0.08}, 1.41^{+0.09}_{-0.09}]$ GHz, $t_0 = [1.35^{+0.35}_{-0.25}, 1.39^{+0.35}_{-0.27}]$ ms, $\chi=[0.05^{+0.01}_{-0.01}, 0.06^{+0.01}_{-0.01}]$, $\beta = [+0.76^{+0.07}_{-0.06}, +0.77^{+0.05}_{-0.07}]$, where $\beta$ is defined as a positive quantity for negative drifting, as per Eq. \ref{eq:nuc}. Note that each parameter has two best-fit values, as the fits were performed for each sub-burst. The leading, narrowest sub-burst is also omitted in the fit, as it is quite faint and does not appear to drift, making it relatively uninformative in the context of this model. We show the fitted model, and the data multiplied by the fitted model in Figure \ref{fig:metzgermodel}. By multiplying the modeled burst with the original data, we are able to better highlight increases in flux density not explicitly accounted for in the toy model, which we will henceforth refer to as ``hot spots''. Variations of this kind could point to the influence of propagation effects like lensing, or other intrinsic effects that induce variability in the burst profile \citep{Beniamini2020, Lu2021}. Our results suggest that the power-law negative drifting observed in FRB 20210813A could be reasonably explained by a relativistic shock propagating into an expanding upstream medium, for which \citet{Metzger2022} predict $\beta \approx 0.2$-$0.7$, 
though we note that  magnetospheric scenarios \citep[e.g., radius-to-frequency mapping;][]{Lyutikov2020} have suggested $\beta \simeq 1$.

\subsubsection{Linear Positive Drifting}\label{subsec:linposdrift}

The ``linear positive drifting'' archetype describes the case where a series of sub-bursts or burst envelope drifts upwards in frequency along a single positive slope in time (also referred to as the ``happy trombone'' effect). We find only one event in the sample, FRB 20201230B, that exhibits this morphology, showing a series of three bright sub-bursts with centroids (regions of maximum flux density) that appear to increase in frequency over time in a linear (or near-linear) fashion. This morphological archetype has been observed in only a handful of FRBs to date \citep[further discussed in \S\protect\ref{subsec:linposdriftdiscuss}]{chime2020b, chime2020, Marthi2020, Main2021, Zhou2022, chime2023b}. We measure the drift rate of FRB 20201230B by performing a least-squares fit of a 2D Gaussian function to the 2D autocorrelation function (ACF) of the dynamic spectrum, and calculate the tilt in its semi-major axis, a standard technique for measuring linear negative drifting behavior \citep{Hessels2019, Pleunis2021}. To best characterize the 2D ACF, we use a modified 2D Gaussian function $G(x, y)$, defined as:

\begin{equation}
\begin{aligned} G(x, y)= & A \exp \left\{-\frac{1}{2}\left[x^2\left(\frac{\cos ^2 \theta}{\sigma_x^2}+\frac{\sin ^2 \theta}{\sigma_y^2}\right)\right.\right. \left.\left.+2 x y \sin \theta \cos \theta\left(\frac{1}{\sigma_x^2}-\frac{1}{\sigma_y^2}\right)+y^2\left(\frac{\sin ^2 \theta}{\sigma_x^2}+\frac{\cos ^2 \theta}{\sigma_y^2}\right)\right]\right\},
\end{aligned}
\end{equation}
where ($x, y$) are the time and frequency lag coordinates, $A$ is the amplitude, $\sigma_x$ and $\sigma_y$ are the standard deviations along the $x$ and $y$ axes, respectively, and $\theta$ is the rotation angle in radians. 

\begin{figure}[ht!]
    \centering
    \subfigure{\includegraphics[width = 0.475\textwidth]{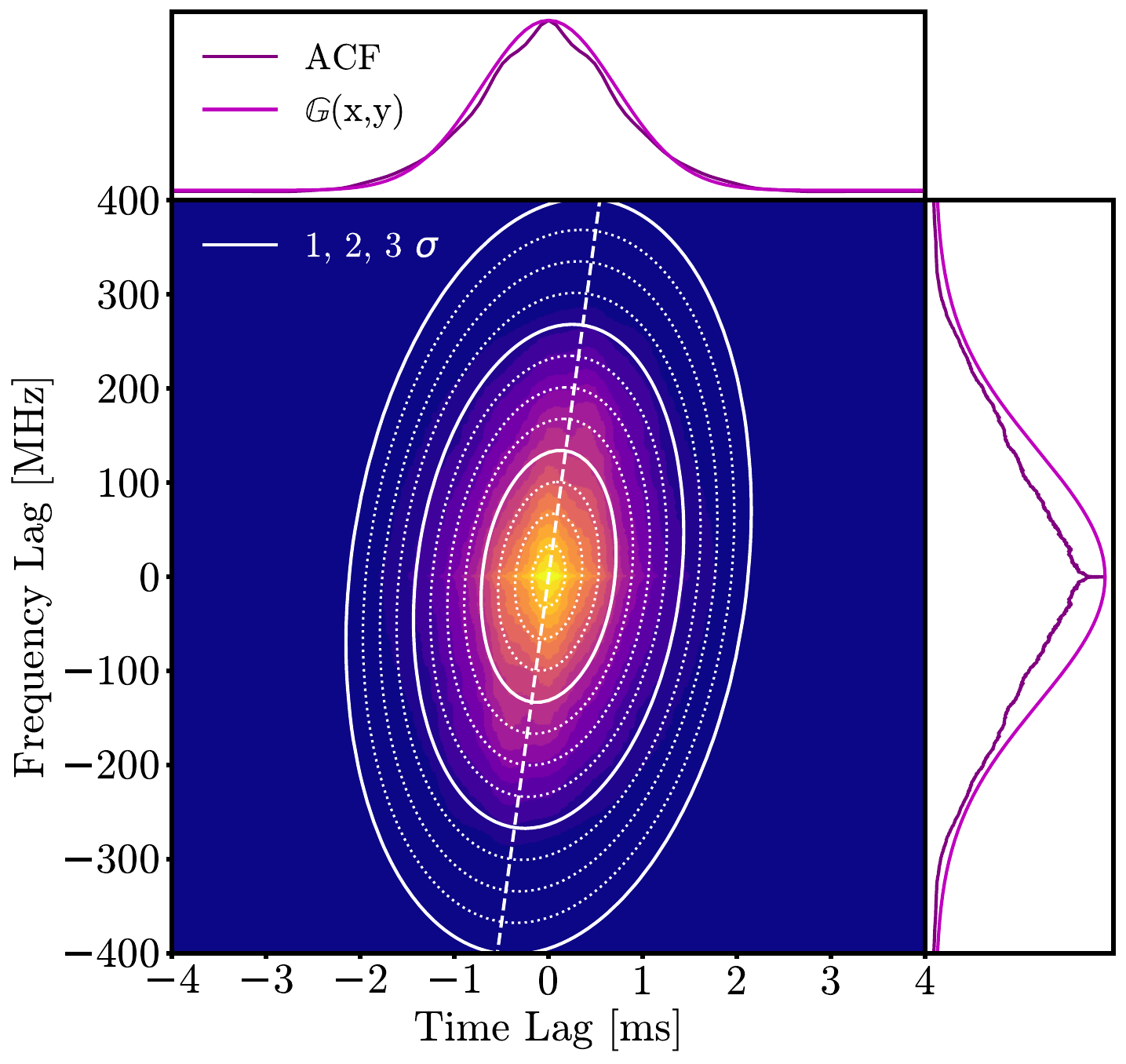}}
    \subfigure{\includegraphics[width = 0.433\textwidth]{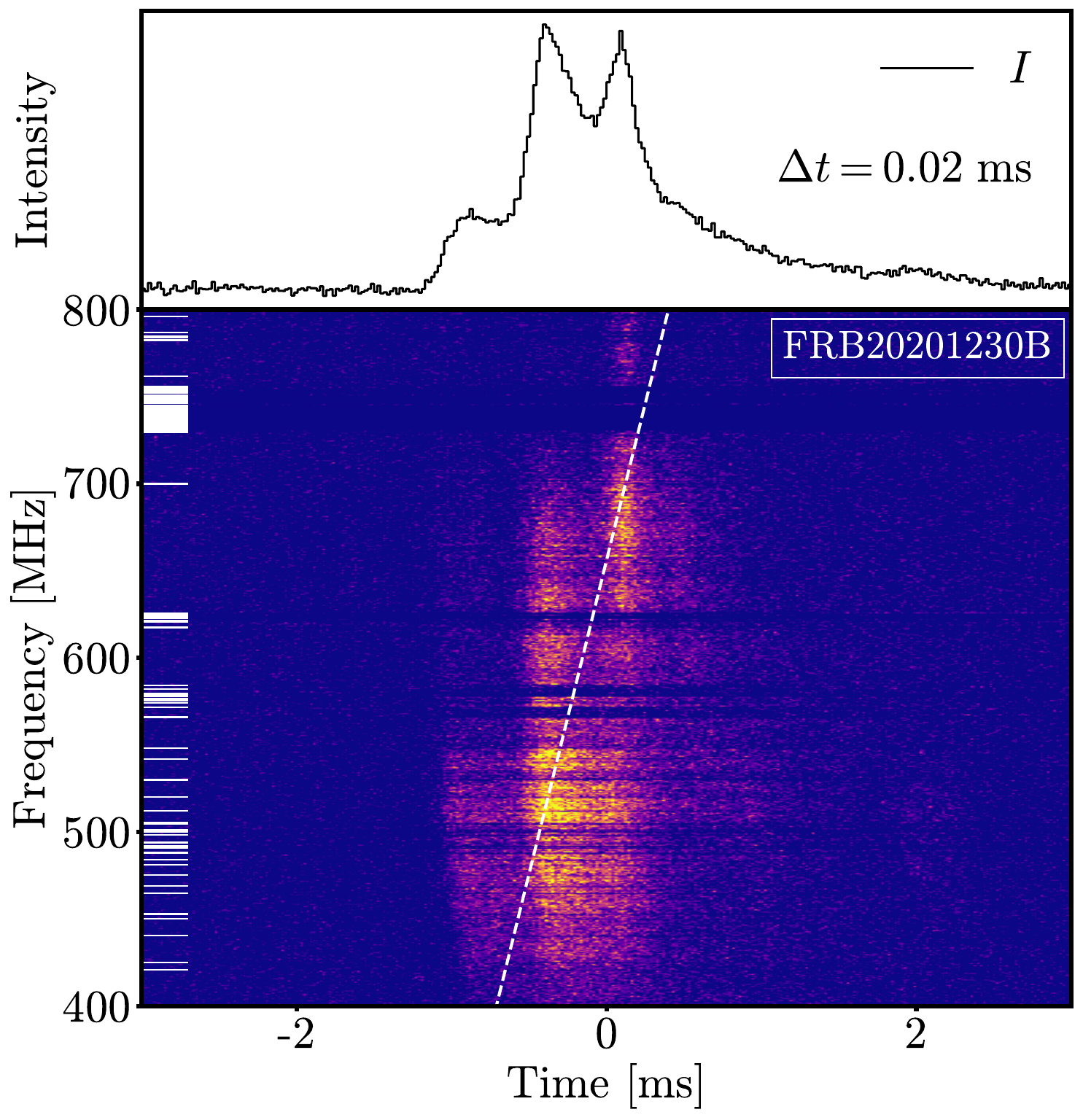}}
    \caption{\textit{Left}: A 2D Gaussian fit to the 2D ACF of event FRB 20201230B; the 1, 2, and 3 $\sigma$ regions of the fitted Gaussian are plotted over the ACF as white solid lines, and the semi-major axis is plotted as a white dashed line. \textit{Right}: The fitted drift rate, represented by the white dashed line, is plotted over a downsampled dynamic spectrum of FRB 20201230B, which better highlights some of the dimmer features.}
    \label{fig:linupdrift}
\end{figure}

The 2D Gaussian fit is shown in Figure \ref{fig:linupdrift}, plotted as contours over the 2D ACF of FRB 20201230B. We also show the best-fit drift rate plotted over the full burst dynamic spectrum. To calculate this drift rate, we simply take the cotangent of the best-fit rotation angle $\mathrm{cot}(\theta)$, and obtain a rate of $+285(50)$ MHz ms$^{-1}$. 

\subsubsection{Power-Law Positive Drifting and DM Variability}\label{subsec:nonlinposdrift}

\begin{figure}[ht!]
    \centering
    \vspace{-2.3cm}
    \subfigure{\includegraphics[width = 0.42\textwidth]{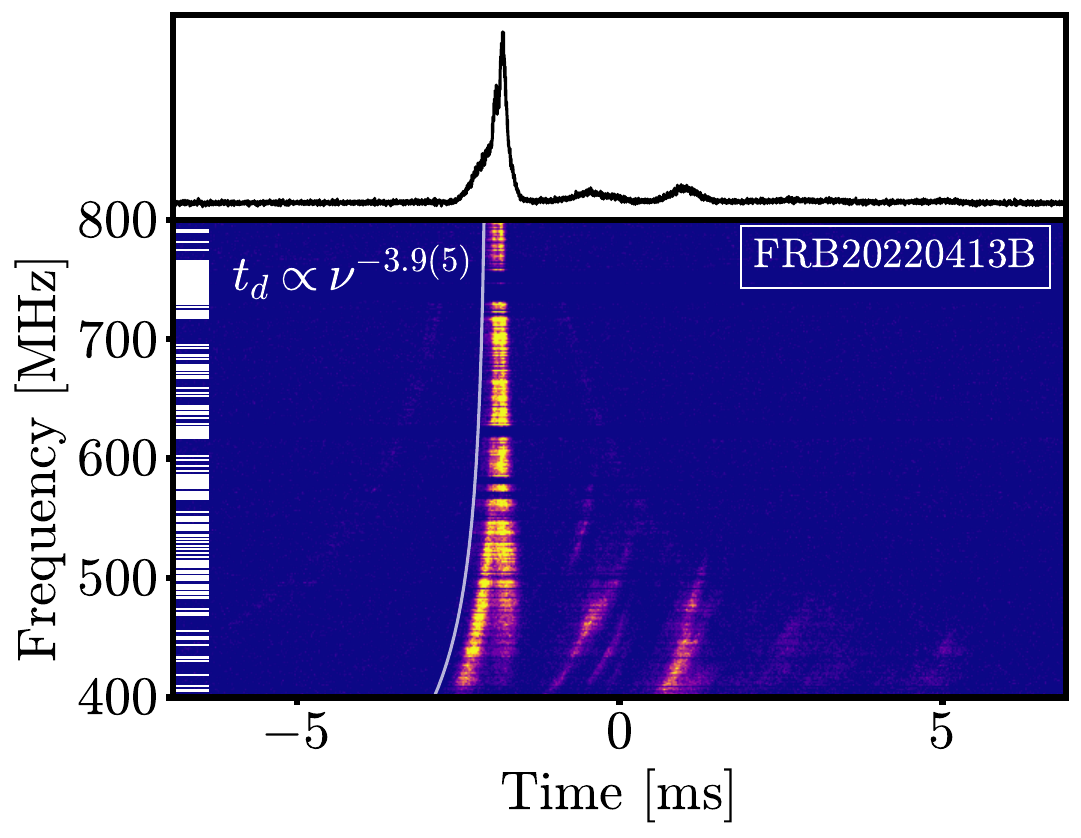}}
    \subfigure{\includegraphics[width = 0.42\textwidth]{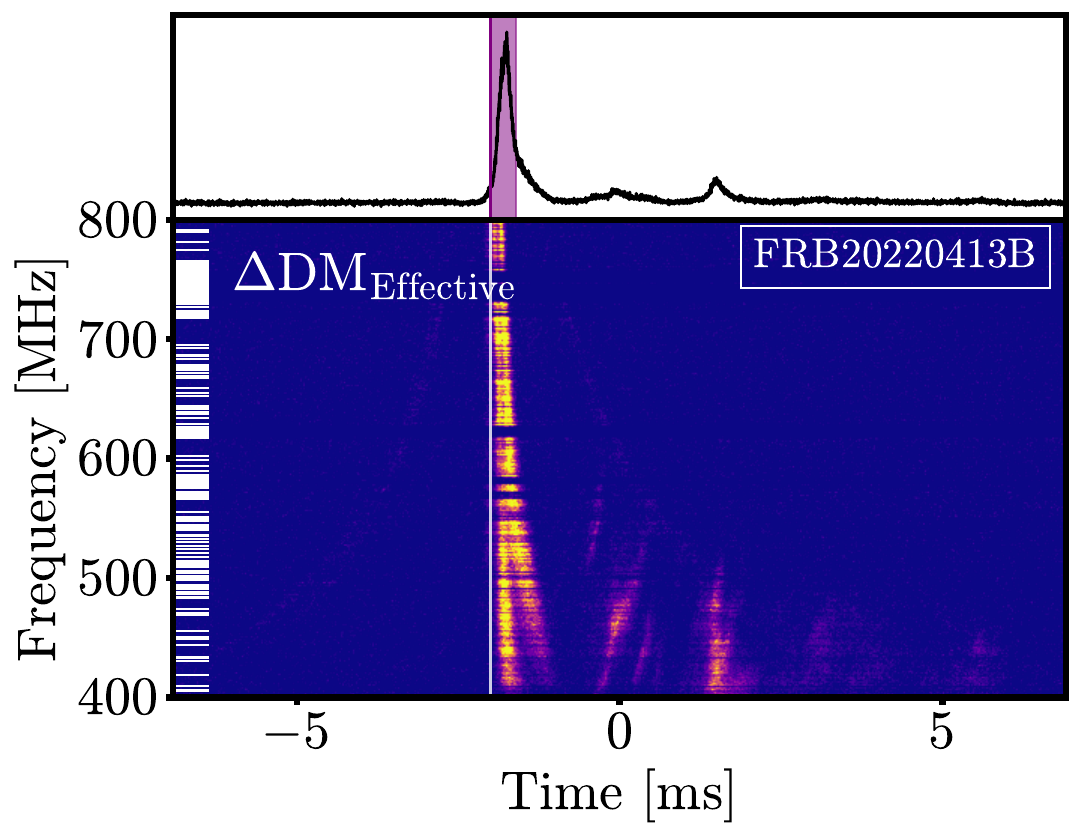}}\\
    \vspace{-0.5cm}
    \subfigure{\includegraphics[width = 0.42\textwidth]{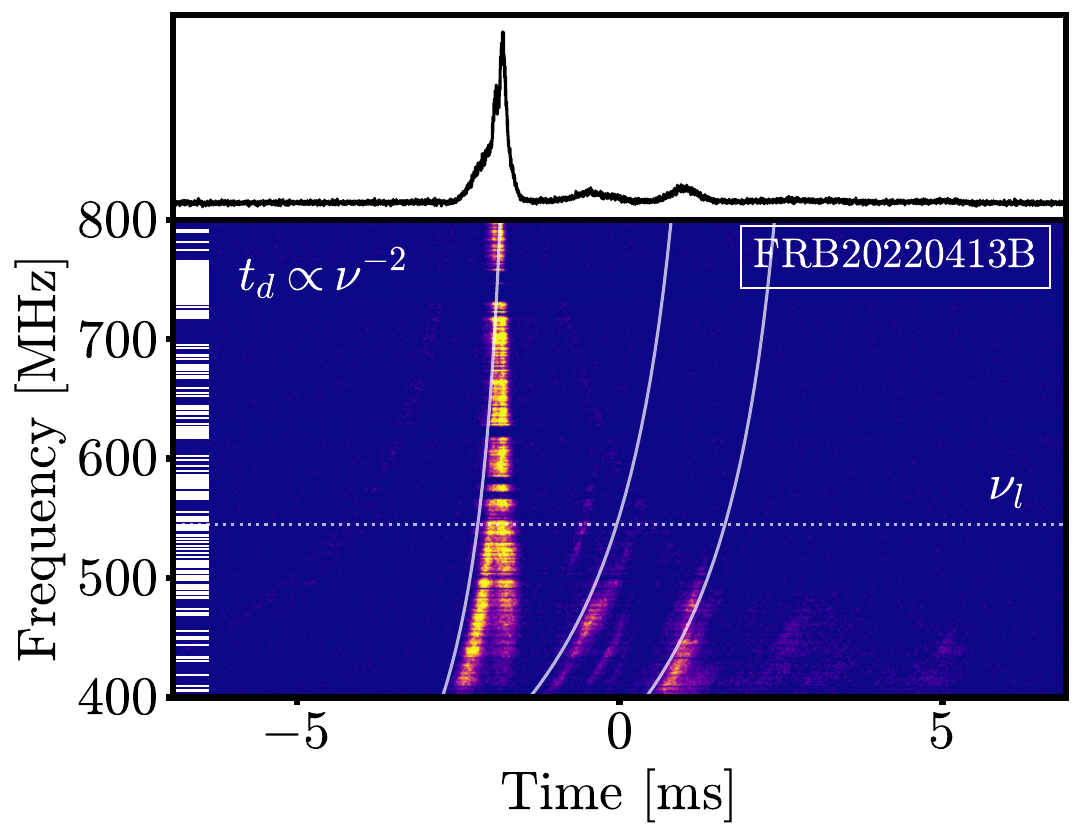}}
    \subfigure{\includegraphics[width = 0.42\textwidth]{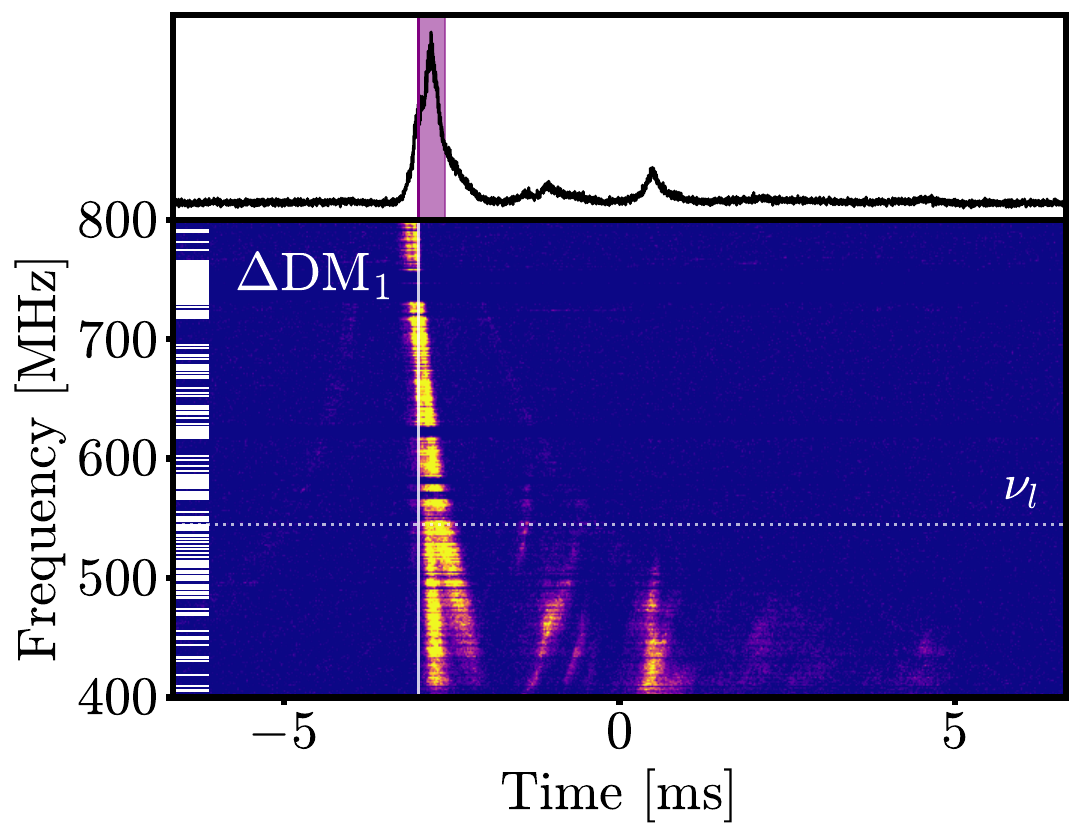}}\\
    \vspace{-0.5cm}
    \subfigure{\includegraphics[width = 0.42\textwidth]{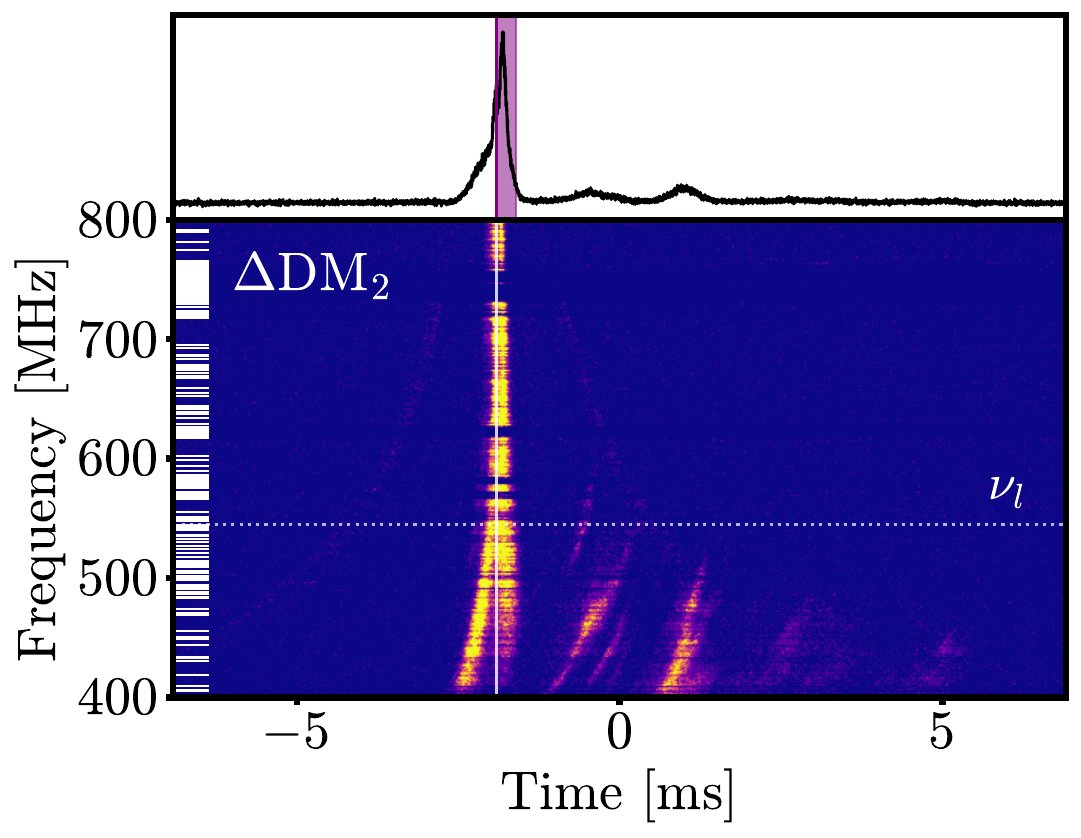}}
    \subfigure{\includegraphics[width = 0.42\textwidth]{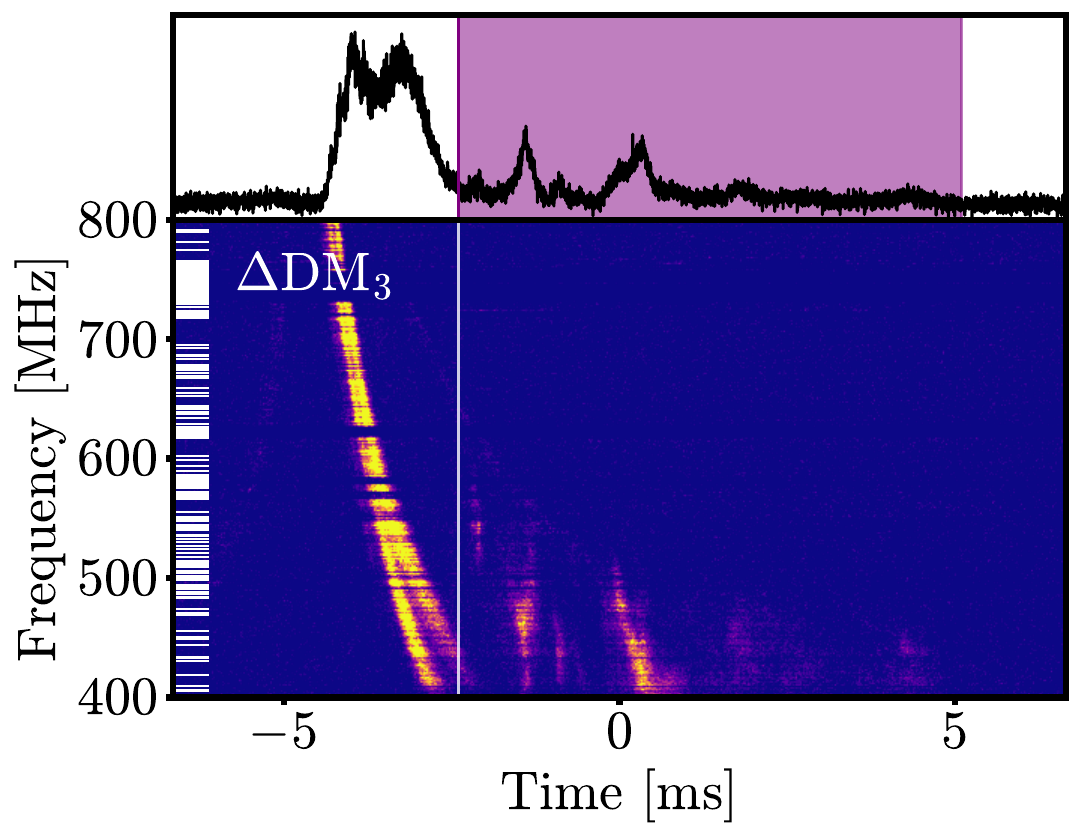}}\\
    \vspace{-0.51cm}
    \hspace{1cm}
    \subfigure{\includegraphics[width = 0.8\textwidth]{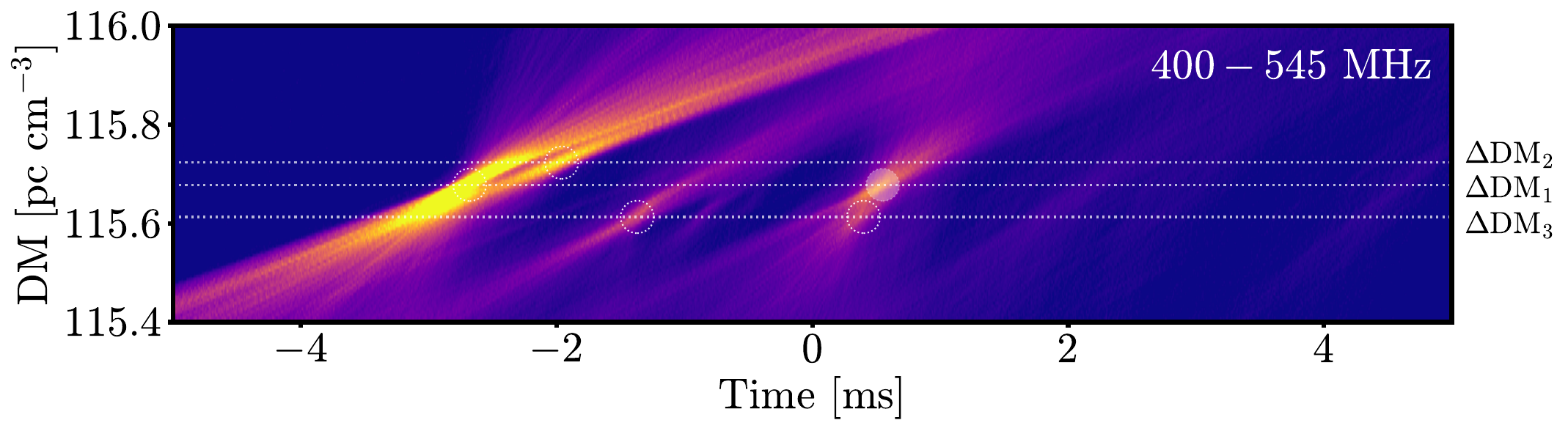}}
    \vspace{-0.5cm}
    \caption{The \textit{upper left} panel shows the dynamic spectrum of event FRB 20220413B, dedispersed to its nominal DM calculated using \texttt{DM\_phase}, where the measured frequency-dependent drift over-plotted as a white solid line. The \textit{upper right} panel shows the dynamic spectrum of FRB 20220413B, re-aligned (``dedispersed'') for a frequency index of $\nu^{-3.9(5)}$. The \textit{middle left} panel shows the dynamic spectrum of FRB 20220413B with DM curves over-plotted as white solid lines and the approximate focal frequency $\nu_{l} \approx$ 545 MHz as a white dotted line. The panels following show dynamic spectra of the same event dedispersed to the coherent power-maximizing $\Delta$DM offsets from the nominal burst DM for respective sub-bursts, highlighted in purple in the timeseries. The lowest panel shows the averaged pulse profile between 400-545 MHz of the band as a function of time and DM, showing clear ``bowtie''-like features that arise from the differential DMs between sub-bursts. The DMs present in the burst are indicated by white dotted lines, whereas the DMs corresponding to respective sub-bursts are specified by white dotted circles. The shaded circle indicates a region where the coherent power-maximizing DM measurement disagrees with the most intense region in the right-most bowtie feature. This feature contains two centroids, the centroid of lowest DM (also highlighted by a white dotted circle) corresponds to the coherent power-maximizing DM of the sub-burst. Figures \protect\ref{fig:moredrift}, \protect\ref{fig:86321776_dmvar}, \protect\ref{fig:169427924_dmvar}, 
    \protect\ref{fig:175128652_dmvar}, and
    \protect\ref{fig:189845329_dmvar} show the $\Delta$DM values for the other events in our sample exhibiting seemingly dispersive drifting, for which values are reported in Table \protect\ref{table:dmvar}.}
    \label{fig:nonlinupdrift}
\end{figure}

\begin{figure}[ht!]
    \centering
    \subfigure{\includegraphics[width = 0.45\textwidth]{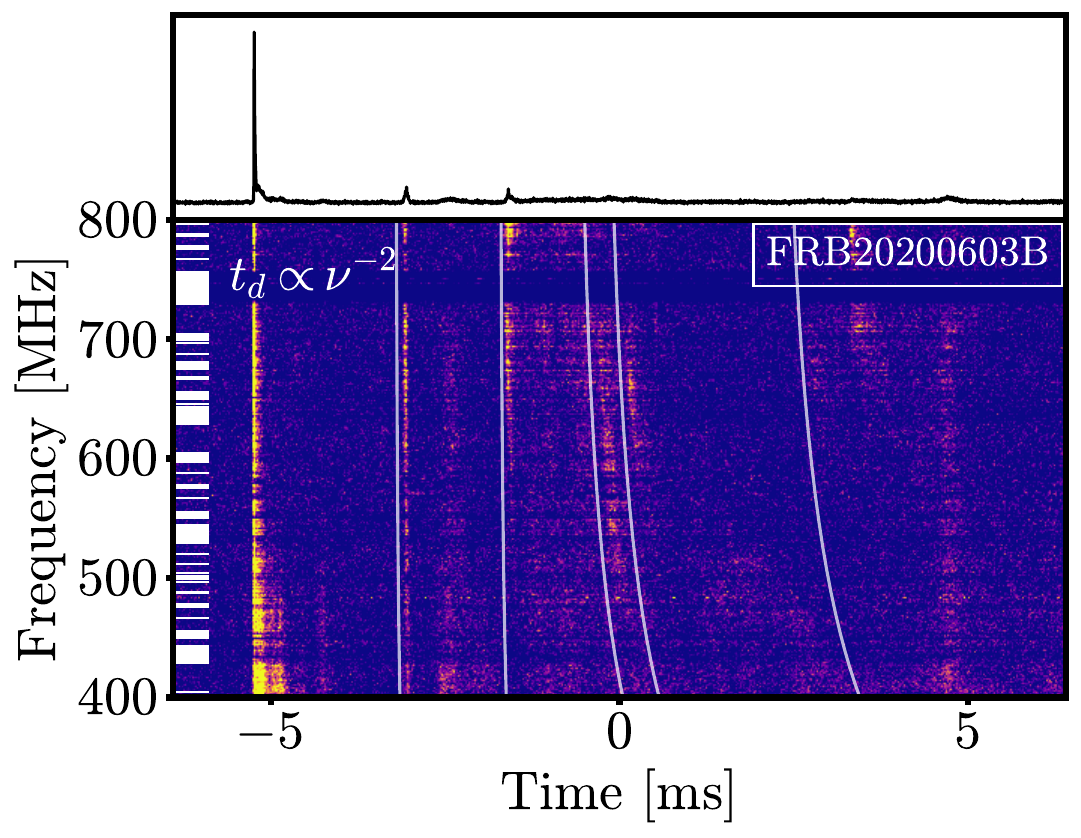}}
    \subfigure{\includegraphics[width = 0.45\textwidth]{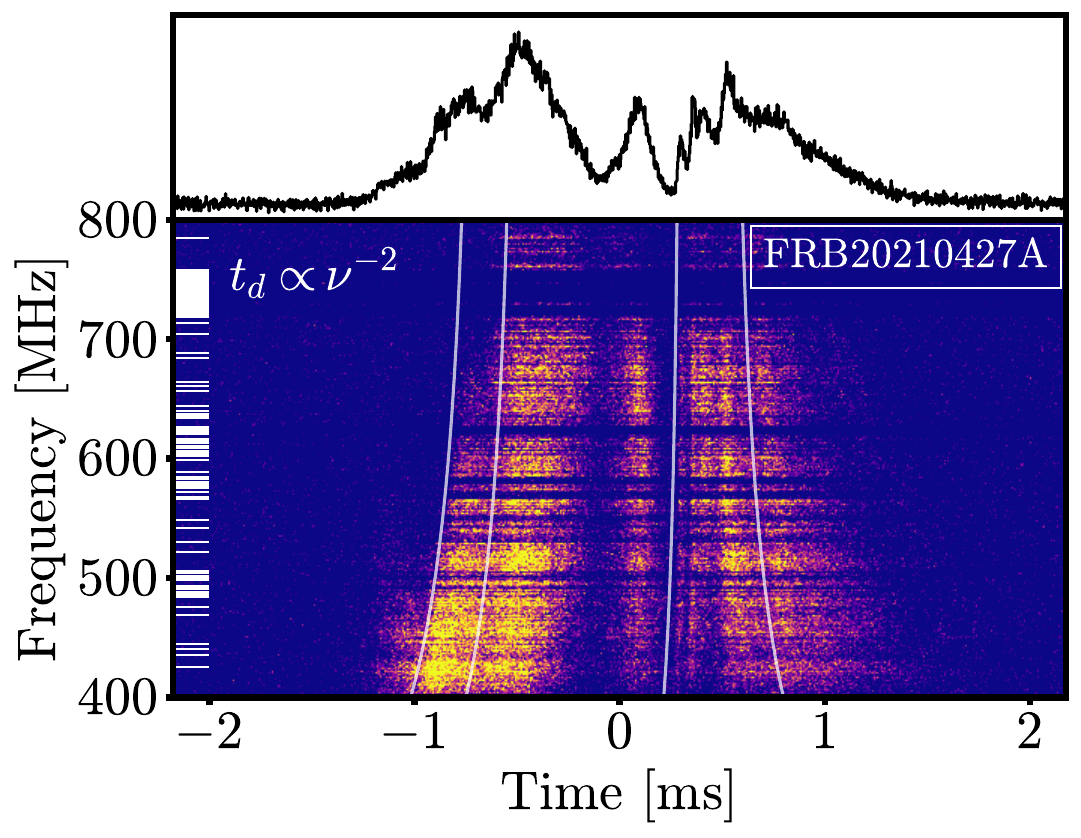}}
    \subfigure{\includegraphics[width = 0.45\textwidth]{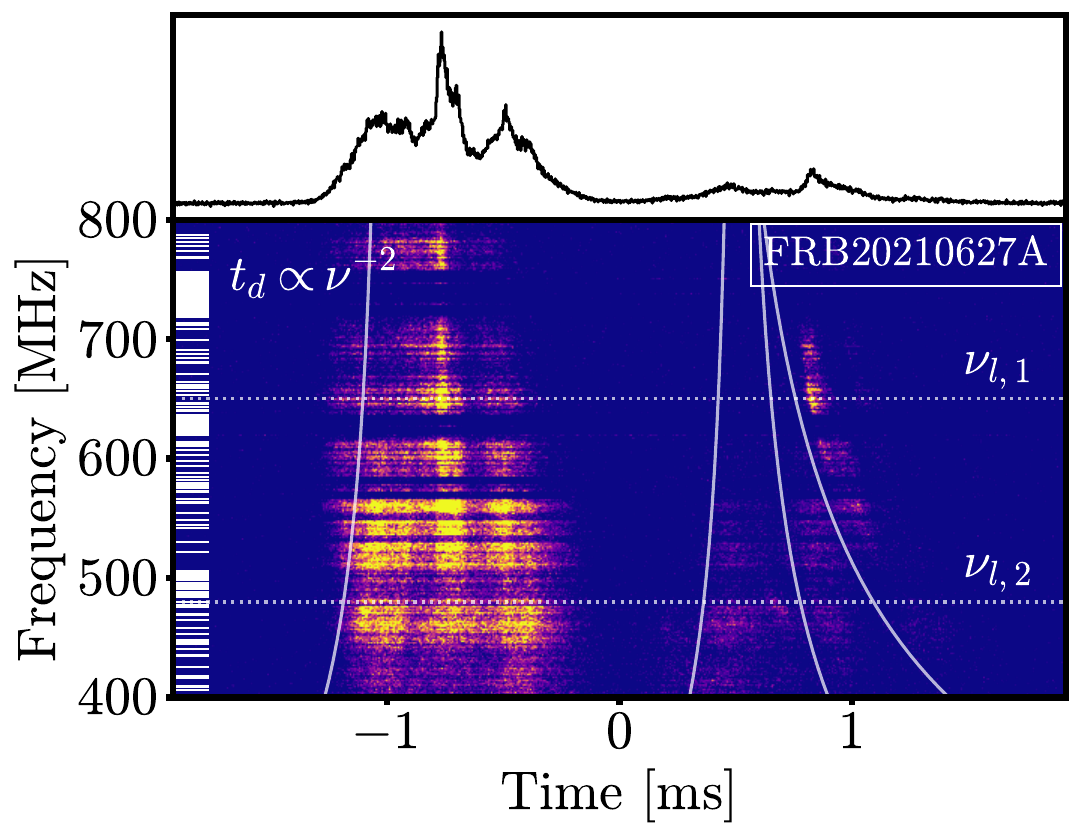}}
    \subfigure{\includegraphics[width = 0.45\textwidth]{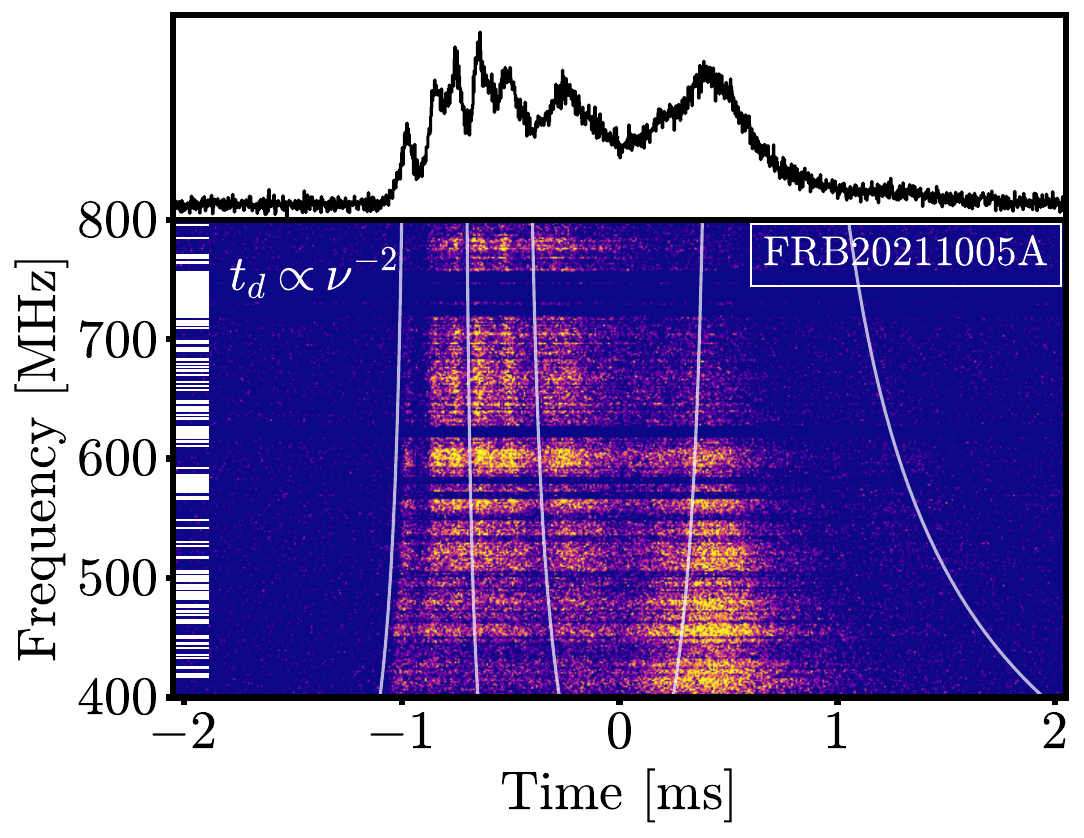}}
    \caption{Dynamic spectra for events FRB 20200603B, FRB 20210427A, FRB 20210627A and FRB 20211005A, with DM curves over-plotted as white lines, highlighting potential DM  variability present in each burst. For FRB 20210627A, we also plot the frequencies at which drifting starts to occur (the so-called ``focal frequencies''; $\nu_{l, 1} \approx 650$ MHz, $\nu_{l, 2} \approx 480$ MHz) as white dotted lines. The color scale of the spectra is more constrained in comparison to those in Figure \protect\ref{fig:burstfig} to better show fainter features. The dispersion measure variations with respect to the nominal DM values of each burst are reported in Table \protect\ref{table:dmvar}.}
    \label{fig:moredrift}
\end{figure}

Perhaps the most remarkable morphological archetype present in our sample is ``power-law positive drifting,'' which describes the case where an individual sub-burst, or entire burst envelope, drifts upwards in frequency as a power-law in time, again after it has been dedispersed to a nominal DM. The clearest example of this phenomenon is exhibited by FRB 20220413B, as shown in Figure \ref{fig:nonlinupdrift}, and will serve as the primary case study for this archetype. Other FRBs in this sample that appear to show power-law positive drifting are FRB 20200711F, FRB 20210427A, FRB 20210627A, and FRB 20211005A (see Figure \ref{fig:moredrift}, as well as \S\protect\ref{apA}). 

Assuming a nominal DM of 115.723(2) pc cm$^{-3}$ for FRB 20220413B, which we measure for the apparently non-drifting upper half (600-800 MHz) of the band using \texttt{DM\_phase}, we proceed to measure the drifting using two techniques.

The first technique is heuristic and makes no assumption about the power-law index of the drifting. Instead we aim to measure the index by iteratively performing a modified form of incoherent dedispersion on burst spectrum across the full band (also performed in \S\protect\ref{subsec:nonlinnegdrift} on FRB 20210813A) as:

\begin{equation}
    t_{d} \simeq 10^6 \mathrm{~ms} \times\left(\nu_1^{-\gamma}-\nu_2^{-\gamma}\right) \times \mathrm{DM_{Effective}}
\end{equation}
where we sample over a range of drifting indices $\gamma$, 
$0 \lesssim \gamma \lesssim 6$, with a dimensionless multiplicative prefactor DM$_{\mathrm{Effective}}$, and $\nu_1$, $\nu_2$ that span 400-800 MHz. Through this iterative process, we search for the index that maximizes the S/N in the time
series. We find that the best-fit drifting index $\gamma$ for the full band is inconsistent with $\nu^{-2}$, and instead follows a $\nu^{-\gamma}$ where $\gamma = 3.9(5)$, as shown in Figure \ref{fig:nonlinupdrift}.


The second technique we use to quantify power-law positive (and negative) drifting assumes that the drifting is dispersive (in agreement with $t_{d} \propto \nu^{-2}$), and occurs due to interactions with compact cold plasma structures of non-uniform density along the line of sight, as might be expected from a plasma lens \citep{Cordes2017}. Given this assumption, we iteratively coherently dedisperse the dynamic spectrum for time-limited regions across the burst for a range of trial DM offsets ($\Delta$DM) between [$-$0.15 pc cm$^{-3}$, 0.15 pc cm$^{-3}$] from the nominal value calculated previously, again using \texttt{DM\_phase}, to search for the DM value that maximizes coherent power in the time-limited region of the spectrum. The ranges in time over which these maximizations are performed are determined by judiciously isolating sub-bursts or sub-burst clusters in time. For certain bursts with narrower band occupations, we limit the frequency range over which we perform dedispersion as well. 

By inspection of the dynamic spectrum of FRB 20220413B, there appears to be a bifurcation in the first sub-burst (a separation into two distinct sub-bursts) that occurs at $\nu_{l} \approx$ 545 MHz \citep[possibly the ``focal frequency'', as described by ][]{Cordes2017}. Below this frequency, the drifting appears to be reasonably consistent with cold plasma dispersion. To show this, we plot DM curves adjacent to the drifting features in accordance with the estimated $\Delta$DM offsets from the nominal value calculated for FRB 20220413B, as shown in Figure \ref{fig:nonlinupdrift}. Under the assumption that this bifurcation arises due to multiple-imaging (a consequence of plasma lensing), we limit the frequency range of the dynamic spectrum to 400-545 MHz and calculate the respective DM values for which the coherent power of the leading sub-burst and trailing sub-bursts is maximized.

We perform the same iterative coherent dedispersion procedure for bursts FRB 20200711F, FRB 20210427A, FRB 20210627A and FRB 20211005A, all of which show both positive and negative power-law drifting. Respective $\Delta$DM values measured for each event are recorded in Table \ref{table:dmvar}. In Figure \ref{fig:moredrift}, we again plot DM curves adjacent to the drifting features to show the respective deviations in DM from the nominal values. Note that, similar to FRB 20220413B, FRB 20210627A also shows possible focal frequencies at $\nu_{l, 1} \approx 650$ MHz, $\nu_{l, 2} \approx 480$ MHz, above which no drifting is observed. The coherently dedispersed dynamic spectra for these events at each $\mathrm{DM}_{\mathrm{nominal}} + \Delta \mathrm{DM}$ value recorded in Table \ref{table:dmvar} are shown in \S\protect\ref{apA} (see Figures \ref{fig:86321776_dmvar}, \ref{fig:169427924_dmvar}, 
\ref{fig:175128652_dmvar}, and
\ref{fig:189845329_dmvar}). This method illustrates that DM variability on microsecond timescales is likely present in a substantial fraction of FRBs in our sample. These data highlight the worthwhile endeavor of characterizing plasma lensing in FRBs through more robust, raw voltage searches of phase coherence between sub-bursts---a smoking gun for lensing phenomena (Kader et al., in prep.).

\begin{table*}[ht]
\caption{The nominal DM values and measured offsets ($\Delta \mathrm{DM}$) for five of the twelve events in the sample, all of which show drifting features that appear consistent with dispersive smearing. FRB 20220413B is shown in Figure \protect\ref{fig:nonlinupdrift}, while the others are shown in the \S\protect\ref{apA}. The $\Delta \mathrm{DM}$ values (all in pc cm$^{-3}$) are listed as they appear across each burst from left to right, and are measured to within errors of $\pm 10\%$.}
\centering  
\begin{tabular*}{1\textwidth}{@{\extracolsep{\fill}}c c c c c c c c c} 
\hline\hline                       
{TNS Name} & $\mathrm{DM}\left(\mathrm{pc~} \mathrm{cm}^{-3}\right)$ & $\Delta \mathrm{DM}_1$ & $\Delta \mathrm{DM}_2$ & $\Delta \mathrm{DM}_3$ & $\Delta \mathrm{DM}_4$ & $\Delta \mathrm{DM}_5$ & $\Delta \mathrm{DM}_6$ \\   
\hline   
FRB 20200603B & $295.0828(4)$ & 0.0 & +0.002 & +0.02 & ... & ... & ... \\
FRB 20210427A & $268.4785(4)$ & $-$0.018 & $-$0.006 & $-$0.002 & 0.0 & +0.005 & ... \\
FRB 20210627A & $299.158(4)$ & $-$0.015 & 0.0 & $-$0.01 & +0.01 & +0.04 & ... \\
FRB 20211005A & $226.1078(3)$ & $-$0.005 & 0.0 & +0.002 & +0.004 & $-$0.008 & +0.05 \\
FRB 20220413B & $115.723(2)$ & $-$0.05 & 0.0 & $-$0.11 & ... & ... & ... \\
\hline\hline  \\                              
\end{tabular*}
\label{table:dmvar}
\end{table*}

\subsection{Microstructure}\label{subsec:micro}

\begin{figure}
  \centering
  \subfigure{\includegraphics[width=0.48\textwidth]{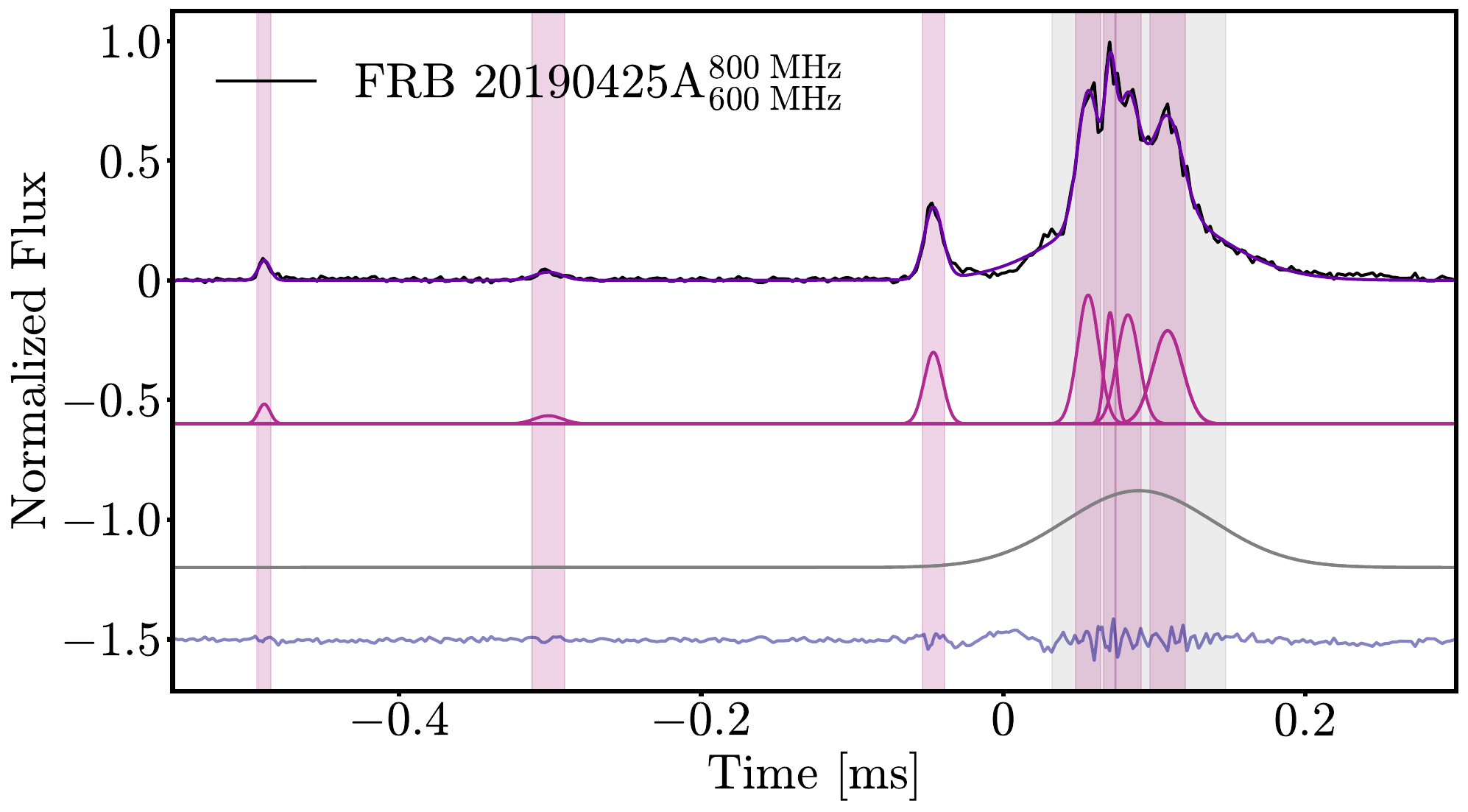}}
  \subfigure{\includegraphics[width=0.48\textwidth]{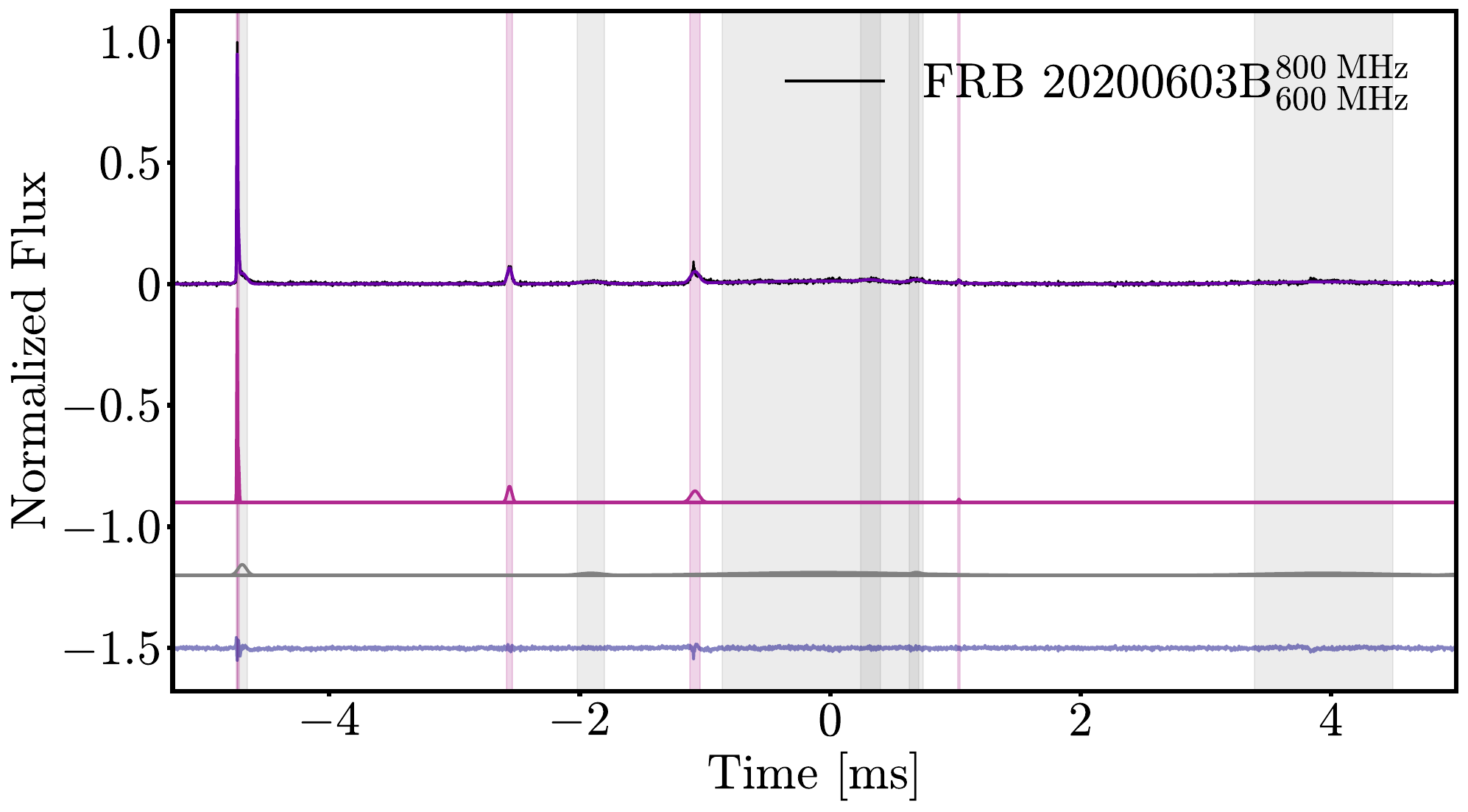}}\\
  \vspace{-0.2cm}
  \subfigure{\includegraphics[width=0.48\textwidth]{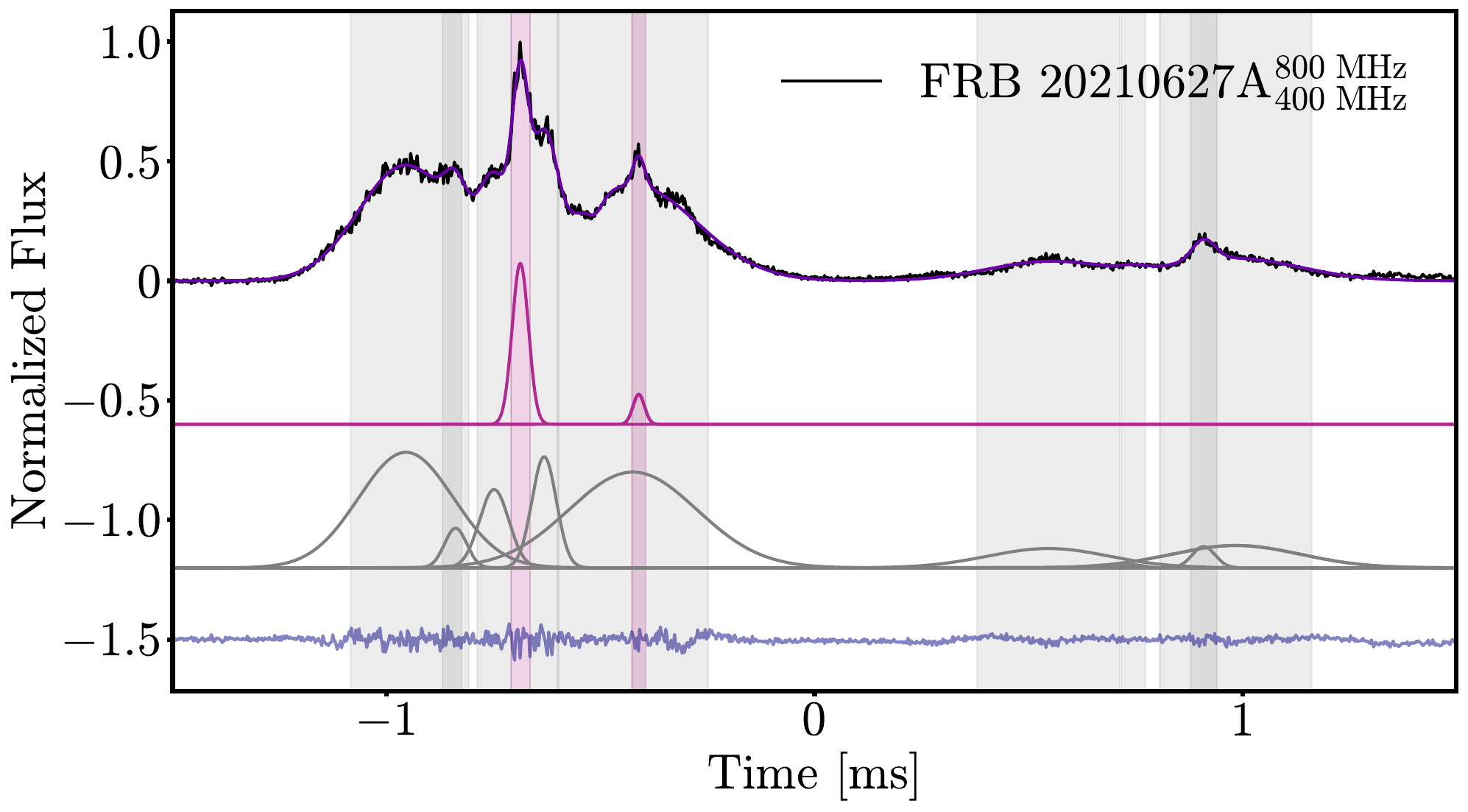}}
  \subfigure{\includegraphics[width=0.48\textwidth]{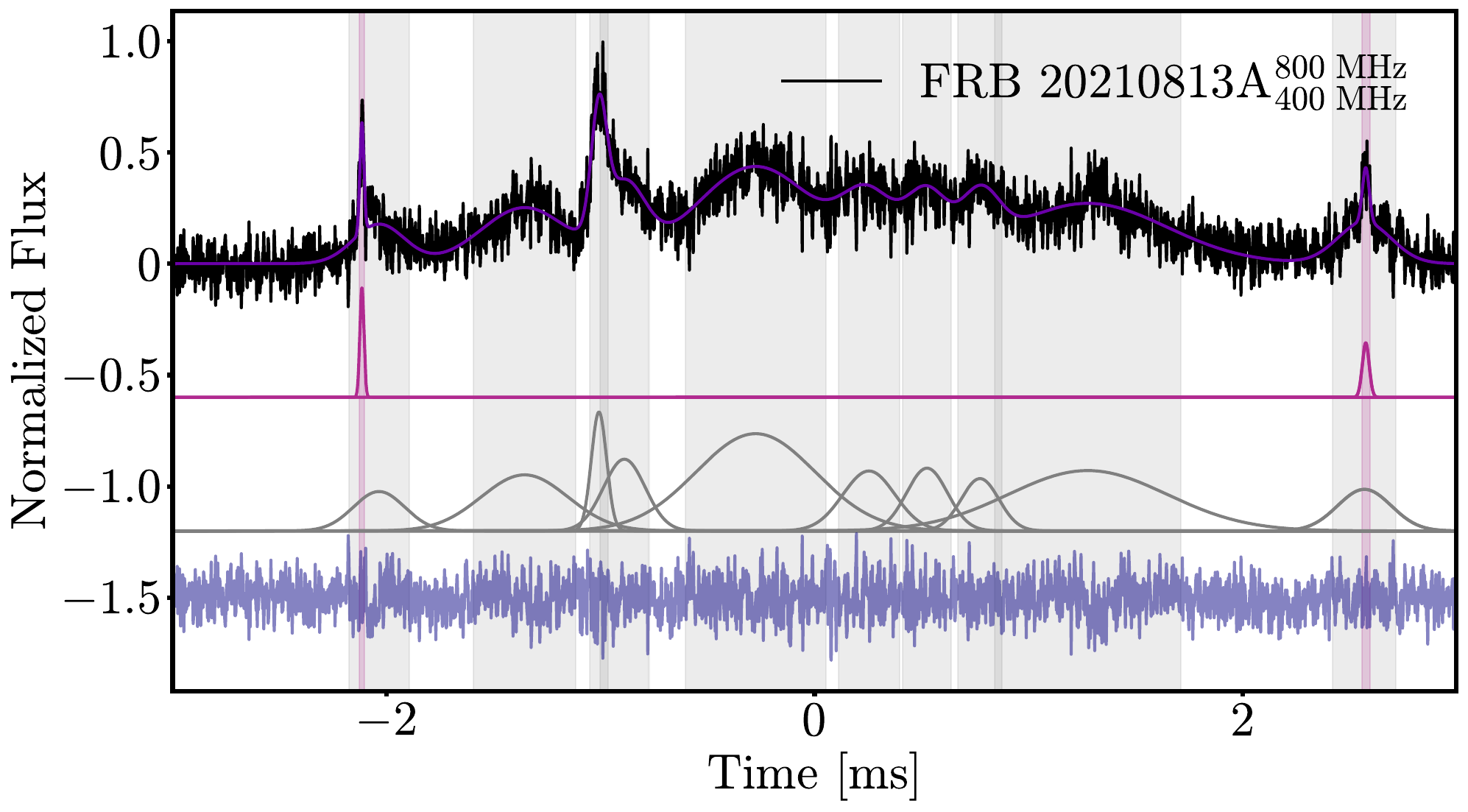}}\\
  \vspace{-0.2cm}
  \subfigure{\includegraphics[width=0.48\textwidth]{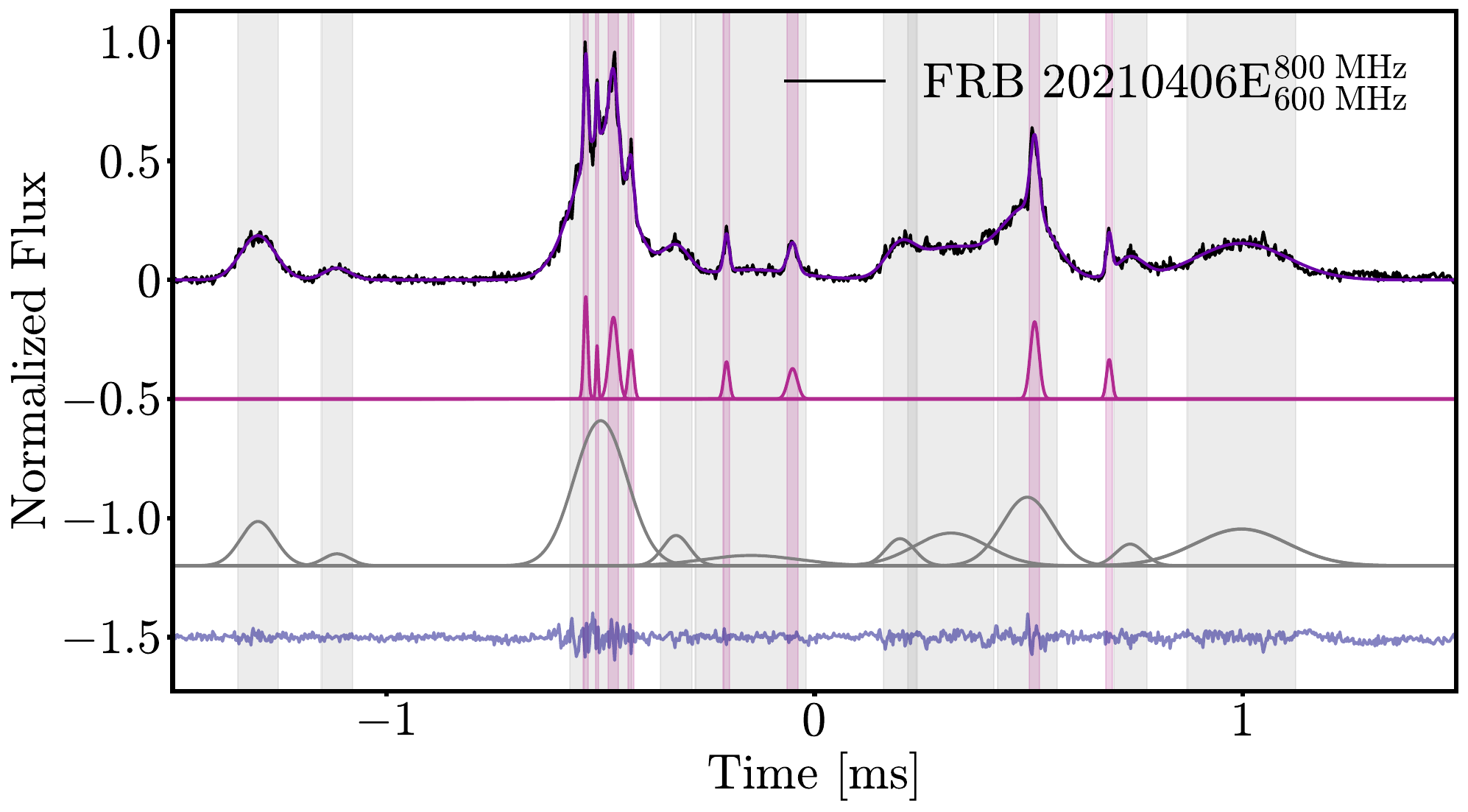}}
  \subfigure{\includegraphics[width=0.48\textwidth]{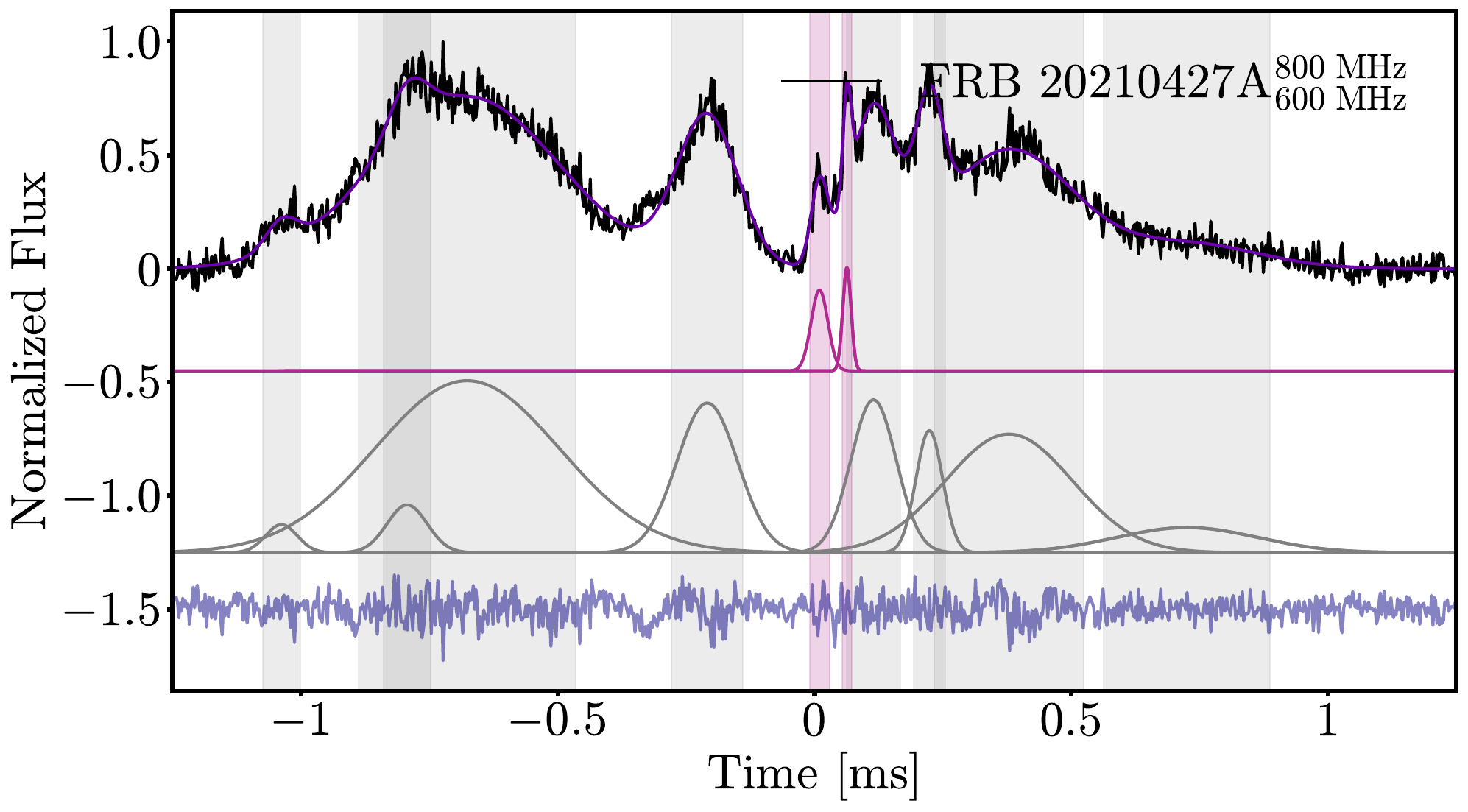}}\\
  \vspace{-0.25cm}
  \hspace{-4.2cm}
  \subfigure{\includegraphics[width=0.485\textwidth]{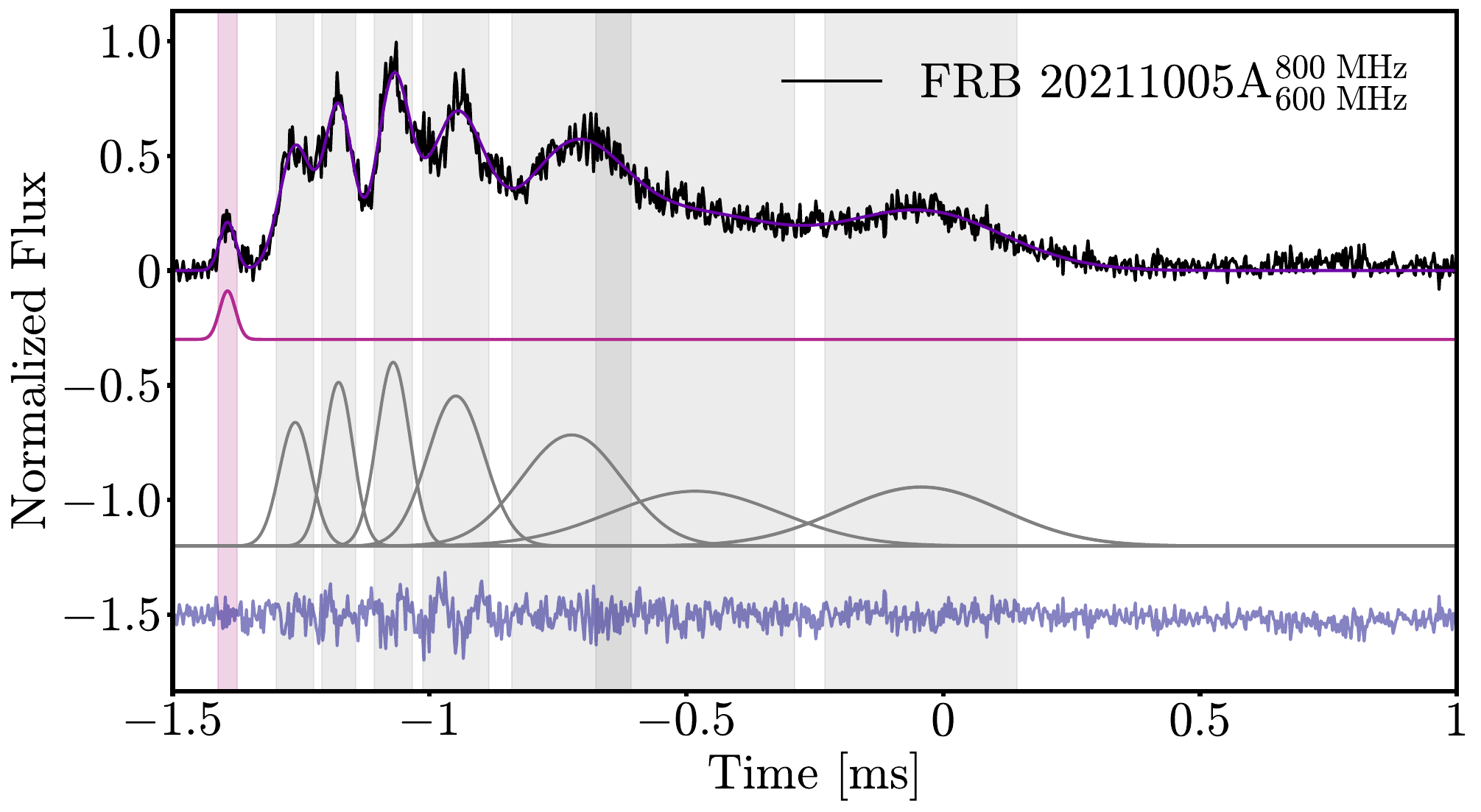}}
  \subfigure{\includegraphics[width=0.25\textwidth]{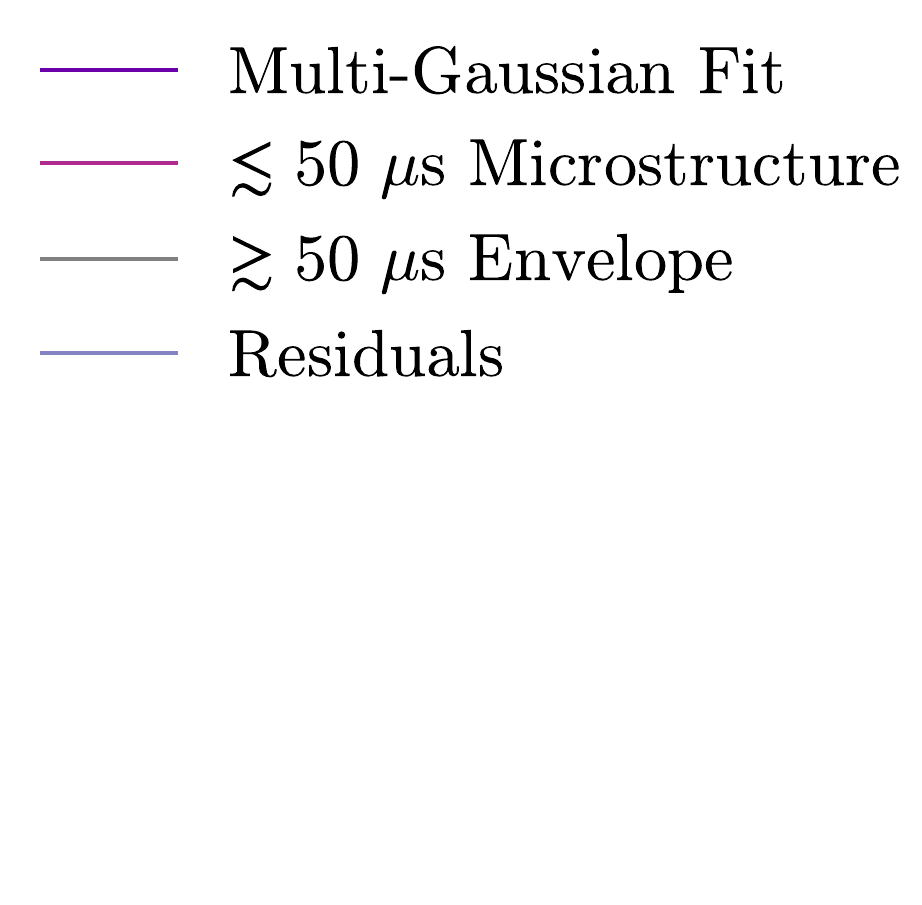}}
  \caption{A microstructure analysis of seven of the twelve bursts the sample. The purple curve plotted over the burst timeseries in each panel shows a multi-Gaussian fit to all features in the burst. Microstructure in five of the seven are significantly masked by flux envelopes in the lower half of the band, the origin of which is not well understood. The magenta curves show the isolated Gaussians from the fit with FWHM $\lesssim 50 ~\mu s$ (i.e., ``microstructure'') while the grey curves show the isolated Gaussian fits with FWHM $\gtrsim 50 ~\mu s$ (i.e., ``envelope''), together forming the full multi-Gaussian fit. The blue curve shows the residuals of the fit. The methods of this analysis are described in \S\protect\ref{subsec:micro}, and the intrinsic widths of $\lesssim 50$-$\mu s$ sub-bursts for each event are reported in Table \protect\ref{table:microgauss}.}
  \label{fig:micro}
\end{figure}

The majority of the bursts in our sample contain narrow $\mu$s features that rival some of the narrowest features seen to date in both repeating \citep{Nimmo2021, Nimmo2022, Majid2021, Snelders2023, Hewitt2023} and non-repeating \citep{Farah2018, Farah2019, Day2020} sources. While complex microstructure appears to be relatively common in repeaters, its prevalence in non-repeaters is not well studied. As many of the bursts in our sample are quite narrow across the full burst profile, we set a limit of $\lesssim 50 ~\mu s$ for the sub-bursts that we consider to be ``microstructure''. In turn, we take all sub-bursts with widths $\gtrsim 50 \; \mu s$ to collectively describe the broader flux distribution or ``envelope'' across the burst. Due to the numerous sub-bursts present in many of the events in this sample, we set this admittedly arbitrary limit to highlight the most strikingly narrow features. We identify microstructure in seven of the twelve bursts in this sample by performing multi-Gaussian fits to the timeseries and ACF measurements for sub-divided regions surrounding individual sub-bursts and sub-burst clusters. For most bursts, we fit the timeseries for the integrated upper half (600-800 MHz) of the observing band rather than the full band, to enhance the S/N of narrow features. We report the measured Gaussian widths of $\lesssim 50~\mu s$ features in Table \ref{table:microgauss}. The fits themselves are shown Figure \ref{fig:micro}.


The microstructure is measured in three steps, which in short reduce to: (1) isolating the upper half of the band if the sub-structure is sufficiently broadband and if broadening (either due to scattering or some other mechanism) is present in the lower half of the band, (2) performing a multi-Gaussian fit to the timeseries using the minimum number of Gaussians that sufficiently characterize the envelope and sub-structure (such that the residuals contain no apparent signal), (3) isolating the identified sub-structures in time, calculating the autocorrelation functions for the specified regions (ACF), and fitting a 1D Lorentzian to the ACF to the validate the previous Gaussian fit.

To ensure that we do not fit for noise spikes, we implement two tests. First, we calculate a 5-$\mu s$ window rolling mean across the burst profile to search for signals that exceed $3\sigma$ in amplitude with respect to the noise statistics measured in the off-pulse region, which we assume to be Gaussian. Second, we measure the maximum width and prominences of noise spikes in the off-pulse region to ensure that the features fit for in the multi-Gaussian profile exceed these values. While these tests do not fully take into account the effects of amplitude-modulated noise, which is expected to be boosted by the signal itself, we assume that these effects are sufficiently ignored by the widths measured from the ACFs. The ACF measurement methods are discussed in \ref{apB} and fits are reported in Table \ref{table:microacf}.

\begin{table*}[ht]
\caption{The measured widths (FWHM) of $\lesssim 50\mu s$ sub-bursts detected in seven events in the sample using a multi-Gaussian fit, as shown in Figure \protect\ref{fig:micro}. The methods used to perform the multi-Gaussian fits are outlined in \S\protect\ref{subsec:micro}.}
\centering  
\begin{tabular*}{1\textwidth}{@{\extracolsep{\fill}}c c c c c c c c c c} 
\hline\hline                       
{TNS Name} & $\Delta \mathrm{t}_1$ & $\Delta \mathrm{t}_2$ & $\Delta \mathrm{t}_3$ & $\Delta \mathrm{t}_4$ & $\Delta \mathrm{t}_5$ & $\Delta \mathrm{t}_6$ & $\Delta \mathrm{t}_7$ & $\Delta \mathrm{t}_8$\\   
\hline   
FRB 20190425A & 9.2(4) & 22(1) & 14.5(1) & 16.6(2) & 8.3(1) & 17.2(4) & 23.3(3) & ... \\
FRB 20200603B & 7.10(2) & 29(1) & 42.5(3) & 16.5(9) & ... & ... & ... & ...\\
FRB 20210406E & 11.1(1) & 6.3(1) & 23.8(2) & 14.3(3) & 15.0(3) & 26.4(5) & 24.0(2) & 15.0(3) \\
FRB 20210427A & 38.5(8) & 19.1(5) & ... & ... & ... & ... & ... \\
FRB 20210627A & 44.6(3) & 31.5(5) & ... & ... & ... & ... & ... & ...\\
FRB 20210813A & 22.3(5) & 38(1) & ... & ... & ... & ... & ... & ...\\
FRB 20211005A & 36.8(9) & ... & ... & ... & ... & ... & ... & ... \\
\hline\hline  \\                              
\end{tabular*}
\label{table:microgauss}
\end{table*}

\subsection{Quasi-Periodicites}\label{subsec:quasiperiod}

\begin{figure}[t]
    \centering
    \subfigure{\includegraphics[width=\textwidth]{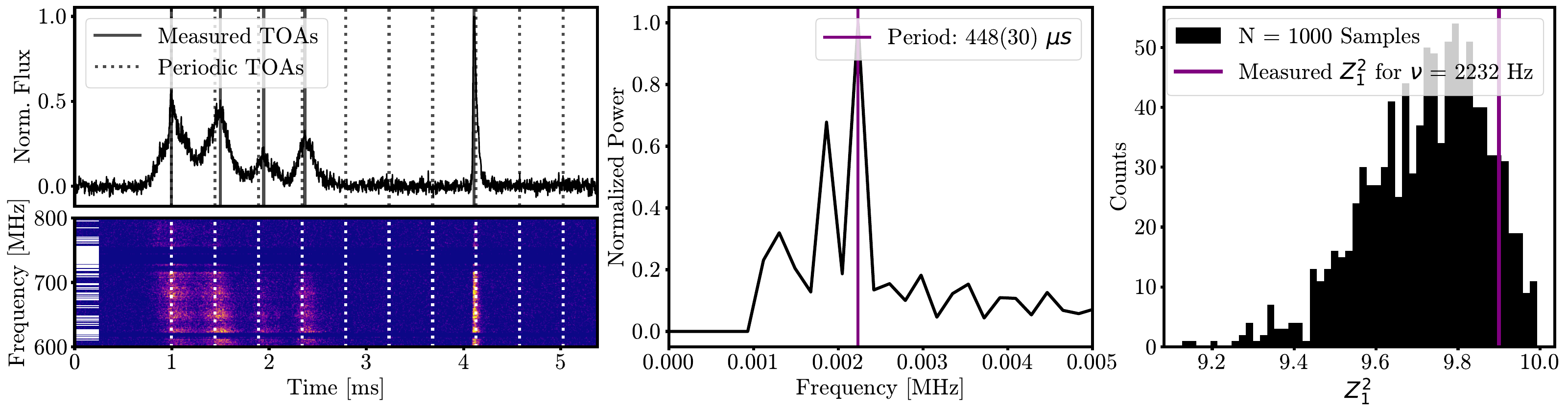}}
    \caption{A diagnostic plot of the power spectrum burst analysis for FRB 20210819A. The \textit{left} panel shows the burst dynamic spectrum for the upper half of the band (600-800 MHz) with a putative period corresponding to the highest peak in the power spectrum, plotted in white dotted lines. The burst timeseries is also shown, with the measured ToAs (solid grey lines) and putative spin period again (dotted grey lines) to highlight the respective ToA offsets. The \textit{middle} panel shows the burst power spectrum, where the peak indicates the best estimate of the period of the burst. The \textit{right} panel shows the results of the Rayleigh significance test, which indicates a statistical significance of 1.1(2)$\sigma$ for the measured period of the burst. The vertical purple line marks the $Z_{1}^{2}$ value of the measured period.}
    \label{fig:period}
\end{figure}

As many of these bursts contain multiple sub-components, we search for periodic or quasi-periodic substructure using the Rayleigh significance test outlined in detail by \citet{chime2021}. To identify candidates in our sample that exhibit plausible quasi-periodic structure, we first calculate the power spectra of the timeseries of each burst and search for prominent peaks in the power spectra. The only event in our sample that appears to show encouraging evidence for a periodicity in its power spectrum is FRB 20210819A, as shown in Figure \ref{fig:period}. As this burst exhibits more complex features below 600 MHz, we choose to isolate the upper-half (600-800 MHz) of the band for the purposes of this analysis, so as to better resolve each individual sub-burst. To evaluate the significance of the periodicity indicated by the most prominent peak in the power spectrum, we use the Rayleigh significance statistic ($Z_1^2$), which has been used previously by \citet{chime2022} to identify the first sub-second periodicity in an FRB, and is commonly used to search for periodicities in high-energy pulsar emission \citep{Buccheri1983, deJager1994}. The Rayleigh statistic is a significance measure for periodicities in irregularly sampled data, defined in general form ($Z_n^2$) as:

\begin{equation}\label{eq:rayleigh}
Z_n^2=\frac{2}{N} \sum_{k=1}^n\left[\left(\sum_{j=1}^N \cos k \phi_j\right)^2+\left(\sum_{j=1}^N \sin k \phi_j\right)^2\right]
\end{equation}
where $N$ is the number of pulses, $n$ is the number of harmonics (for which we choose $n = 1$, in accordance with the Rayleigh test), and $\phi_j$ represents the pulse phases. 
We perform a blind periodicity search using Eq. \ref{eq:rayleigh} by randomly generating time-of-arrival (ToA) differences, drawn from a uniform probability distribution \citep[see][for details; the same exclusion parameters are applied]{chime2022}. We generate $10^3$ random ToA distributions with the same number of components and duration as FRB 20210819A, sufficient to uncover a notable statistical significance, and apply the Rayleigh test to each distribution, shown in Figure \ref{fig:period}. Finally, we compare the $Z_1^2$ values obtained for each simulation to that obtained for the ToAs measured for FRB 20210819A. The most significant (quasi-)period indicated by the power spectrum of FRB 20210819A is measured to be $448(40) ~\mu$s ($2232(230)$ Hz), also shown in Figure \ref{fig:period}. 
When we apply the Rayleigh significance test, we find that the significance of this periodicity is only 1.1(2)$\sigma$, insufficient to indicate a detection.

We further note that the quasi-periodicity likely exceeds the maximum estimated rotation frequency possible for a neutron star under currently known equations of state, suggesting it is unlikely to be attributed to neutron star rotation \citep{lp16}.

\section{Discussion}\label{sec:discussion}

\subsection{Revisions to Time-Frequency Drifting Archetypes}\label{subsec:archetypes}

The events in our sample offer an opportunity to revisit our current understanding of the FRB morphologies, and possible archetypes, first proposed by \citet{Pleunis2021}. These include ``simple broadband,'' ``simple narrowband,'' ``temporally complex,'' and ``downward drifting''. As seen in previous investigations of high-resolution FRB data \citep{Farah2018, Farah2019, Hessels2019}, bursts previously considered ``simple'' may only be simple by virtue of the resolution at which the data are stored. On $\mu$s timescales, many FRBs are seen to be complex. Based on the bursts in our sample, it is clear that ``drifting'' in FRBs can deviate from the linear negative drifting, or ``downward drifting'' archetype, thus motivating us to 
suggest new categories of drifting archetypes, as described in \S\protect\ref{subsec:measurearchetypes} above. We discuss possible physical origins for each archetype below.

\subsubsection{Power-Law Negative Drifting and the Nebula Toy Model}\label{sec:nonlinnegdriftdiscuss}


Of the twelve events in our sample, three show clear indications of power-law negative drifting: FRB 20210813A, FRB 20210627A, and FRB 20210427A, as described in \S\protect\ref{subsec:nonlinposdrift}. Only FRB 20210813A, however, appears to disagree with dispersive smearing, suggesting that the \citet{Metzger2022} toy model may be relevant. 

\citet{Metzger2022} provide a useful tool to compare the observed burst properties with theoretical predictions from other FRB models that invoke both relativistic shock and magnetospheric emission mechanisms \citep[e.g.][]{Lyutikov2020, Sridhar2021}. 
While the overall shape of the flux density envelope is explained by the model, it does not 
obviously account for sub-structure. \citet{Metzger2022} suggest that sub-structure may originate from propagation effects such as plasma lensing. We entertain this possibility by multiplying the model from the data and plotting the result in the right-most panel of Figure \ref{fig:metzgermodel} to highlight the ``hot spots'' in flux density that may arise due to amplifications by an intervening plasma lens (i.e., caustics). While we cannot conclusively extract the lens position or geometry, it motivates extending this model to include such effects. 

Evaluated in the context of the \citet{Metzger2022} model, the best-fit values for the drifting index $\beta = [+0.76^{+0.07}_{-0.06}, +0.77^{+0.05}_{-0.07}]$ for the two spectral components in FRB 20210813A seem to imply emission by a decelerating relativistic shock propagating into an expanding upstream medium. For this scenario, \citet{Metzger2022} estimate an index range of $\beta \simeq 0.2-0.7$. The precise range of drifting indices depends, however, on the density gradient $n(r) \propto r^{-k}$ of the upstream medium in the radial direction, which is unknown in this case, but is thought to lie in the range of $0 \lesssim k \lesssim 2$. 

\subsubsection{Linear Positive Drifting}\label{subsec:linposdriftdiscuss}
 
While FRB 20201230B is not the first FRB to show linear positive drifting, it is the first to show linear positive drifting between sub-bursts separated by $\lesssim 1$ ms. Linear positive drifting is not nearly as common as linear negative drifting in FRBs, and certainly not often seen in non-repeater events, but it has been detected in bursts from six other FRBs of which the authors are aware\footnote{The number of FRBs showing linear positive drifting may be higher, but as this feature is not widely reported, it is challenging to gauge its prevalence in the literature.}, two of which were seen by CHIME/FRB. The first detection was in a burst emitted by FRB 20180916B, which contained two components separated in time by $\sim$60 ms, exhibiting a drift rate of $\sim +$1.6 MHz ms$^{-1}$ \citep{chime2020}. The second was seen in the detection of two bursts emitted by SGR 1935+2154 \citep{chime2020b}, separated in time by $\sim$29 ms. The drift rate of the latter is more challenging to measure as the bursts extend beyond the CHIME/FRB band, but if the centroids were taken to be the lower and upper band limits (400 and 800 MHz, respectively), the drift rate would be close to $\sim$13 MHz/ms. The third instance of drifting, though not explicitly reported by the authors, was again observed in a burst from FRB 20180916B by \citet{Marthi2020} with the uGMRT array, showing a sub-burst separation of order $\sim$ 20 ms between $\sim 600$-$650$ MHz, hence a drift rate of $\sim +2.5$ MHz ms$^{-1}$. The fourth instance of drifting, again not explicitly reported, can be seen in a burst from FRB 20201124A observed with the Effelsberg telescope by \citet{Main2021}, which shows a sub-burst separation of order $\sim 8$ ms between $\sim 1300$-$1425$ MHz, indicating a drift rate of $\sim +15$ MHz ms$^{-1}$. The fifth instance of linear positive drifting was seen in three separate events, again of FRB 20201124A, by \citet{Zhou2022} with the FAST telescope during an extremely active period. The specific drift rates were not measured, the dynamic spectra suggest that the drifting sub-burst centroids span $\gtrsim 10$ ms in time and $\sim 1100$-$1400$ MHz in frequency, giving a rough upper limit of $\sim +30$ MHz ms$^{-1}$. Finally, the most recent instance of linear positive drifting was observed in FRB
20190630D by \citet{chime2023b}. While the drift rate has not yet been carefully measured, the dynamic spectrum appears to suggest a frequency span of $\sim 200$ MHz, and a time span of $\sim 2$-$3$ ms, resulting in an estimated drift rate of $\sim +66$-$100$ MHz ms$^{-1}$.

Overall, the drift rates of these events are notably less dramatic than that seen in FRB 20201230B, which we measured to be $+285(50)$ MHz/ms. This difference is due to the sub-millisecond separations of each sub-burst in FRB 20201230B.

Notably, linear positive drifting has already been seen in some radio pulsars \citep[e.g.][]{Bilous2022}. One possible mechanism by which linear positive drifting can occur is via radius-to-frequency mapping in the magnetospheres of 
neutron stars, both
pulsars and magnetars \citep{Ruderman1975, Manchester1977, Beloborodov2017, Lyutikov2020, Bilous2022}. This occurs when charged particle bunches move along curved magnetic field lines in the magnetosphere, which rotate around the magnetic axis, allowing different bunches to cross the line of sight at different times, producing a series of sub-bursts that drift upward or downward in frequency, the direction dependent on the geometry of the dipolar field and the viewing angle \citep{Bilous2022}. The rate of the drifting is thought to depend on the rotation rate of the loop and the curvature radius of the magnetic field lines. The assumption is that the plasma in the magnetosphere produces coherent emission with frequencies that depend on the height of the emitting region (higher frequencies originating at lower heights). The fact that FRB 20201230B also exhibits a moderate PA swing of approximately 80 degrees, favors the hypothesis that emission originates from within the magnetosphere.  However, there are major challenges with FRBs emerging from neutron star magnetospheres.
\citet{bel21,bel23} argue that such luminous radio waves get damped in the magnetosphere, interacting with plasma particles, accelerating them to high energies at the expense of the wave energy.  If so, FRBs could not escape a magnetosphere. On the other hand, \citet{qkz22} argue that in the open field line region, the likely high outward plasma speed and the alignment of the field with the radio wave propagation direction mean the expected interaction between them is reduced, such that full damping is avoided. 


\subsubsection{Power-Law Positive Drifting and DM Variability}

There are several possible explanations for power-law positive drifting observed in FRB 20200711F, FRB 20210427A, FRB 20210627A, FRB 20211005A, and FRB 20220413B. Notably, this phenomenon bears some similarity to a subclass of Type III solar radio bursts \citep{Alvarez1973, Lyutikov2019b, Vedantham2020} which are generated by stimulated electron beams propagating along open magnetic field lines toward the stellar surface rather than away, the latter of which would lead to power-law negative drifting, as we see in the model proposed \citet{Metzger2022}. Comparisons between FRBs and solar radio bursts have been discussed previously by \cite{Hewitt2023}.

The frequency dependence of the power-law positive drifting features exhibited by events in this sample seems to agree with $t_{d} \propto \nu^{-2}$, and may therefore be explainable by propagation through a dispersive medium, such as a discrete plasma structure (or lens) with varying electron column densities. 
Plasma lensing occurs when electron over- or under-densities in a plasma structure along the line of sight magnify or de-magnify the radio signal from a distant source. In addition to magnification, lensing can lead to multiple-imaging. This occurs when radio waves propagate through a region of plasma with a density gradient, such as a clump or sheet, which leads to the formation of caustics (images) below a specific focal frequency. Below this frequency, images are expected to experience differential delays in time, dependent on the region of the lens that is being probed. These delays ought to be both dispersive, reflecting the DM depth of the region, and geometric, due to the differential path lengths traversed by each image. Hence, we expect multiply imaged FRBs to exhibit multiple burst components of different DM values.

Plasma lensing can occur in local magneto-ionic environments, where some FRBs are known to reside \citep[such as FRB 20121102A;][]{Michilli2018}, or at larger distances from the source. It has been suggested as a dominant agent in shaping the morphologies of radio pulses from certain Galactic pulsars in binary systems \citep{Main2018, Li2023, Lin2023}, as well as the Galactic Center magnetar, PSR J1745$-$2900 \citep{Pearlman2018}, which is believed to be embedded in a highly magneto-ionic environment as well.


Currently, plasma lensing has yet to be unambiguously demonstrated in FRBs, partially due to the still limited sample of well-localized FRBs measured at high time resolution. Plasma lensing has been suggested, however, as a possible contributor to the morphologies of some FRBs, such as FRB 20121102 \citep{Gajjar2018, Michilli2018, Hessels2019, Platts2021}. The persistence of downward drifting morphologies in bursts from FRB 20121102 without a comparable occurrence of upward drifting, however, calls this claim into question. Lensing has also been suggested as a possible cause for quasi-periodic temporal modulations seen in select bursts from FRB 20180916B \citep{Nimmo2021}, though these features could also be explained by self-modulation \citep{Sobacchi2020}. 

The disparate degrees of seemingly dispersive drifting between multiple sub-bursts observed in FRB 20220413B, FRB 20200603B, FRB 20210427A, FRB 20210627A and FRB 20211005A, as highlighted by the DM curves plotted over drifting features in their respective spectra (see Figure \ref{fig:nonlinupdrift}, \ref{fig:moredrift} and Table \ref{table:dmvar}), seem to suggest that plasma lensing may play an important roll in shaping the morphologies of these bursts.
Dispersive drifting of this kind would be expected in the case of multiple imaging. The simplest lens model that explains the upward drifting morphology of FRB 20220413B by plasma lensing, is one that invokes a convergent (i.e. underdense) Gaussian lens, and assumes that the emission is probing different transverse regions of the lens, provided the lens has a non-zero effectively velocity with respect to the source. If this scenario is valid, the varying electron densities in each region of the lens would cause the multiple images to drift upward in accordance with the specific DM depth ($\Delta$DM$_{\l}$) of that region. Such scenarios have been modeled previously by \citet{Platts2021}. We measure these $\Delta$DM$_{\l}$ values in \S\protect\ref{subsec:linposdrift} and report them in Table \ref{table:dmvar}. 

As FRB 20220413B shows the clearest example of apparently dispersive delays below a visually apparent focal frequency, we will explore explore the possibility of plasma lensing being the dominant agent in shaping the morphology using conventional lensing theory outlined in \citet{Cordes2017}.

The condition that must be satisfied in order to produce multiple images below a focal frequency $\nu_l$ can be defined using Eq. 8 from \cite{Cordes2017}:

\begin{equation}
\nu_l \approx 39.1 ~\mathrm{GHz} \times \frac{|\Delta\mathrm{DM}_{l}|^{1 / 2}}{a}\left(\frac{d_{\mathrm{sl}} d_{\mathrm{lo}} / d_{\mathrm{so}}}{1 \mathrm{kpc}}\right)^{1 / 2} ,
\end{equation}
where $|\Delta\mathrm{DM}_{l}|$ is the DM depth of the lens at a specific region (in pc cm$^{-3}$), $\nu_{l}$ is focal frequency (i.e., the frequency at which multiple-imaging, or drifting is observed in GHz), $d_{\mathrm{sl}}$ is the distance from the source to the lens (in pc), $d_{\mathrm{so}}$ is the distance from the source to the observer (in Gpc), and $a$ is the characteristic lens scale (in AU). We assume a conservative geometry in this relation, placing the lens close to the source, such that $d_{\mathrm{sl}} = 1$ kpc and $d_{\mathrm{so}}/d_{\mathrm{sl}} \sim 10^6$. 
Of course, these values describe just one of many possible lensing geometries.


Under these assumptions, we can estimate the lower-limit of $a_{\mathrm{AU}}$ required for multiple-imaging to occur at $\nu_{l} \approx 545$ MHz (the focal frequency), where $| \Delta \mathrm{DM}_{l}| \approx 0.05$ pc cm$^{-3}$ (measured between the first and second sub-burst, see Figure \ref{fig:nonlinupdrift}) to be 

\begin{align}
    a \approx 16 ~\mathrm{AU} \times \Biggl(\frac{|\Delta \mathrm{DM}_{l}|}{0.05 ~\mathrm{pc ~cm^{-3}}}\Biggr)^{1 / 2}\Biggl(\frac{0.545~ \mathrm{GHz}}{\nu_{l}}\Biggr)\Biggl(\frac{d_{\mathrm{sl}} d_{\mathrm{lo}} / d_{\mathrm{so}}}{1 \mathrm{kpc}}\Biggr)^{1 / 2},
\end{align}
a reasonable estimate for the characteristic scale of the lens, as plasma lenses in the Milky Way are typically only a few AU across \citep{Bannister2016}.

To evaluate if the effects present in FRB 20220413B and other events truly originate from lensing, however, we can look for phase-coherence between the sub-bursts to assess whether these are in fact multiply imaged. This is possible given the phase-preserving nature of baseband raw voltages and is currently under study (Kader et al., in prep.). 



\subsection{Microstructure} \label{subsec:microdiscuss}

In the magnetospheric reconnection model, FRBs are generated via coherent curvature radiation or inverse Compton scattering by charged bunches moving along curved magnetic field lines. In this model, the burst duration is determined by the light-crossing time of the emission region, which is typically of order $\sim 10~ \mu s$ for a neutron star radius of order $\sim 10$ km. The burst morphology can be influenced by the plasma density distribution and the magnetic field configuration in the magnetosphere, as well as by propagation effects in the magnetosphere and the interstellar medium \citep{Cordes2017}. 

In the outflow model, however, FRBs are produced by synchrotron maser emission from relativistic shocks driven by magnetar flares \citep{Metzger2019,Beloborodov2019}. The burst duration, in turn, is determined by the shock crossing time of the outflow shell, which must be much longer than the light-crossing time of the central neutron star. The morphology may then be influenced by some combination of the shock dynamics, the outflow geometry, and the ambient medium. These models involve coherent emission mechanisms. As estimated by \citet{Metzger2019}, bursts emitted via relativistic shocks can span roughly $0.01$-$10$ ms in duration---values extending past the light crossing time of the neutron star by the propagation of the shock. Hence narrower features favor magnetospheric emission, if the burst durations are indeed intrinsic \citep{Beniamini2020}.

The narrowest sub-burst measured in our sample appears in FRB 20210406E with a width $\Delta t =6.3(1)$~$\mu s$ (consistent to within $\sim 1 ~\mu s$ as measured by the fitting the ACF; see \S\protect\ref{apB} and Tables \ref{table:microgauss}, \ref{table:microacf}).  Assuming $d_e \lesssim c \Delta t$, the upper limit on the diameter of the emission region, $d_e$, inferred from this would be $\approx$ 1.9 km. All of the narrowest sub-structures in the bursts that we measured suggest emission regions of order kilometers in size. This estimate neglects relativistic effects such as aberration and retardation \citep[e.g.][]{Blaskiewicz1991}.  



Alternatively, 
plasma lensing \citep{Cordes2017} and self-modulation \citep{Sobacchi2020} can also produce microstructure due to propagation through a turbulent, possibly magnetized plasma along the line of sight. This plasma can either reside in the interstellar medium (ISM) of the host galaxy or in the circumburst environment surrounding the FRB source \citep[such as a supernova remnant or a companion star;][]{Main2018}. 

\section{Conclusion}\label{sec:conclusion}


In this work, we have shown that thus far non-repeating FRBs can exhibit complex time-frequency structure, opening up a possible bridge between the morphological characteristics of both repeating and non-repeating FRBs. We have used this sample of events to further inform and refine our understanding of such characteristics, particularly as they relate to drifting. Furthermore, we suggest a new framework for classifying and interpreting drifting phenomenologies that goes beyond the linear negative drifting behavior observed in many repeating FRBs. We divide this framework into four morphological classes or ``archetypes,'' and consider both intrinsic emission mechanisms and extrinsic propagation effects that may be responsible for them. The most promising emission mechanisms considered in this paper include the synchrotron maser process and magnetospheric modulations of curvature radiation, expected to originate from compact objects such as magnetars. We further highlight events that show apparent frequency-dependent magnifications and dispersion measure variablity over time, which favor extrinsic propagation effects like plasma lensing. While this study includes just twelve newly discovered FRBs, the recent expansion in publicly available baseband raw voltage data for both apparent non-repeaters and repeaters by CHIME/FRB \citep{chime2023b} will allow for more detailed explorations of the proposed archetype classification framework, and enable more rigorous tests and constraints of the models outlined in this paper, as well as others. 

 \clearpage
\section{Acknowledgements}\label{sec:acknowledgements}

\begin{acknowledgments}

We acknowledge that CHIME is located on the traditional, ancestral, and unceded territory of the sylix (Okanagan) people.

We thank the Dominion Radio Astrophysical Observatory, operated by the National Research Council Canada, for gracious hospitality and expertise. CHIME is funded by a grant from the Canada Foundation for Innovation (CFI) 2012 Leading Edge Fund (Project 31170) and by contributions from the provinces of British Columbia, Québec and Ontario. The CHIME/FRB Project is funded by a grant from the CFI 2015 Innovation Fund (Project 33213) and by contributions from the provinces of British Columbia and Québec, and by the Dunlap Institute for Astronomy and Astrophysics at the University of Toronto. Additional support was provided by the Canadian Institute for Advanced Research (CIFAR), McGill University and the McGill Space Institute via the Trottier Family Foundation, and the University of British Columbia. The Dunlap Institute is funded through an endowment established by the David Dunlap family and the University of Toronto. Research at Perimeter Institute is supported by the Government of Canada through Industry Canada and by the Province of Ontario through the Ministry of Research \& Innovation. The National Radio Astronomy Observatory is a facility of the National Science Foundation (NSF) operated under cooperative agreement by Associated Universities, Inc. FRB research at UBC is supported by an NSERC Discovery Grant and by the Canadian Institute for Advanced Research. The baseband instrument for CHIME/FRB is funded in part by a CFI John R. Evans Leaders Fund grant to IHS.

This research used the Canadian Advanced Network For Astronomy Research (CANFAR) operated in partnership by the Canadian Astronomy Data Centre and The Digital Research Alliance of Canada with support from the National Research Council of Canada the Canadian Space Agency, CANARIE and the Canadian Foundation for Innovation.

JTF would like to thank the Fulbright Canada Foundation for funding this work and all members of the CHIME/FRB Collaboration at the Trottier Space Institute at McGill (tsi.mcgill.ca) for their gracious mentorship.\allacks

\end{acknowledgments}

\vspace{5mm}
\facilities{CHIME/FRB}

\software{astropy \citep{astropy2018}; scipy \citep{scipy2020}, bilby \citep{Ashton2019}, dynesty \citep{Speagle2020}}

\bibliography{refs}
\bibliographystyle{aasjournal}

\appendix

\section{Additional DM Variability Measurements}\label{apA}

\begin{figure}[h!]
    \subfigure{\includegraphics[width = 0.31\textwidth]{86321776_full_DMvar.pdf}}
    \subfigure{\includegraphics[width = 0.31\textwidth]{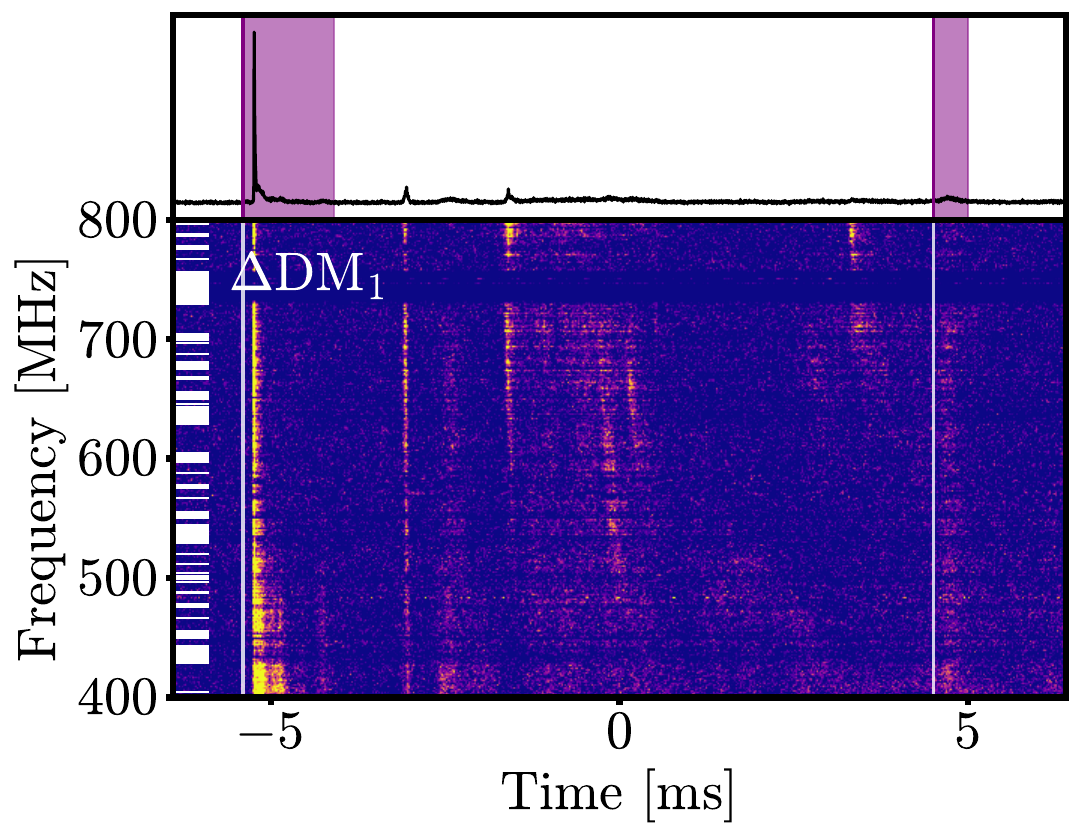}}
    \subfigure{\includegraphics[width = 0.31\textwidth]{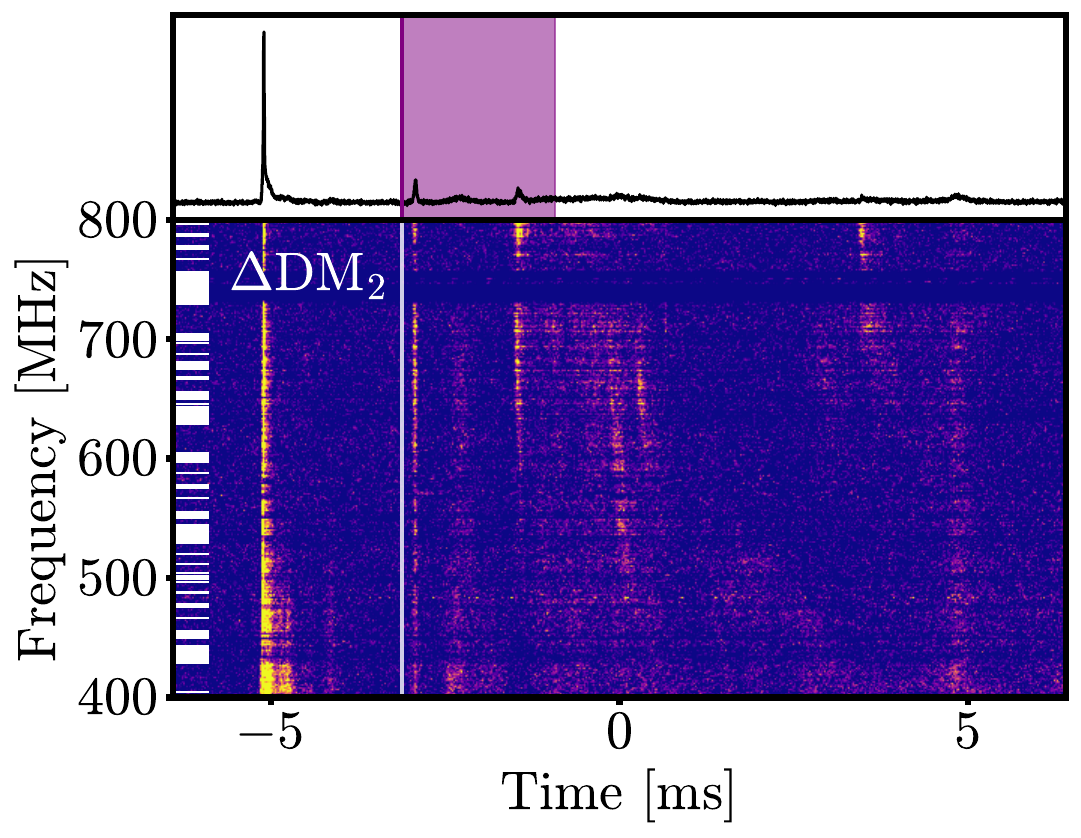}}\\
    \subfigure{\includegraphics[width = 0.31\textwidth]{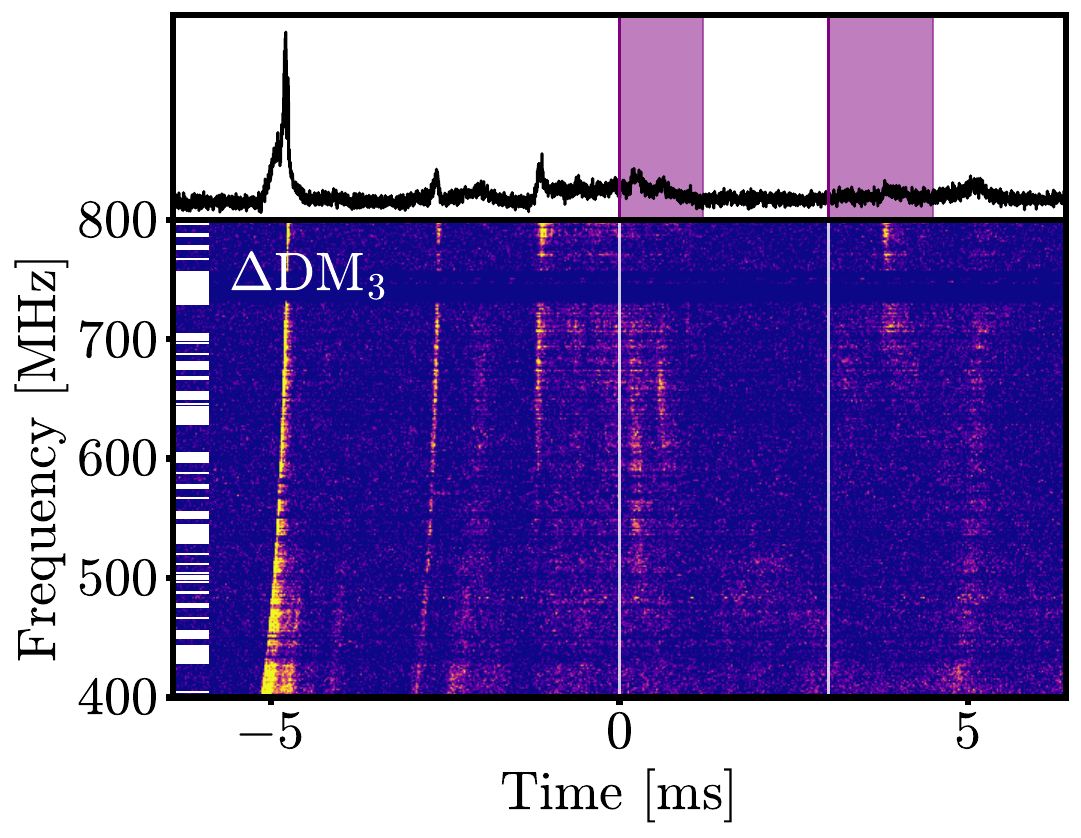}}
    \caption{The \textit{upper left} panel shows the dynamic spectrum of FRB 20200603B with DM curves over-plotted as white solid lines. The panels following show dynamic spectra of the same event dedispersed to the coherent power-maximizing $\Delta$DM offsets from the nominal DM for respective sub-bursts, highlighted in purple in the timeseries. The DMs to which the full burst is dedispersed are ordered by sub-burst ToA.}
    \label{fig:86321776_dmvar}
\end{figure}

\begin{figure}[h!]
    \subfigure{\includegraphics[width = 0.3\textwidth]{169427924_full_DMvar.pdf}}
    \subfigure{\includegraphics[width = 0.3\textwidth]{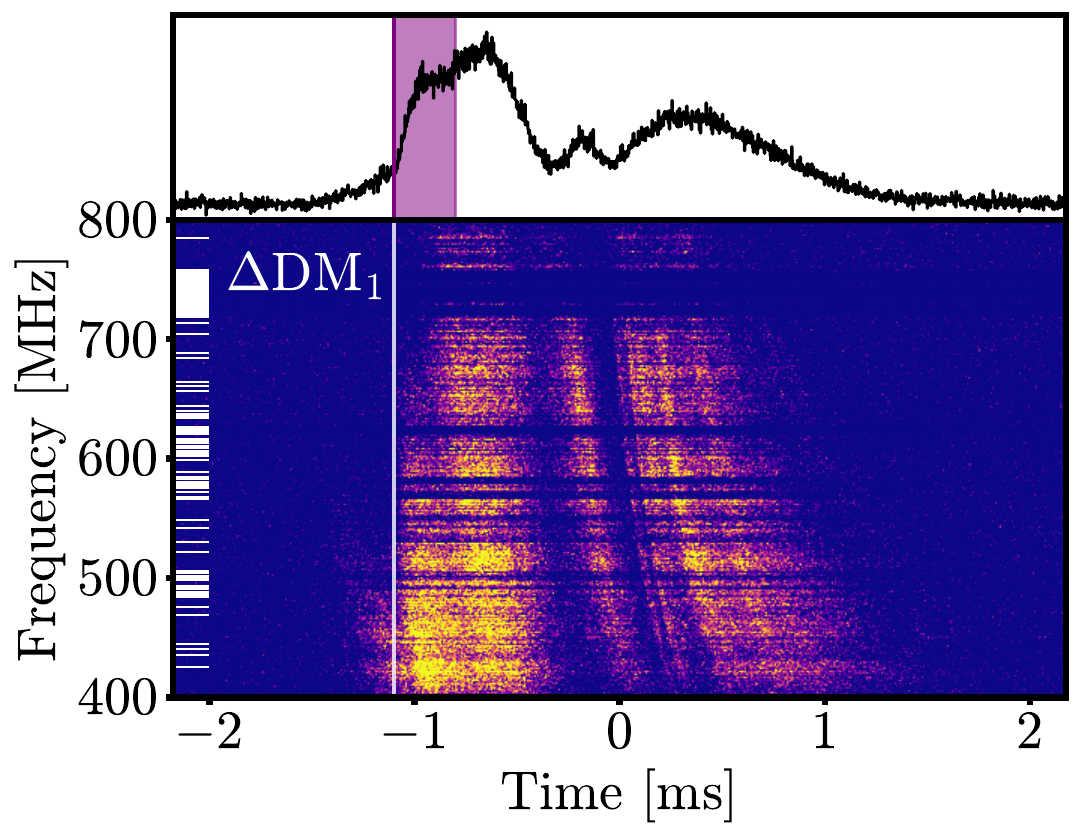}}
    \subfigure{\includegraphics[width = 0.3\textwidth]{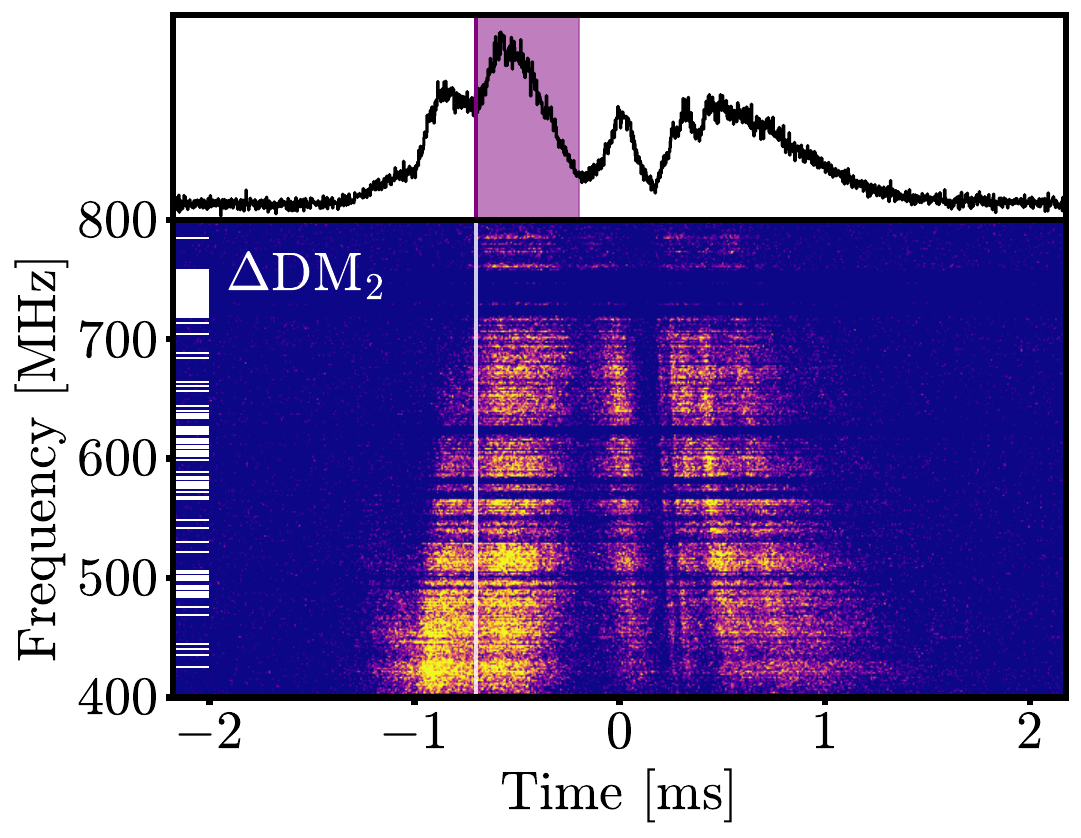}}\\
    \subfigure{\includegraphics[width = 0.3\textwidth]{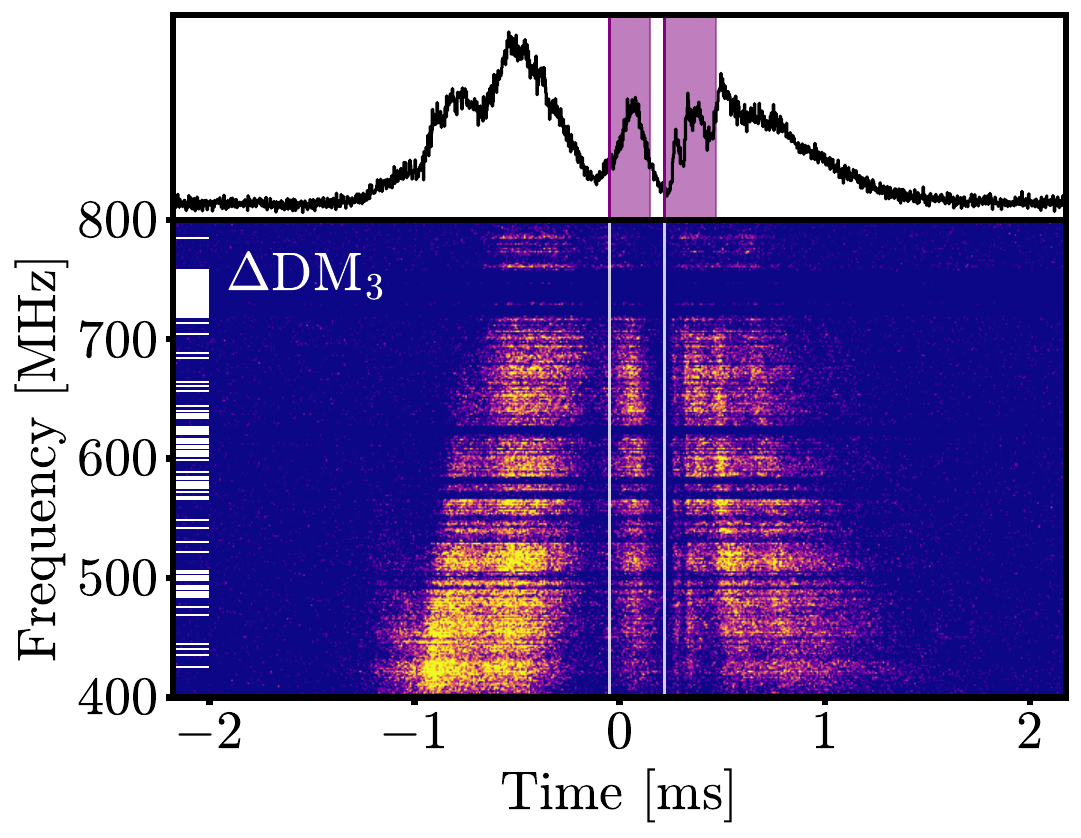}}
    \subfigure{\includegraphics[width = 0.3\textwidth]{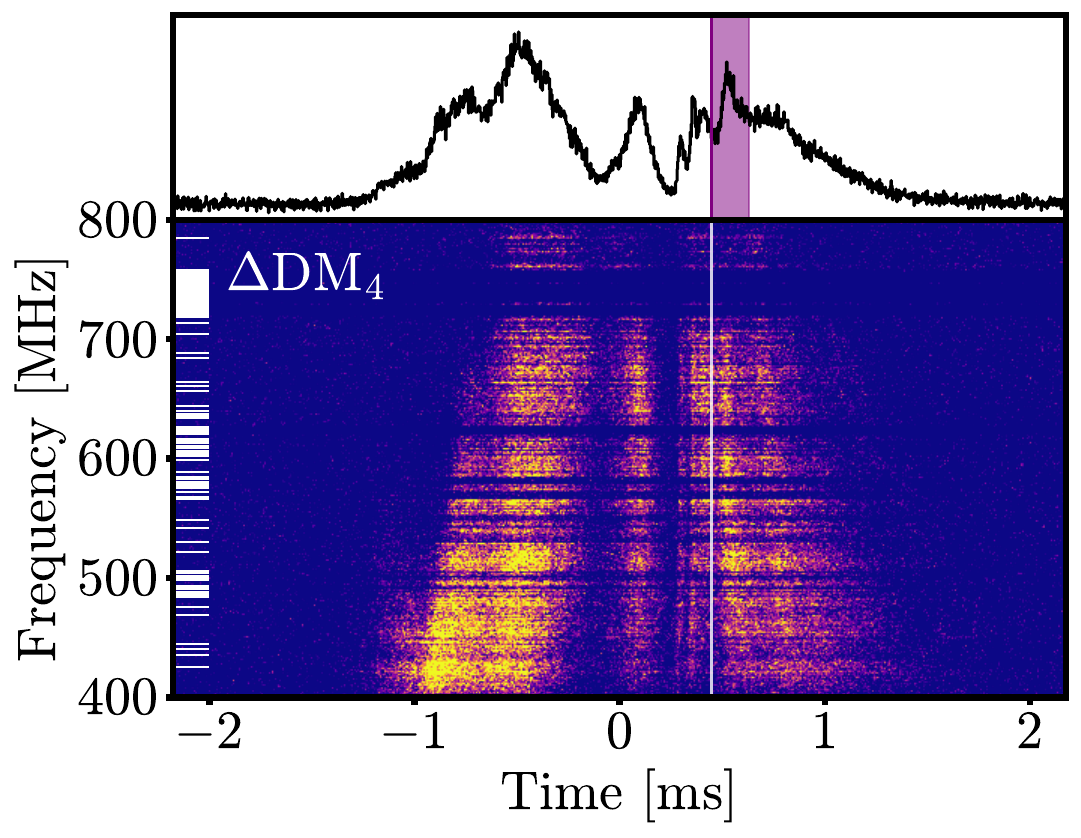}}
    \subfigure{\includegraphics[width = 0.3\textwidth]{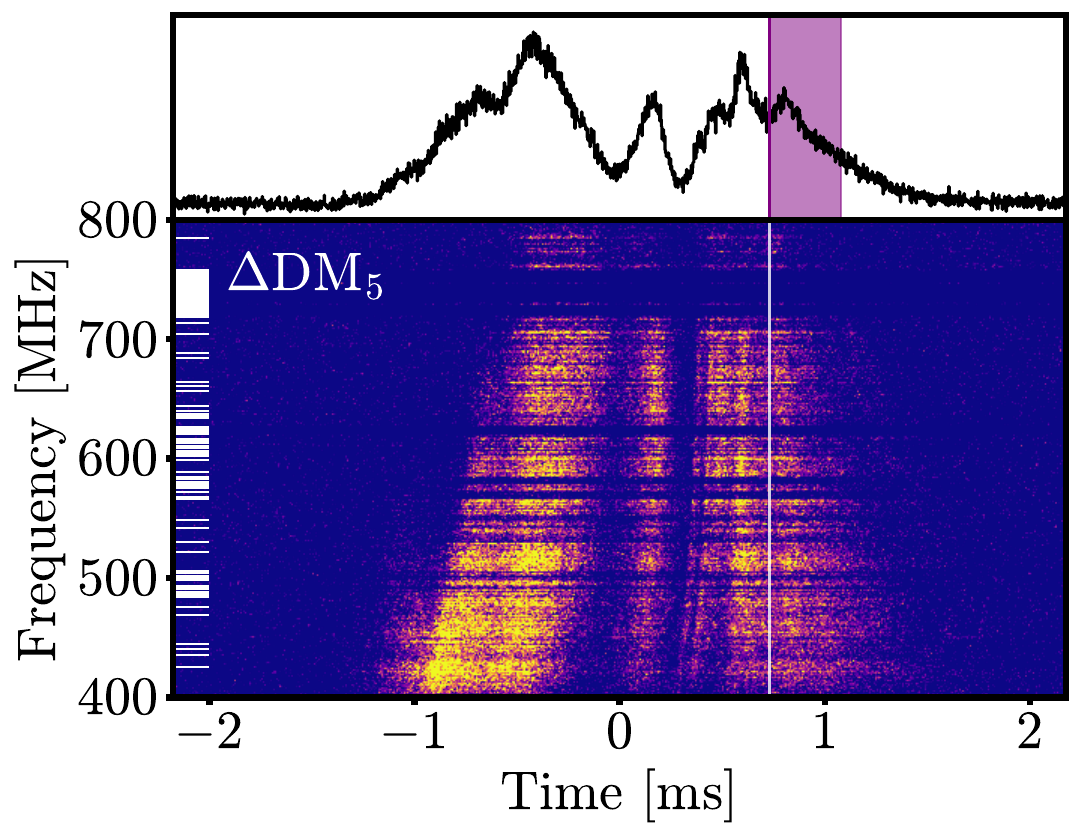}}\\
    \caption{The \textit{upper left} panel shows the dynamic spectrum of FRB 20210427A with DM curves over-plotted as white solid lines. The following panels show dynamic spectra of the same event dedispersed to the coherent power-maximizing $\Delta$DM offsets from the nominal DM for respective sub-bursts, highlighted in purple in the timeseries. The DMs to which the full burst is dedispersed are ordered by sub-burst ToA.}
    \label{fig:169427924_dmvar}
\end{figure}

\begin{figure}[h!]
    \subfigure{\includegraphics[width = 0.3\textwidth]{175128652_full_DMvar.pdf}}
    \subfigure{\includegraphics[width = 0.3\textwidth]{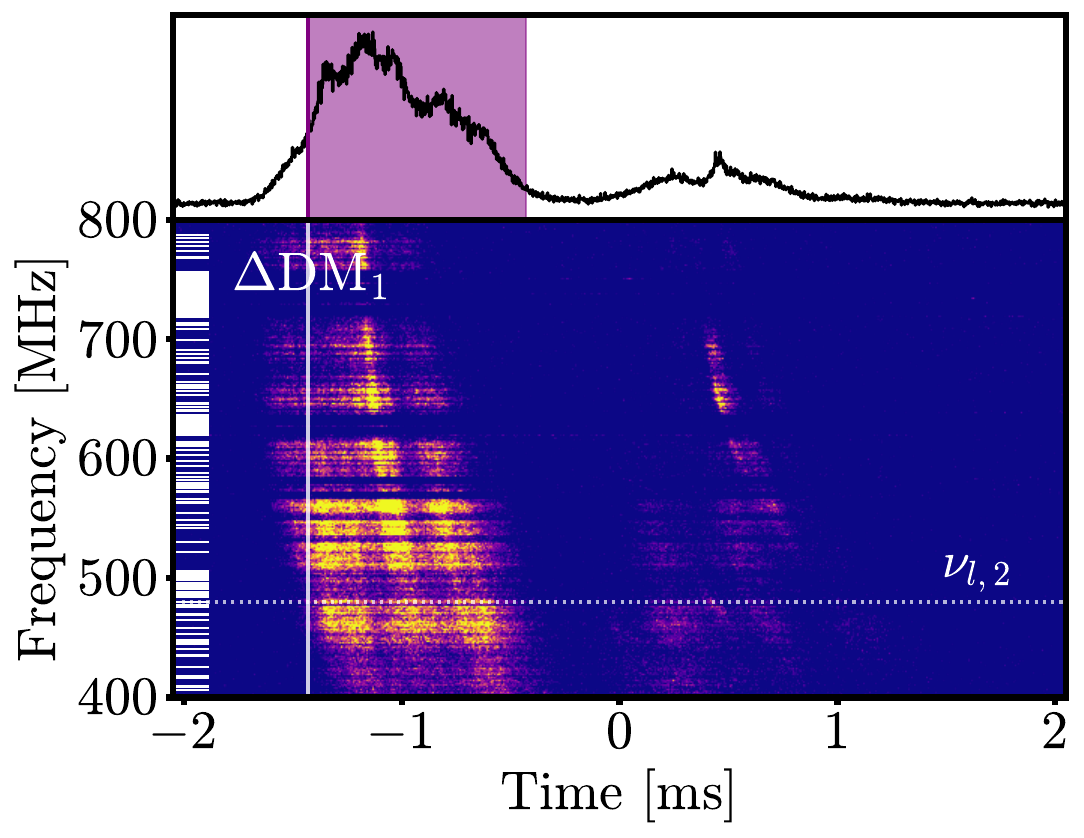}}
    \subfigure{\includegraphics[width = 0.3\textwidth]{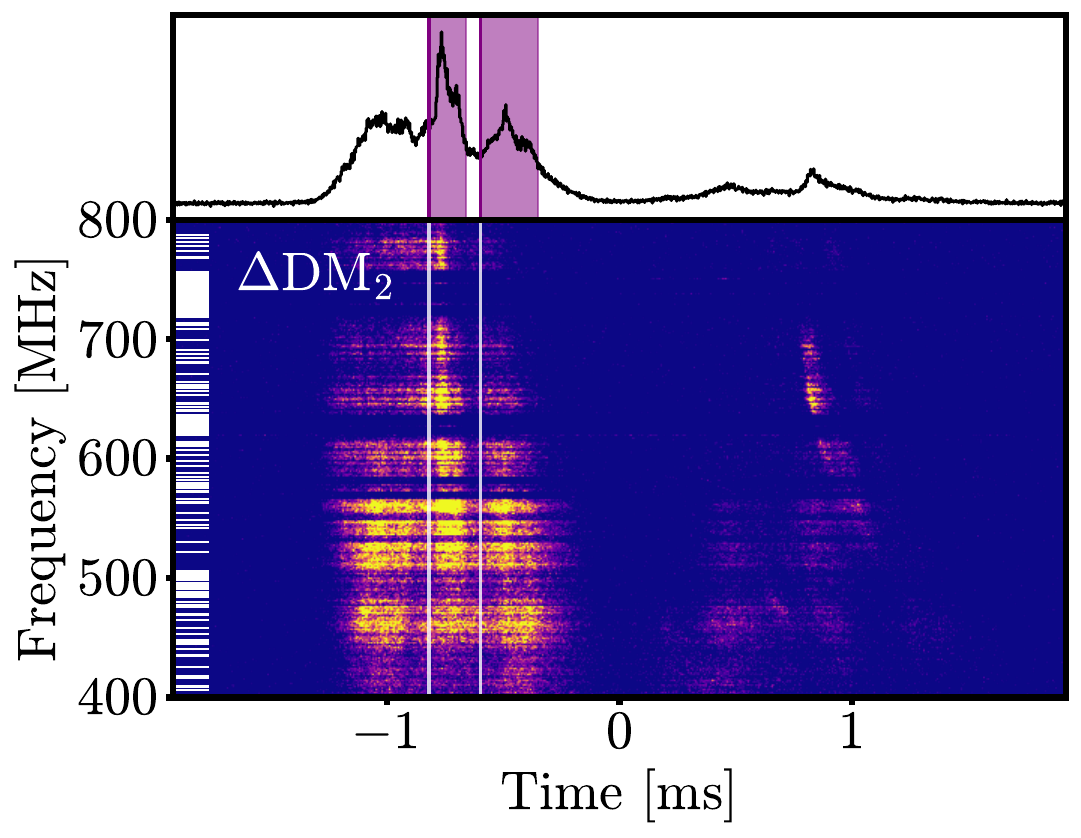}}\\
    \subfigure{\includegraphics[width = 0.3\textwidth]{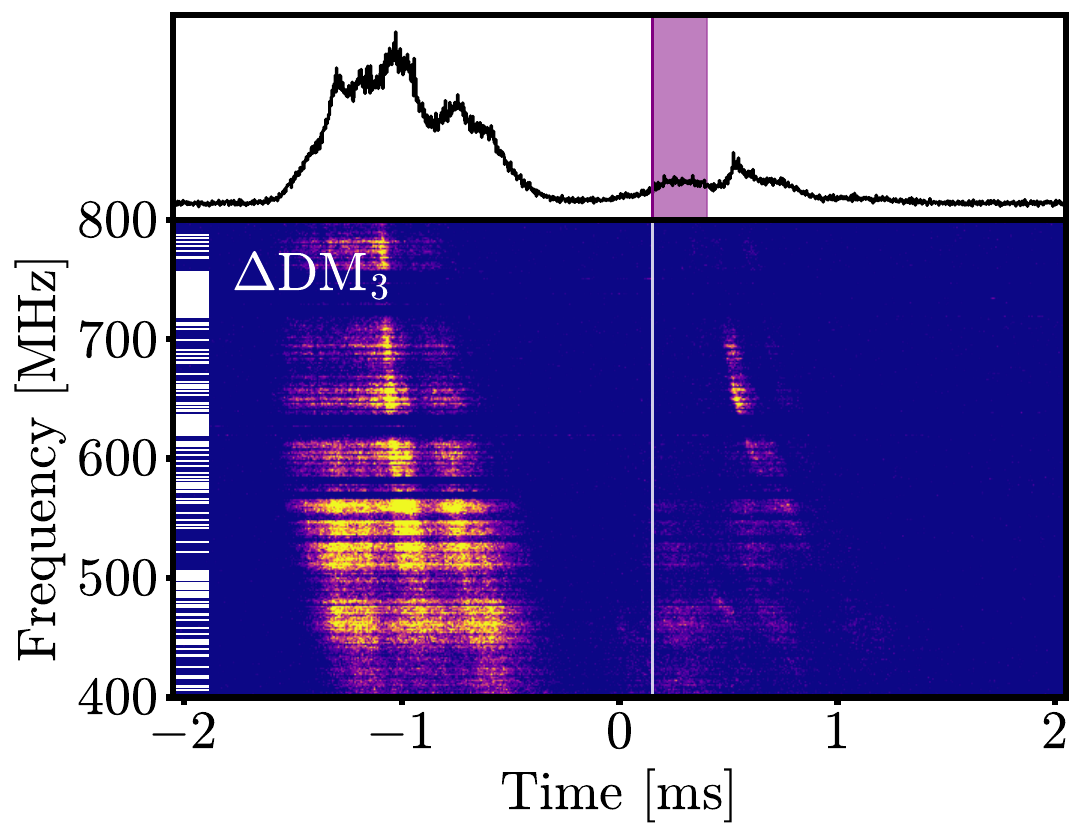}}
    \subfigure{\includegraphics[width = 0.3\textwidth]{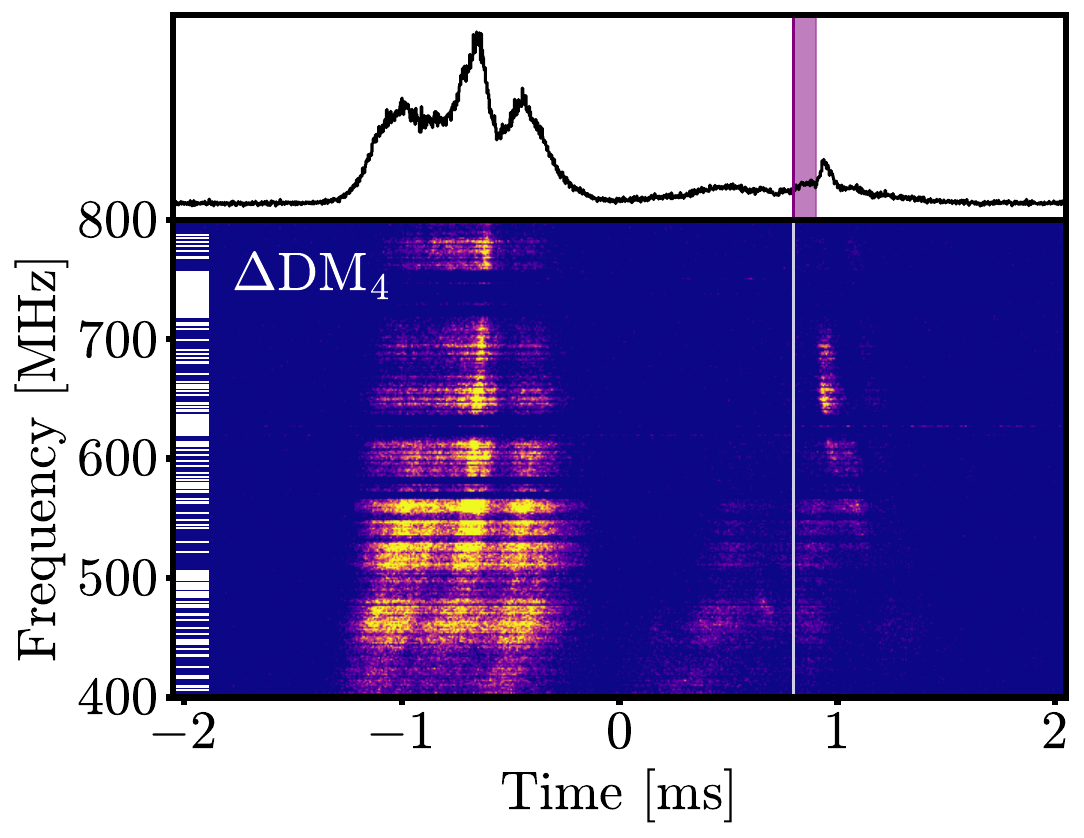}}
    \subfigure{\includegraphics[width = 0.3\textwidth]{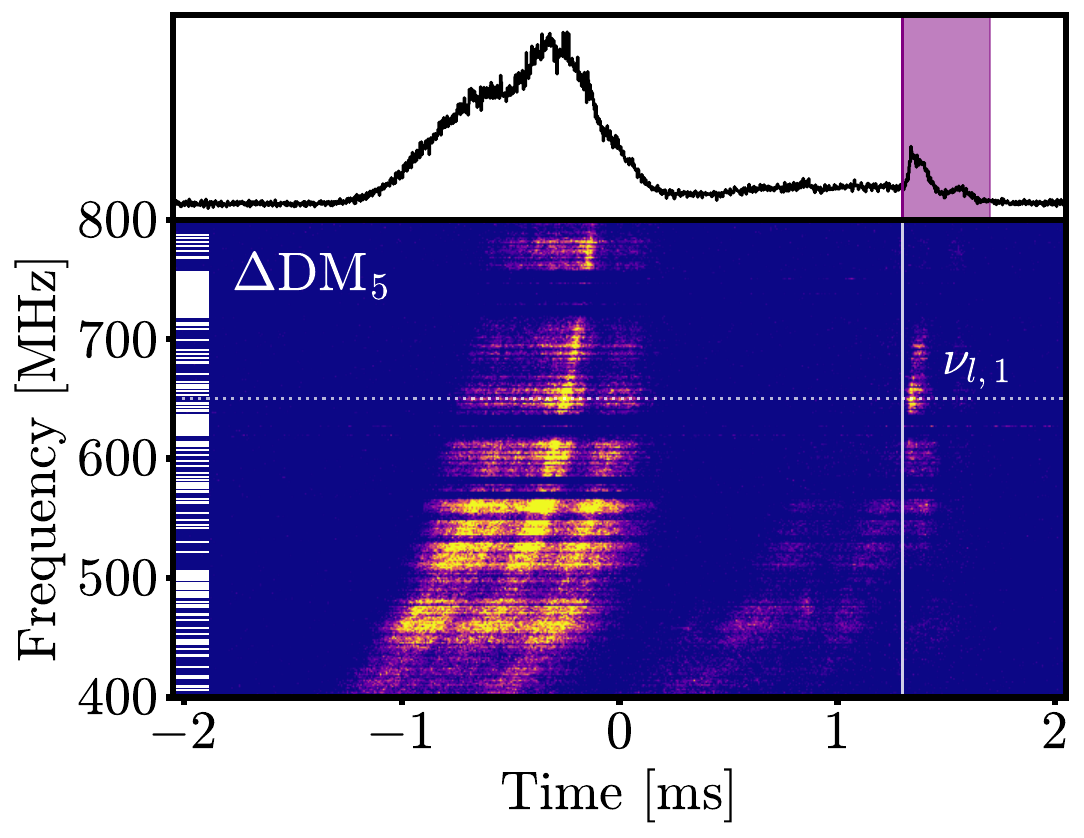}}
    \caption{The \textit{upper left} panel shows the dynamic spectrum of FRB 20210627A with DM curves over-plotted as white solid lines. The following panels show dynamic spectra of the same event dedispersed to the coherent power-maximizing $\Delta$DM offsets from the nominal DM for respective sub-bursts, highlighted in purple in the timeseries. The DMs to which the full burst is dedispersed are ordered by sub-burst ToA. Note that in measuring $\Delta \mathrm{DM}_{1}$, the coherent power has been maximized for the sub-band below a focal frequency of $\nu_{l,2} \approx 480$ MHz, at which point drifting away from the nominal DM is visible, and that $\Delta \mathrm{DM}_{5}$ is measured similarly below a focal frequency of $\nu_{l,1} \approx 650$ MHz.}
    \label{fig:175128652_dmvar}
\end{figure}

\begin{figure}[h!]
    \subfigure{\includegraphics[width = 0.3\textwidth]{189845329_full_DMvar.pdf}}
    \subfigure{\includegraphics[width = 0.3\textwidth]{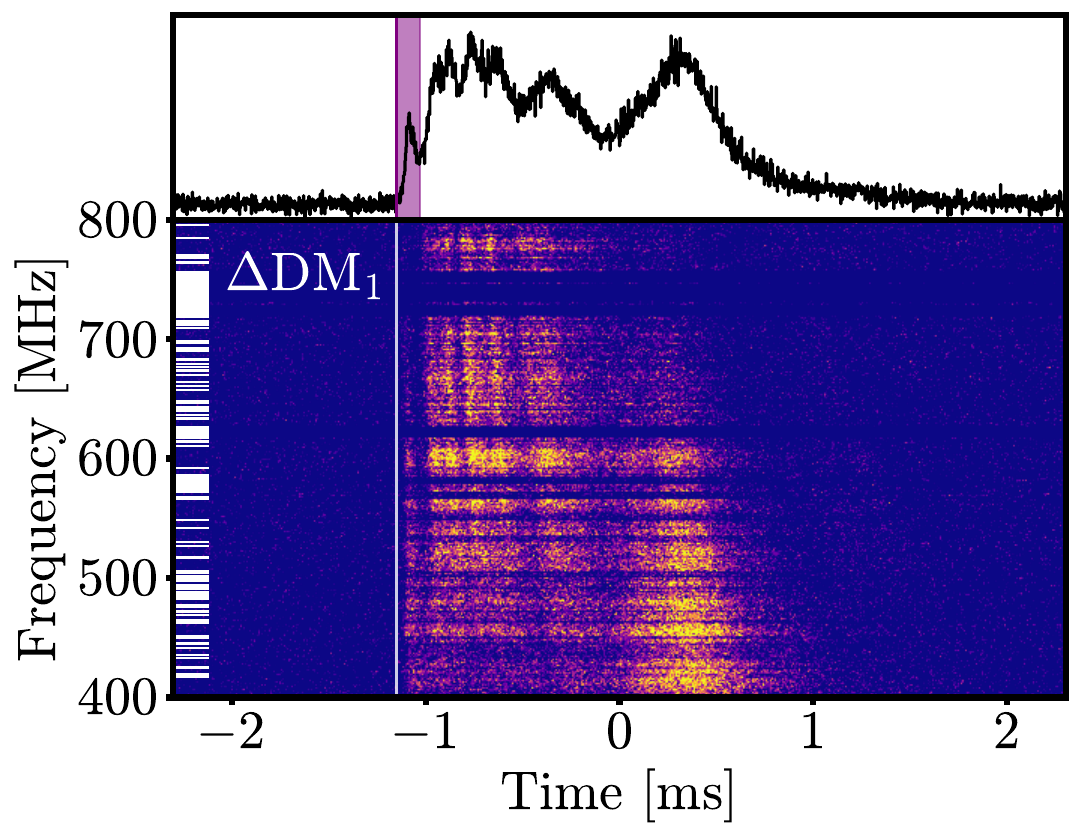}}
    \subfigure{\includegraphics[width = 0.3\textwidth]{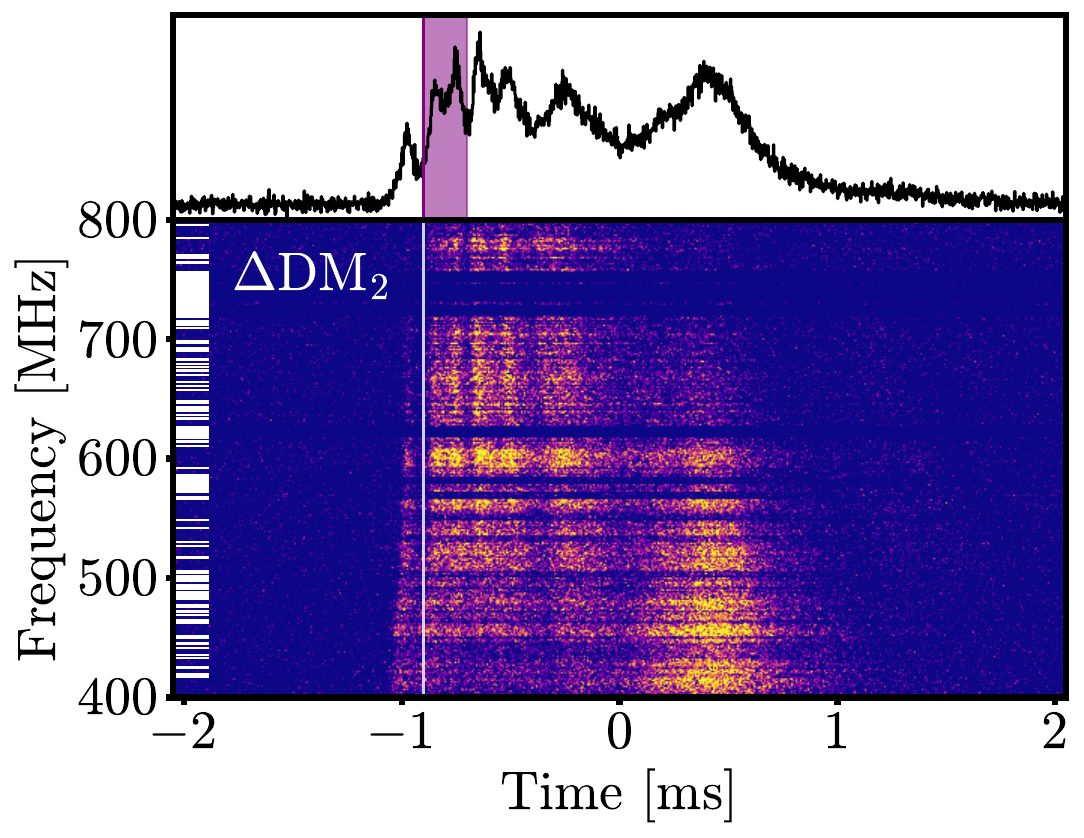}}\\
    \subfigure{\includegraphics[width = 0.3\textwidth]{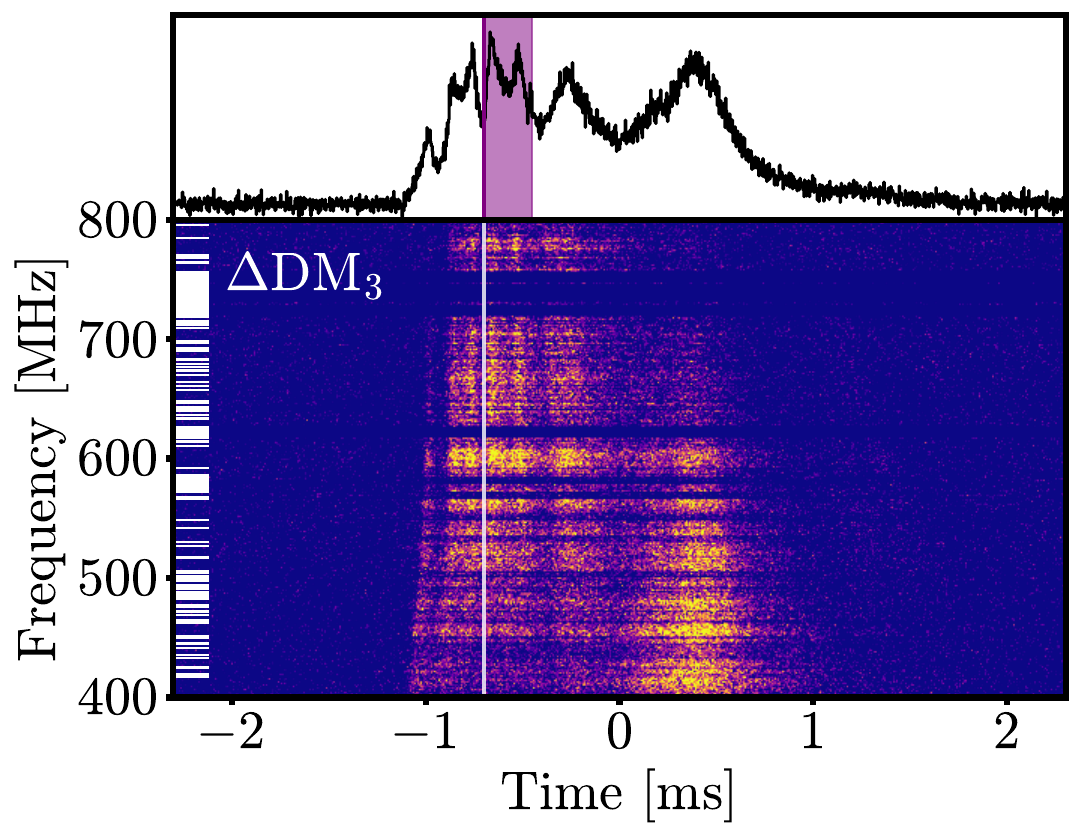}}
    \subfigure{\includegraphics[width = 0.3\textwidth]{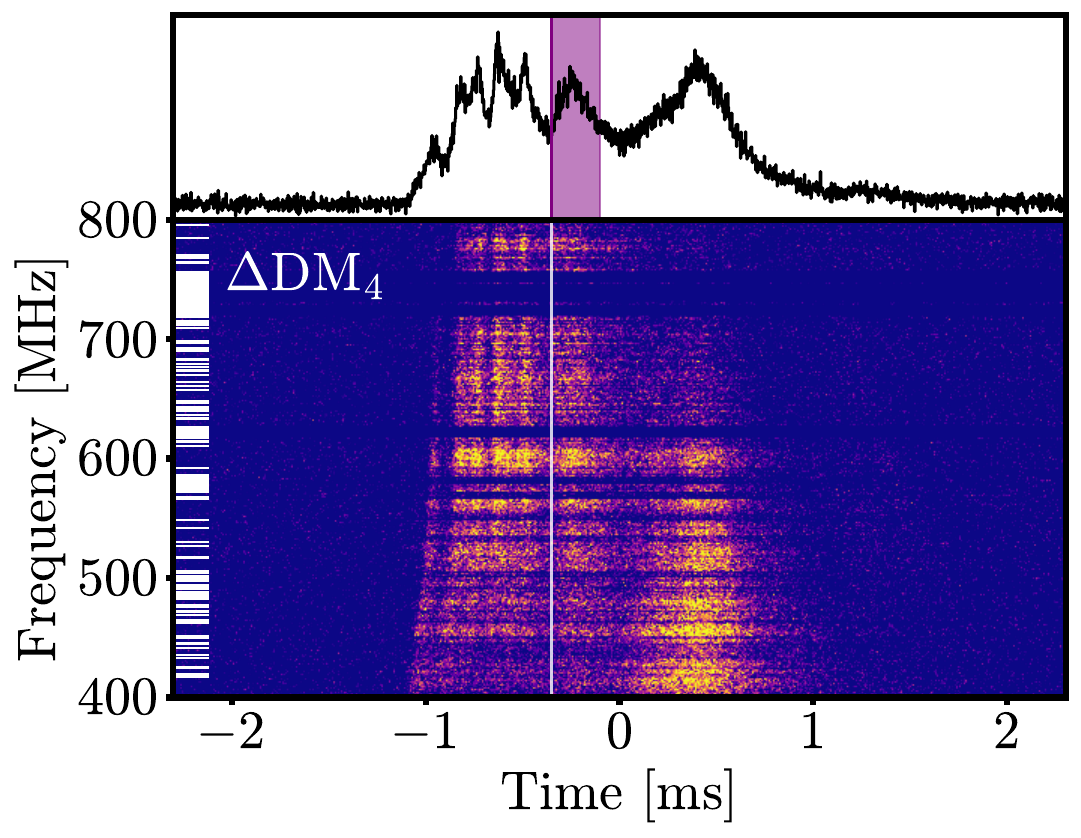}}
    \subfigure{\includegraphics[width = 0.3\textwidth]{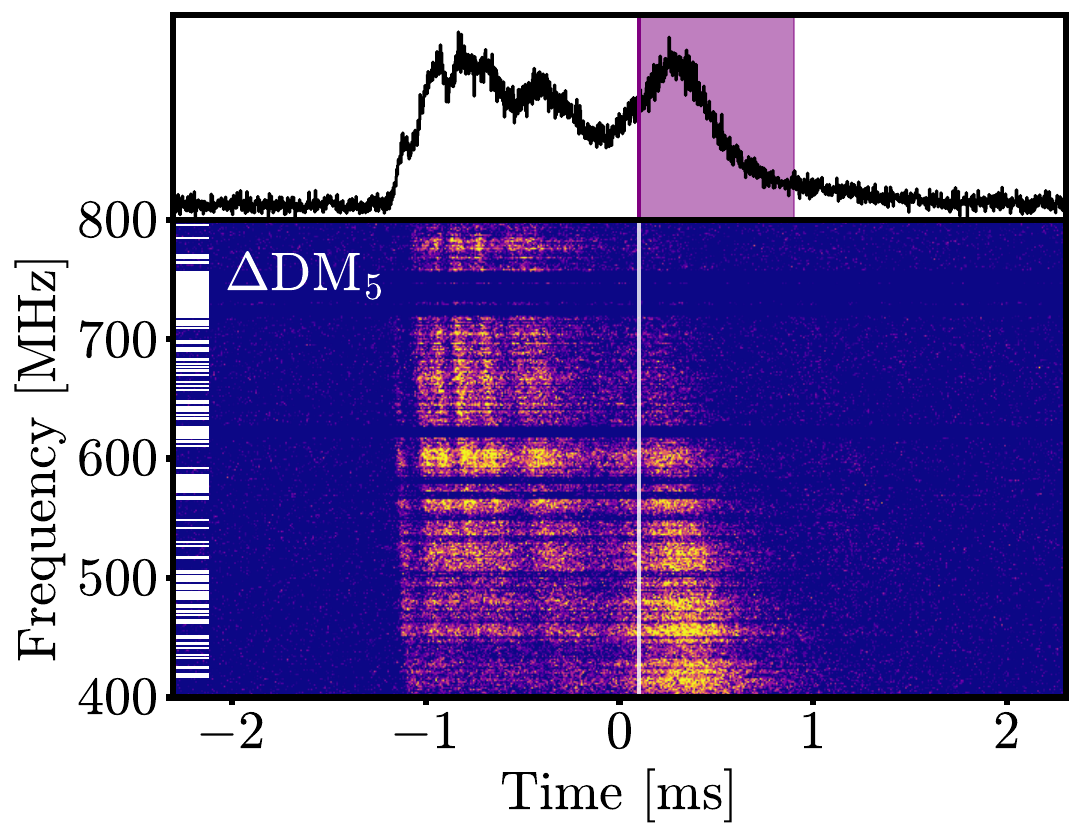}}\\
    \subfigure{\includegraphics[width = 0.3\textwidth]{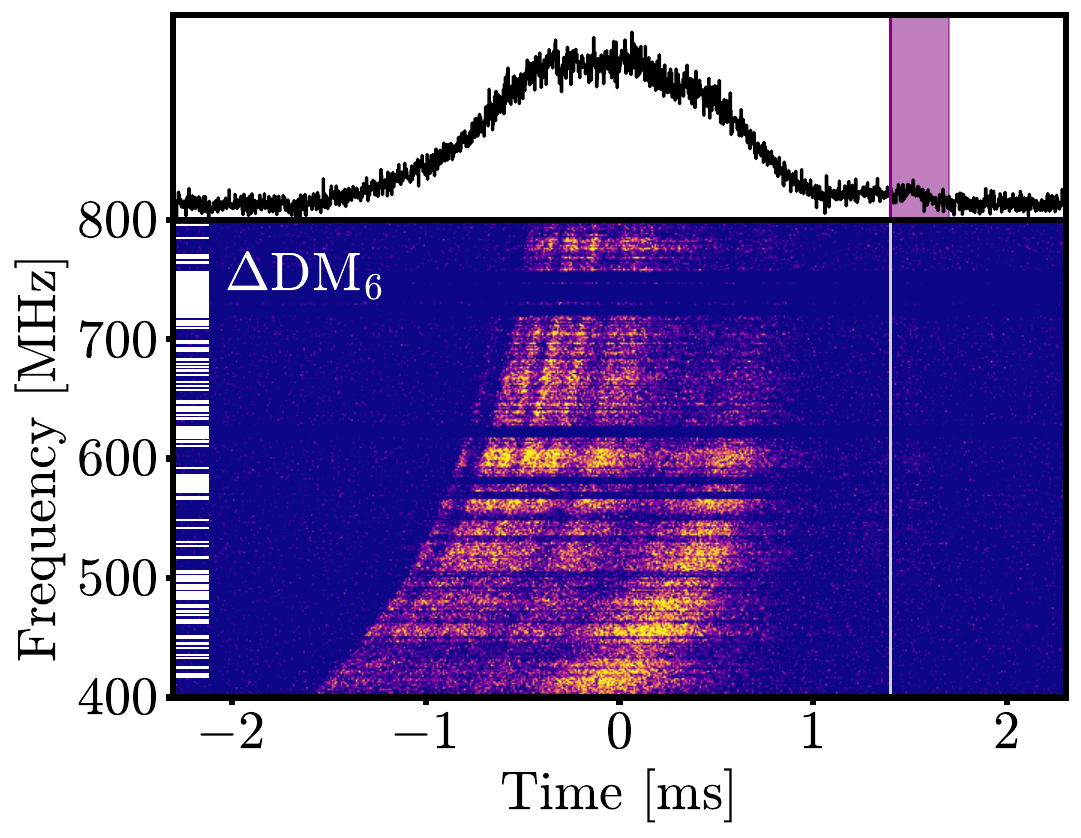}}
    \caption{The \textit{upper left} panel shows the dynamic spectrum of FRB 20211005A with DM curves over-plotted as white solid lines. The following panels show dynamic spectra of the same event dedispersed to the coherent power-maximizing $\Delta$DM offsets from the nominal DM for respective sub-bursts, highlighted in purple in the timeseries. The DMs to which the full burst is dedispersed are ordered by sub-burst ToA.}
    \label{fig:189845329_dmvar}
\end{figure}

Repeating the analysis in \S\protect\ref{subsec:nonlinposdrift}, we estimate dispersion measure variations in the dynamic spectra of FRB 20200603, FRB 20210427A, FRB 20210627, and FRB 20211005A for event-specific time-limited regions across the burst. 
With the exception of 
FRB 20210627A, we assume a focal frequency of 800 MHz, as the drifting in all other bursts appears to set in at or beyond this frequency. For FRB 20210627A, however, there appear to be two potential focal frequencies within the observing band, manifesting at $\nu_{l, 2} \approx$ 480 MHz for the first burst cluster and $\nu_{l, 1} \approx$ 650 MHz for the second. Hence we limit the frequency range over which we perform dedispersion to a range between 400 MHz and the focal frequency, as we did for FRB 20220413B. Below the focal frequency, drifting appears consistent with cold plasma dispersion. To highlight this, we plot DM curves adjacent to the drifting features in accordance with the estimated $\Delta$DM offsets from the nominal value, as shown in Figures \ref{fig:86321776_dmvar}, \ref{fig:169427924_dmvar}, \ref{fig:175128652_dmvar}, and \ref{fig:189845329_dmvar}, respectively. $\Delta$DM values measured for each event are recorded in Table \ref{table:dmvar}.

\clearpage
\section{Measuring Microstructure with Autocorrelation Functions}\label{apB}


Building on the analysis in \S\protect\ref{subsec:micro}, we re-measure the widths (FWHM) of $\lesssim 50 ~\mu s$ features (i.e., ``microstructure'') in seven of the twelve bursts in this sample. We now estimate sub-burst widths by calculating the ACF of each sub-burst or sub-burst cluster within a time-limited region of the full burst profile, and fitting a 1D Lorentzian function to the ACF. Prior to fitting, we mask the zero-lag spike in the ACF. Sub-burst clusters are defined as regions in the burst timeseries where peaks in intensity are not well-separated in time, but still show $\gtrsim 3 \sigma$ variability across an underlying envelope of intensity with respect to noise in the off-pulse region.

\begin{figure}[h!]
    \subfigure{\includegraphics[width = 0.24\textwidth]{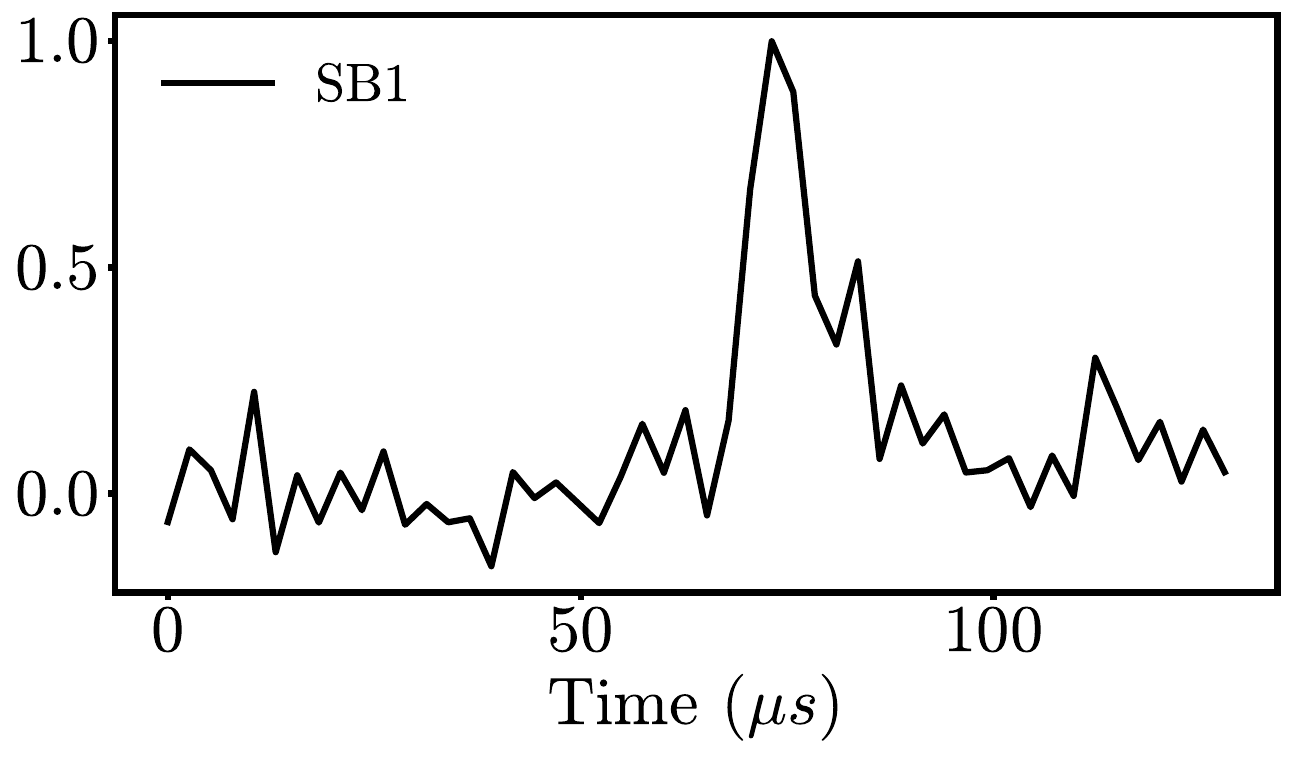}}
    \subfigure{\includegraphics[width = 0.24\textwidth]{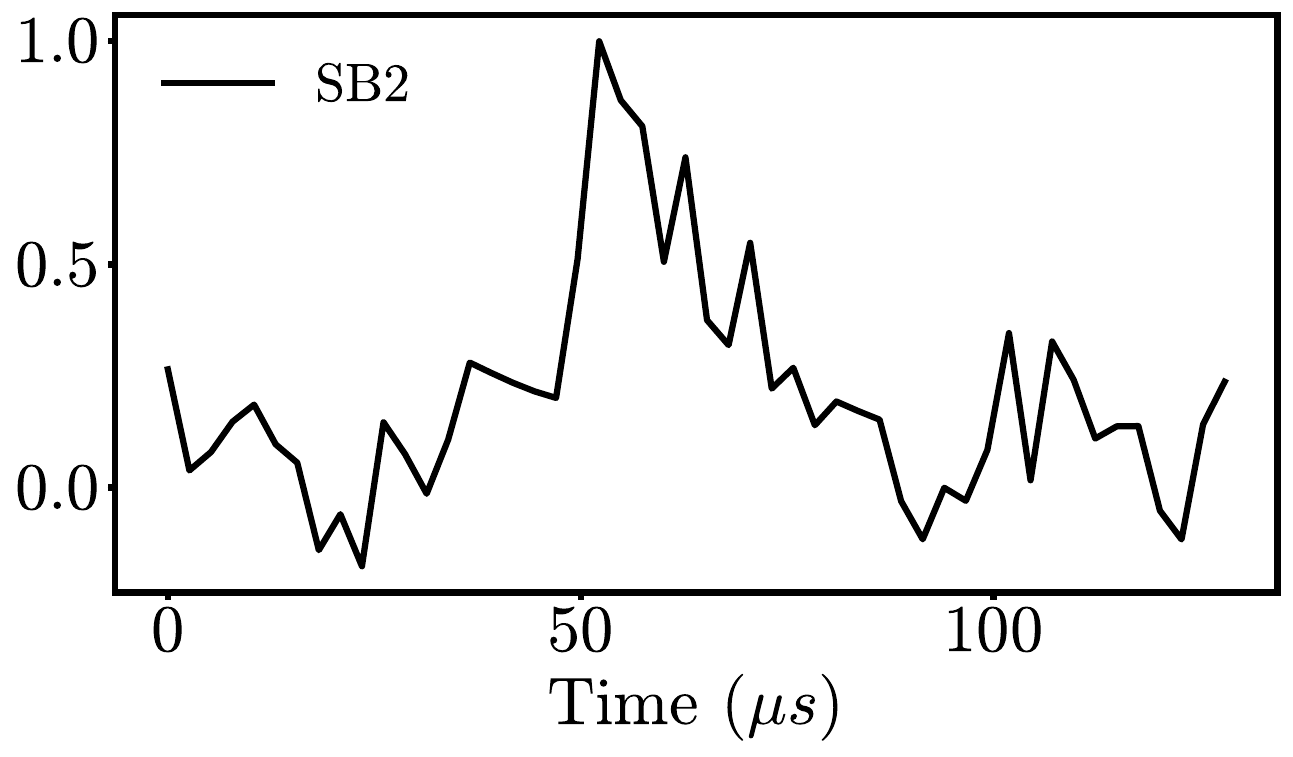}}
    \subfigure{\includegraphics[width = 0.24\textwidth]{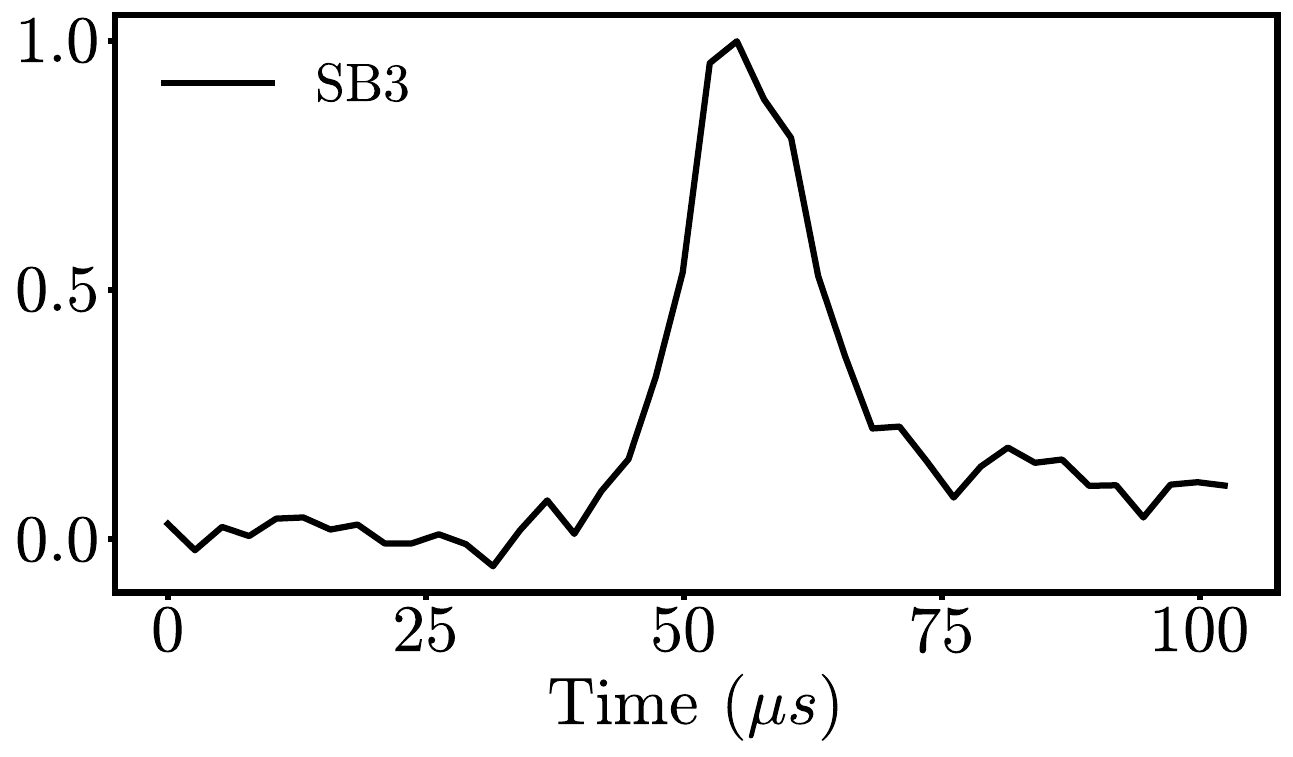}}
    \subfigure{\includegraphics[width = 0.24\textwidth]{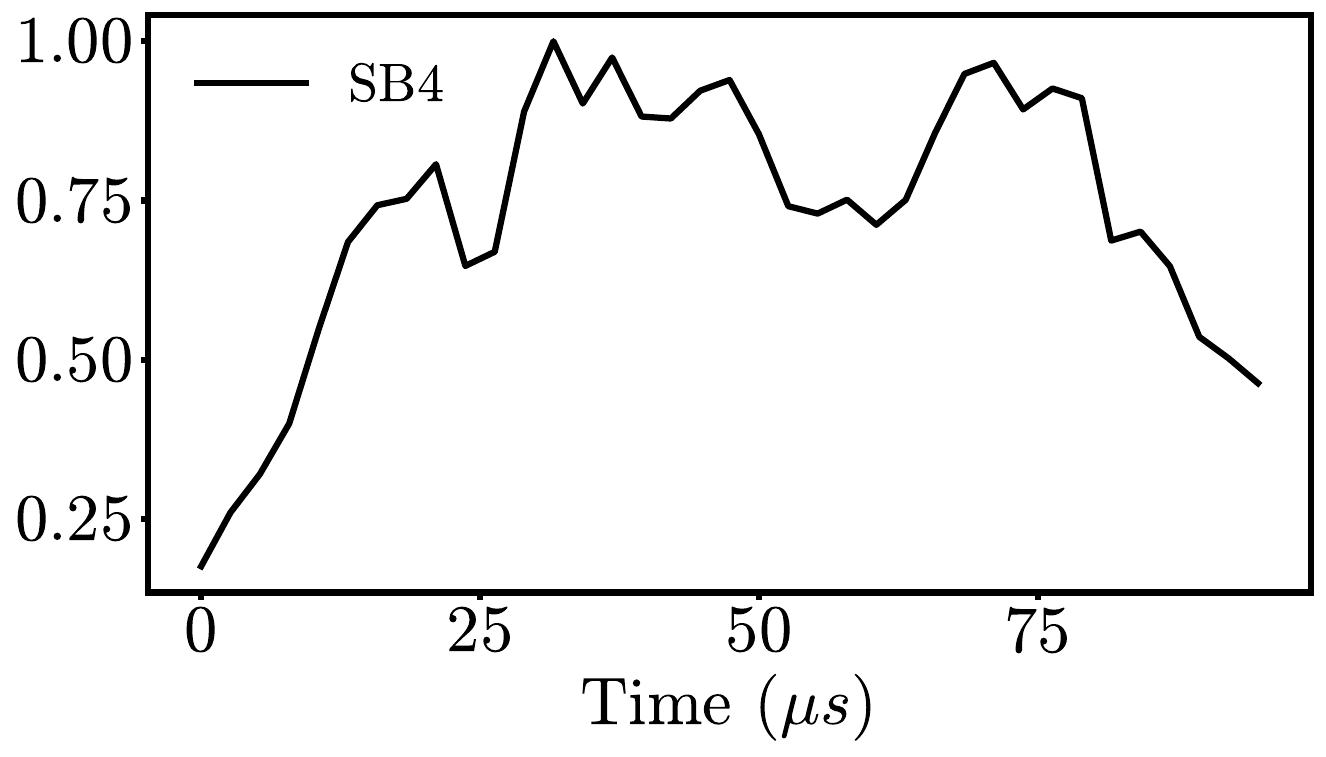}}\\
    \subfigure{\includegraphics[width = 0.24\textwidth]{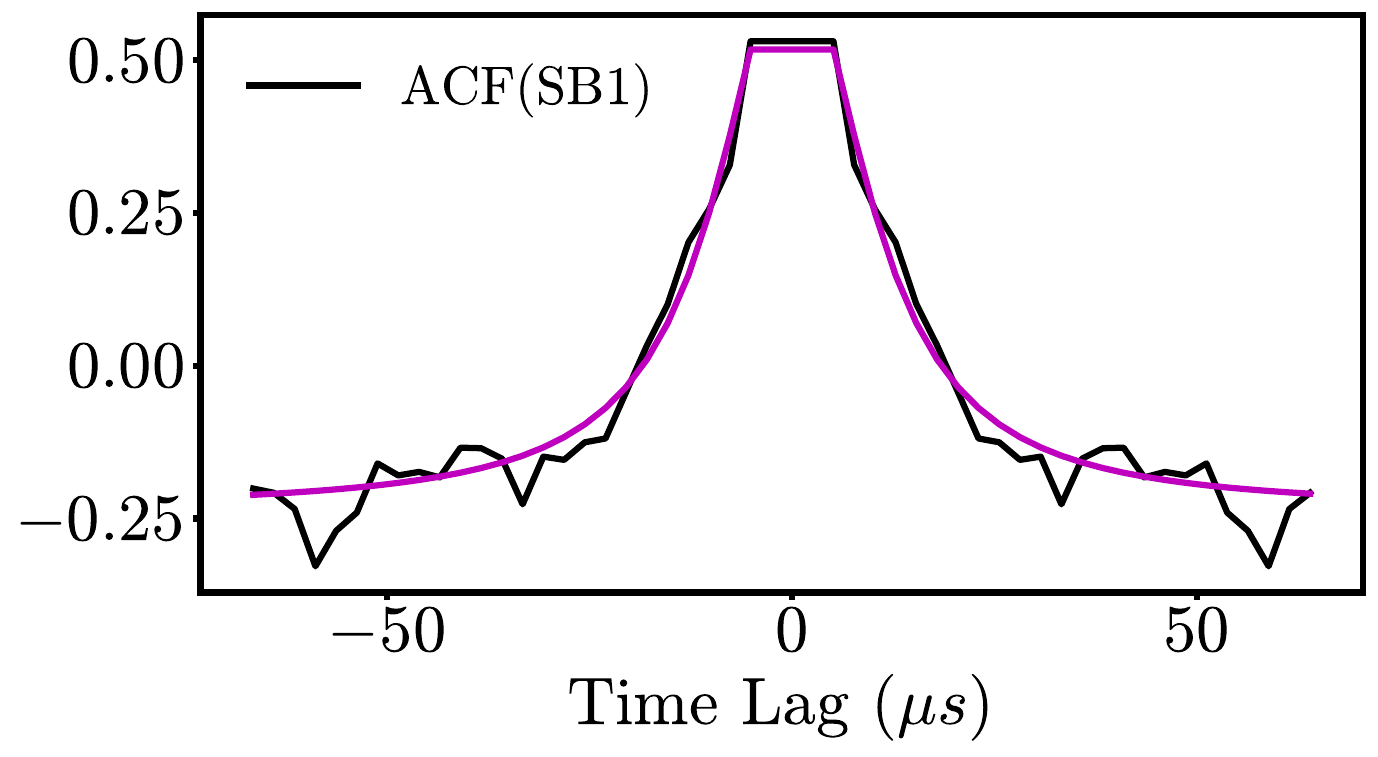}}
    \subfigure{\includegraphics[width = 0.24\textwidth]{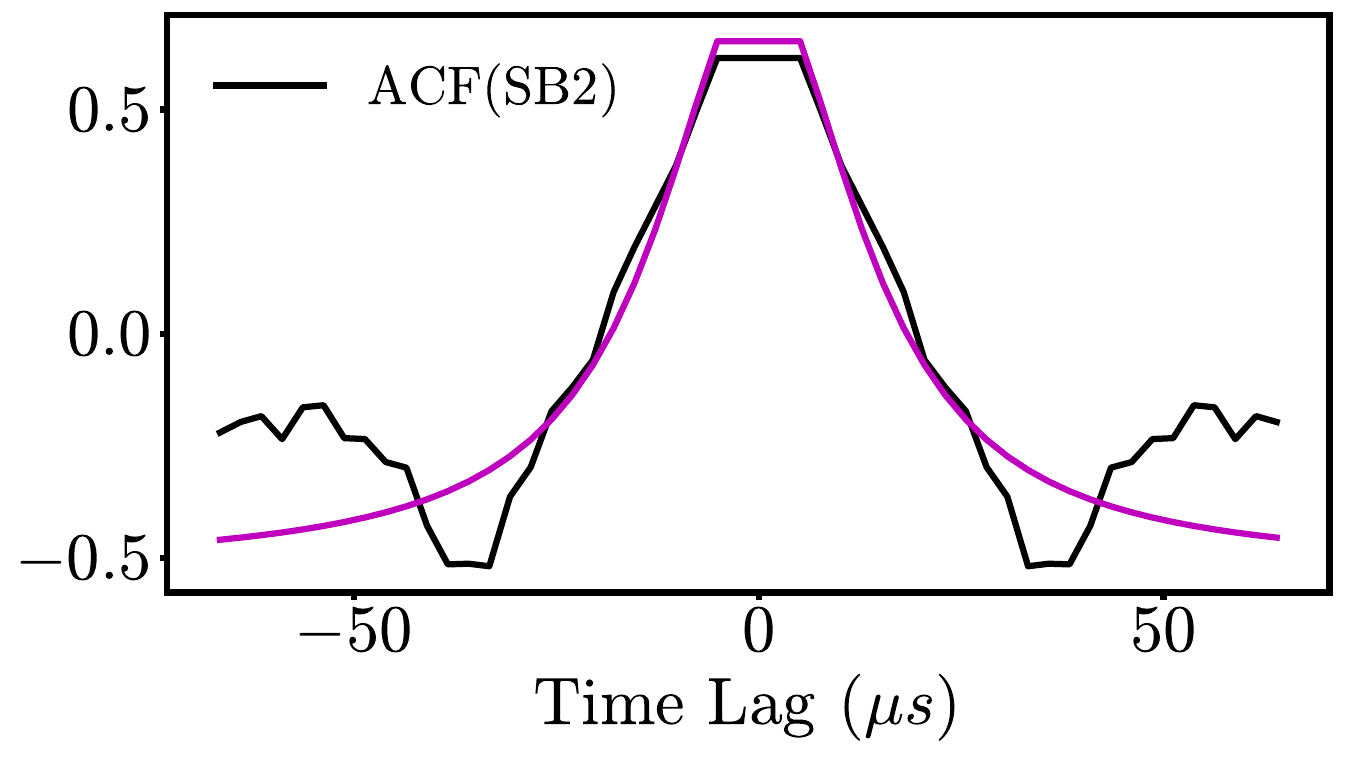}}
    \subfigure{\includegraphics[width = 0.24\textwidth]{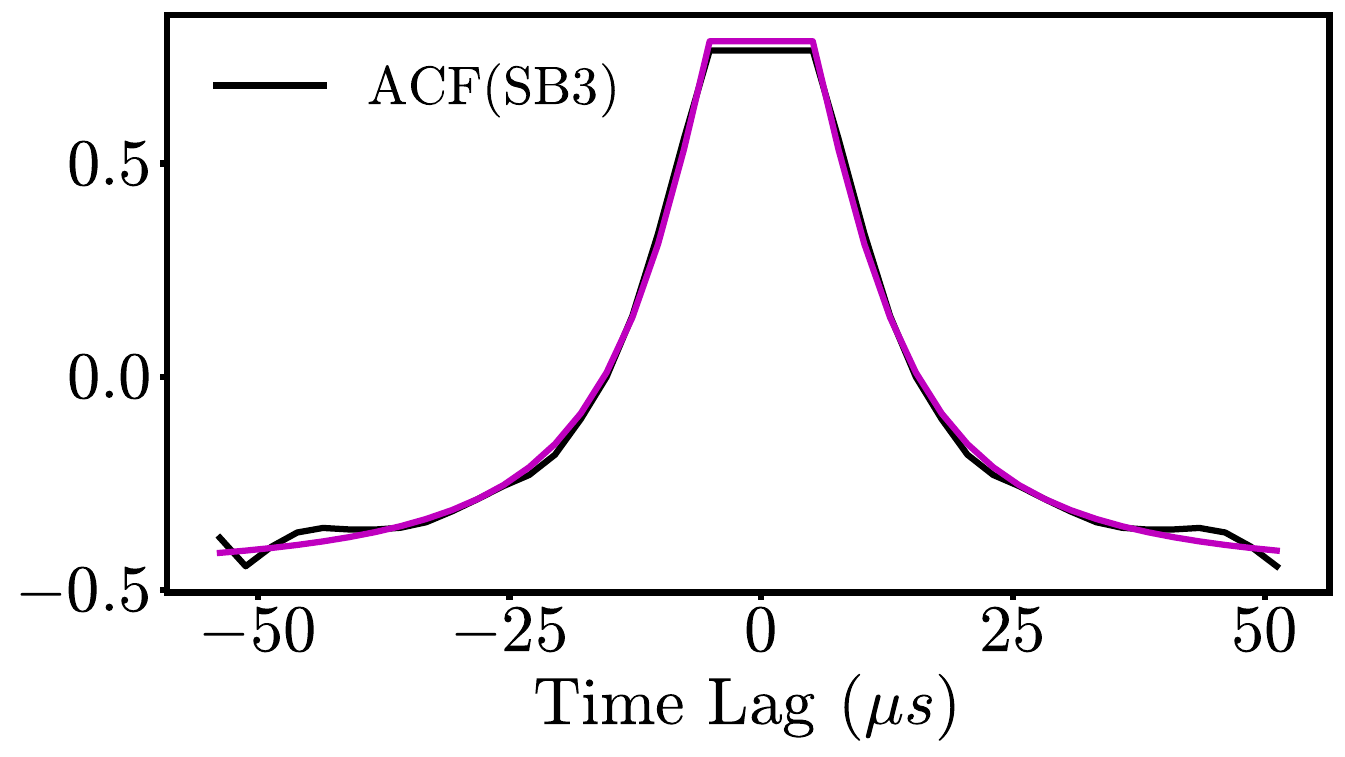}}
    \subfigure{\includegraphics[width = 0.24\textwidth]{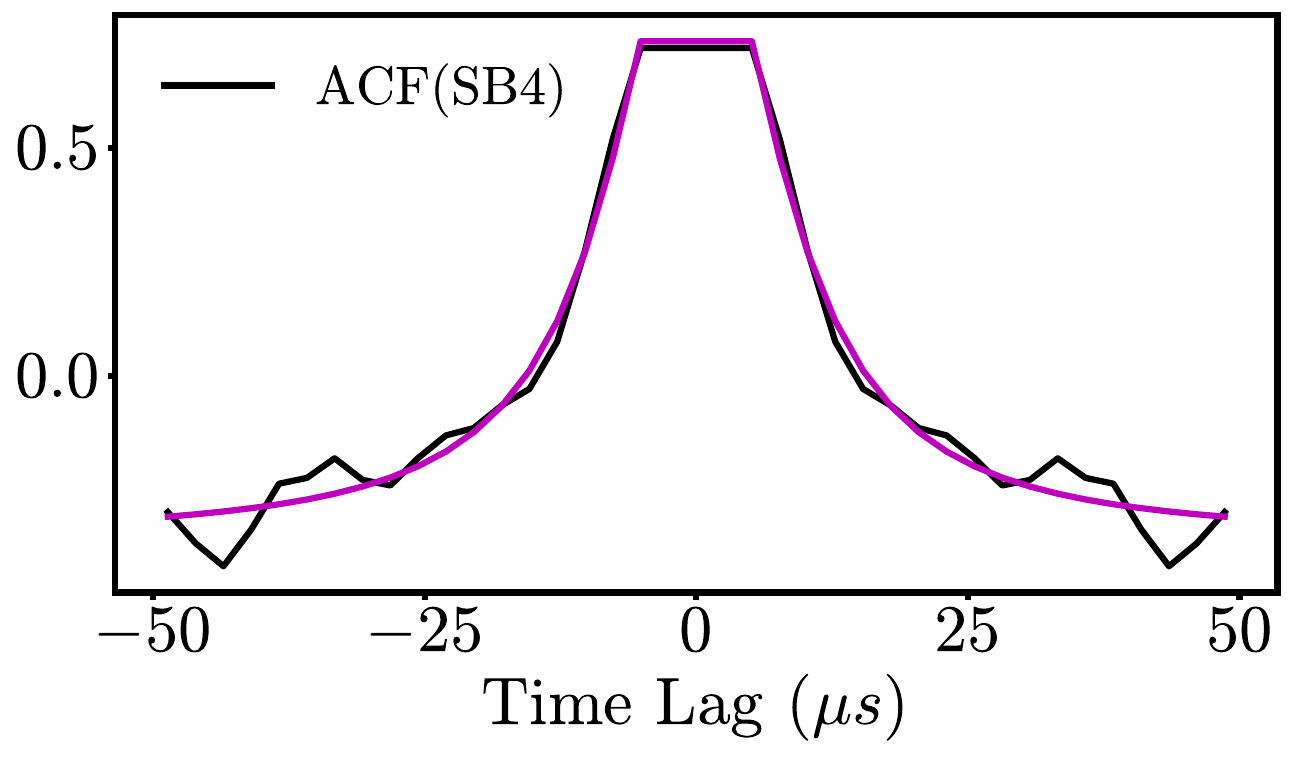}}\\
    \caption{Time-limited regions across the FRB 20190425A timeseries, isolating four regions containing either a sub-burst (in the three left-most panels) or sub-burst cluster (in the right-most panel). The ACFs are calculated and fitted with 1D Lorentzian functions to confirm that the widths are comparably narrow to those quoted in Table \ref{table:microgauss}.}
    \label{fig:37888771_micro}
\end{figure}

The measured widths for each sub-burst or sub-burst cluster are presented in Table \ref{table:microacf}. Figure \ref{fig:37888771_micro} shows an example of the respective timeseries and Lorentzian fits to the ACFs of sub-bursts and sub-burst clusters in the burst profile of FRB 20190425A. 
Certain width measurements obtained using this method are slightly broader than those measured in \S\protect\ref{subsec:micro} (see Table \ref{table:microgauss}), primarily due to the inability to completely isolate sub-bursts in clustered regions while still ensuring a reasonable fit to the ACF. The widths do, however, agree to well within an order of magnitude of those measured by a multi-Gaussian fit, validating the measurement technique outlined in \S\protect\ref{subsec:micro}, and confirming the intrinsically narrow ($\lesssim 50 ~\mu s$) timescales of these features.

\begin{table*}[ht]
\caption{The intrinsic widths (FWHM) of $\lesssim 50$-$\mu s$ sub-bursts observed in seven events in the sample. The widths are derived from 1D Lorentzian fits to ACFs calculated for individual sub-bursts and sub-burst clusters in time. An example of the time-limits placed on both isolated and clustered sub-bursts is shown in Figure \protect\ref{fig:37888771_micro}. Due to the clustering of sub-bursts in certain events, there are fewer $\Delta t$ measurements per event than in Table \protect\ref{table:microgauss}, as the multi-Gaussian fits performed in \S\protect\ref{subsec:micro} are better-suited to disentangle clusters into their individual components.} 
\centering  
\begin{tabular*}{1\textwidth}{@{\extracolsep{\fill}}c c c c c c c c} 
\hline\hline                       
{TNS Name} & $\Delta \mathrm{t}_1$ & $\Delta \mathrm{t}_2$ & $\Delta \mathrm{t}_3$ & $\Delta \mathrm{t}_4$ & $\Delta \mathrm{t}_5$ & $\Delta \mathrm{t}_6$\\   
\hline   
FRB 20190425A & 10.8(7) & 15(2) & 10.1(2) & 9.0(6) & ... & ... \\
FRB 20200603B & 9.7(2) & 29(1) & 49.2(7) & 17(7) & ... & ... \\
FRB 20210406E & 7.4(9) & 19(4) & 12.9(4) & 24.9(8) & 24.7(4) & 14.9(6) & \\
FRB 20210427A & 33(2) & ... & ... & ... & ... & ... \\
FRB 20210627A & 44.7(2) & 43.7(5) & ... & ... & ... & ...\\
FRB 20210813A & 18(2) & 20(3) & ... & ... & ... & ...\\
FRB 20211005A & 29(1) & ... & ... & ... & ... & ... \\
\hline\hline  \\                              
\end{tabular*}
\label{table:microacf}
\end{table*}

\clearpage
\section{Polarimetry}\label{apC:pol}

The CHIME/FRB polarization pipeline is part of the analysis stage of the CHIME/FRB baseband system, 
described in \S\protect\ref{sec:observations}.
Since baseband data retain full Stokes parameters and phase information, we are able to measure Faraday Rotation Measure (RM) using both RM synthesis and $QU$-fitting \citep[see methods outlined in][]{Mckinven2021}. RM synthesis is a non-parametric Fourier-like transformation method that gives an initial estimate of the RM. $QU$-fitting is a parametric technique that fits the modulations of Stokes $Q$ and $U$ using a nested sampling algorithm, which offers more flexibility than RM synthesis by the inclusion of parameters characterizing instrumental polarization. 

The model described by \citet{Mckinven2021} has 4 parameters: linear polarization fraction ($p$), polarization angle at zero wavelength ($\phi_0$), RM, and cable delay between the two linear polarizations ($\tau$; due to the relative path length differences between the two linear feeds of each antenna). However, this model was found to not entirely describe the instrumental phase between the X and Y polarizations. When unaccounted for, this differential phase can result in sign ambiguities in the RM as detailed in \citet{Mckinven2021}. The residual instrumental polarization can be adequately fit by including additional wavelength-independent parameters that account for the phase offset between the $X$ and $Y$ polarizations at zero frequency, $\phi_{\rm{lag}}$. Thus, in the polarization analysis shown here, we include 5 additional parameters on top of the original model to help constrain the RM sign. This more complex model considers the variation of linear and circular polarization fractions across bands. We include (1) a wavelength-independent parameter that acts as a constant offset ($\phi_{\rm{lag}}$), (2) a non-zero circular polarization fraction that follows a power law ($p_\nu$), (3) an exponent for the power law of $p$ ($\gamma_L$), (4) an exponent for $p_\nu$ ($\gamma_L$), and (5) a gain difference that considers the leakage from Stokes $I$ to $Q$. Furthermore, we limit the bounds of the cable delay and $\phi_{\rm{lag}}$ by constraining them to ranges of values they normally fall into. This method allowed us to confidently determine the RMs and linear polarization fractions of seven FRBs in our sample (see Table \ref{table:polprop}), all of which have $QU$-fitting results that are consistent with the signs obtained from RM synthesis.

\begin{table*}[b]
\caption{Polarization properties for selected bursts, including: measured RM ($\mathrm{RM}_{\text {obs }}$), the RM contribution expected from the Milky Way ($\mathrm{RM}_{\mathrm{MW}}$), and linear polarization ($\mathrm{L/I}$). The RM reported here is corrected for an ionospheric contribution of $+0.1$ rad m$^{-2}$. Estimates of the Galactic contribution to RM are drawn from the Faraday Sky map of \protect\citet{Hutschenreuter2022}, which infers $\mathrm{RM}_{\mathrm{MW}}$ values based on measurements of polarized extragalactic sources.}
\centering  
\begin{tabular*}{1.\textwidth}{@{\extracolsep{\fill}}c c c c c} 
\hline\hline     
{TNS Name} & {$\mathrm{RM}_{\mathrm{obs}}\left(\mathrm{rad} ~\mathrm{m}^{-2}\right)$} & {RM$_{\textrm{MW}}$ (rad m$^{-2}$)} & $\langle L/I\rangle$ & \\  
\hline   
FRB 20190425A & +57.30$_{-0.01}^{+0.01}$ & +48.6 $\pm$ 14.2 &  0.946$_{-0.003}^{+0.003}$ & \\
FRB 20191225A & $-$328.06$_{-0.02}^{+0.02}$ & $-$26.7(8.7) & 0.577$_{-0.002}^{+0.002}$ & \\
FRB 20200603B & ... & $-$18.7 $\pm$ 5.6 & ... & \\
FRB 20200711F & ... & +9.0 $\pm$ 1.7 & ... &   \\
FRB 20201230B & +80.63$_{-0.04}^{+0.04}$ & +92.1 $\pm$ 17.4 & 0.354$_{-0.002}^{+0.002}$ &  \\
FRB 20210406E & +77.609$_{-0.01}^{+0.01}$ & +17.7 $\pm$ 22.5 & 0.887$_{-0.002}^{+0.002}$ & \\
FRB 20210427A & ... & +5.3 $\pm$ 8.4 & ... & \\
FRB 20210627A & $-$28.327$_{-0.01}^{+0.01}$&$-$4.9 $\pm$ 3.5 & 0.82$_{-0.01}^{+0.01}$ &  \\
FRB 20210813A & ... & $-$25.4 $\pm$ 30.4 & ... &  \\
FRB 20210819A & +32.23$_{-0.04}^{+0.04}$ & $-$7.2 $\pm$ 2.8 & 0.340$_{-0.001}^{+0.001}$ & \\
FRB 20211005A & +17.24$_{-0.02}^{+0.02}$ & 17.0 $\pm$ 5.5 & 0.887$_{-0.002}^{+0.002}$ & \\
FRB 20220413B & ... & +36.4 $\pm$ 7.2 & ... & \\
\hline\hline  \\  
\end{tabular*}
\label{table:polprop}
\end{table*}

\end{document}